\documentclass[journal]{IEEEtran}

\usepackage{graphicx, amssymb, amsmath, amsfonts, amsthm}
\usepackage{subfigure, mathtools, bm, dsfont, siunitx, cite}

\newcommand{\E}{\mathbb{E}}
\newcommand{\Pt}{P_\text{tx}}
\newcommand{\Nt}{\widetilde{N}}
\newcommand{\PP}{\mathbb{P}}

\newcommand{\s}{\sigma}

\DeclareSIUnit{\belmilliwatt}{Bm}
\DeclareSIUnit{\belisotropic}{Bi}
\DeclareSIUnit{\dBm}{\deci\belmilliwatt}
\DeclareSIUnit{\dBi}{\deci\belisotropic}

\allowdisplaybreaks
\IEEEoverridecommandlockouts
\title{Wireless Information and Energy Transfer\\ in the Era of 6G Communications}

\author{Constantinos Psomas, \IEEEmembership{Senior Member, IEEE}, Konstantinos Ntougias, \IEEEmembership{Member, IEEE}, Nikita Shanin, \IEEEmembership{Graduate Student Member, IEEE}, Dongfang Xu, \IEEEmembership{Member, IEEE}, Kenneth Mayer, \IEEEmembership{Student Member, IEEE}, Nguyen Minh Tran, \IEEEmembership{Graduate Student Member, IEEE}, Laura Cottatellucci, \IEEEmembership{Member, IEEE}, Kae Won Choi, \IEEEmembership{Senior Member, IEEE}, Dong In Kim, \IEEEmembership{Fellow, IEEE}, Robert Schober, \IEEEmembership{Fellow, IEEE}, and Ioannis Krikidis, \IEEEmembership{Fellow, IEEE}
	
\thanks{C. Psomas, K. Ntougias, and I. Krikidis are with the Department of Electrical and Computer Engineering, University of Cyprus, Nicosia, Cyprus (email: \{psomas, ntougias.konstantinos, krikidis\}@ucy.ac.cy).\\
N. Shanin, K. Mayer, L. Cottatellucci, and R. Schober are with the Institute for Digital Communications, Department of Electrical Engineering, Friedrich-Alexander Universität Erlangen-Nürnberg, Germany (email: \{nikita.shanin, kenneth.m.mayer, laura.cottatellucci, robert.schober\}@fau.de).\\
D. Xu is with the Academy of Interdisciplinary Studies, The Hong Kong University of Science and Technology (email: eedxu@ust.hk).\\
N. M. Tran was with the Department of Electrical and Computer Engineering, School of Information and Communication Engineering, Sungkyunkwan University, Suwon, Korea, at the time of writing this paper. He is now with the Department of Wireless Communications, Faculty of Electronics and Telecommunications, VNU - University of Engineering and Technology, Hanoi, Vietnam (email: minhtran.nguyen@vnu.edu.vn).\\
K. W. Choi, and D. I. Kim are with the Department of Electrical and Computer Engineering, Sungkyunkwan University, Suwon, Korea (email: kaewonchoi@skku.edu, dikim@skku.ac.kr).

This work has received funding from the European Research Council (ERC) under the European Union's Horizon 2020 research and innovation programme (Grant agreement No. 819819) and the European Union's Horizon Europe programme (ERC, WAVE, Grant agreement No. 101112697). Views and opinions expressed are however those of the author(s) only and do not necessarily reflect those of the European Union or the European Research Council Executive Agency. Neither the European Union nor the granting authority can be held responsible for them. This work was also funded by the German Ministry for Education and Research (BMBF) under the program of ``Souverän. Digital. Vernetzt.'' joint project 6G-RIC (Project-ID 16KISK023) and the Deutsche Forschungsgemeinschaft (DFG, German Research Foundation) under projects SFB 1483 (Project-ID 442419336, EmpkinS). Furthermore, this work was supported by the National Research Foundation of Korea (NRF) grant funded by the Korean Government (MSIT) under Grants 2021R1A2C2007638 and 2022R1A4A1033830.}}

\begin{document}
	
\maketitle
	
\begin{abstract}
Wireless information and energy transfer (WIET) represents an emerging paradigm which employs controllable transmission of radio-frequency signals for the dual purpose of data communication and wireless charging. As such, WIET is widely regarded as an enabler of envisioned 6G use cases that rely on energy-sustainable Internet-of-Things (IoT) networks, such as smart cities and smart grids. Meeting the quality-of-service demands of WIET, in terms of both data transfer and power delivery, requires effective co-design of the information and energy signals. In this article, we present the main principles and design aspects of WIET, focusing on its integration in 6G networks. First, we discuss how conventional communication notions such as resource allocation and waveform design need to be revisited in the context of WIET. Next, we consider various candidate 6G technologies that can boost WIET efficiency, namely, holographic multiple-input multiple-output, near-field beamforming, terahertz communication, intelligent reflecting surfaces (IRSs), and reconfigurable (fluid) antenna arrays. We introduce respective WIET design methods, analyze the promising performance gains of these WIET systems, and discuss challenges, open issues, and future research directions. Finally, a near-field energy beamforming scheme and a power-based IRS beamforming algorithm are experimentally validated using a wireless energy transfer testbed. The vision of WIET in communication systems has been gaining momentum in recent years, with constant progress with respect to theoretical but also practical aspects. The comprehensive overview of the state of the art of WIET presented in this paper highlights the potentials of WIET systems as well as their overall benefits in 6G networks.
\end{abstract}

	
\section{Introduction}\label{sec:intro}
Wireless connectivity has been pivotal in boosting the economy and improving the quality of life over the last decades~\cite{Eurostat,COVID}. The 5th Generation (5G) standard for radio communications represents the latest milestone in the evolution of wireless networks. Its recent advent, which has been jointly driven by technological advances and market needs, resulted in the introduction of a multitude of Internet-of-Things (IoT) services across diverse vertical industries, ranging from agriculture to manufacturing \cite{5GUseCases2020}. The digital transition of these industrial sectors has enhanced both their productivity and their operational efficiency~\cite{WEFWP}. This has been translated into financial gains as well as cost and energy consumption savings.

Nevertheless, as societies become increasingly data-centric and automation-dependent, the demand for supporting advanced use cases with extreme requirements that exceed the 5G capabilities and expand digitalization to new verticals, such as transportation~\cite{liu20236g} and energy, is rapidly growing. Furthermore, 5G struggles to effectively cope with the accumulating effects of societal challenges, such as the digital divide~\cite{DigitalDivide}, climate change~\cite{ClimateChange}, and the energy crisis~\cite{EnergyCrisis}, since its design has not focused on meeting sustainability requirements and societal values. The inability of 5G to fully realize the digitalization vision and address the rising sustainability issues and societal needs calls for a paradigm shift centered around disruptive technological enablers. Consequently, several Research \& Innovation initiatives and strategic programs on 6G technology have been launched around the world, e.g., in the USA \cite{NGA}, China \cite{IMT-2030}, Japan \cite{RCRJapan}, South Korea \cite{RCRKorea}, India \cite{6GIndia}, and the European Union \cite{SNS,Hexa-X,Hexa-X-II}, although the initial 6G network deployment activities are expected to begin only around 2030.

6G will target substantial performance improvements over its predecessor in terms of data rates, latency, connections density, coverage, energy efficiency, and reliability, in order to support novel service classes and empower additional verticals. Also, it will aim at accelerating the green transition and steering the digital transformation of the economy and the society towards sustainability, in line with key United Nations’ Sustainable Development Goals (SDG) \cite{SDG} and relevant environmental policies, such as the European Green Deal \cite{EGD}. Specifically, 6G will integrate aerial and satellite networks with terrestrial networks \cite{6GHua}, in order to offer seamless, ubiquitous connectivity in a cost-effective manner, thereby mitigating the socio-economic imbalances attributed to the digital divide. Moreover, next-generation networks will embed artificial intelligence (AI)/machine learning (ML) into their architecture \cite{6GHua}. AI-driven optimization can minimize the energy consumption of 6G networks, hence increasing their energy efficiency~\cite{AIEE,AIEE2}. Furthermore, 6G will support ultra-massive connectivity, extreme reliability, and sub-ms latency, thus paving the way to a number of IoT applications beyond the 5G scope \cite{6GHua}, such as smart grid, smart utility management, smart cities, and intelligent transportation systems. These services enable real-time tracking and optimization of energy distribution and utility usage, pollution monitoring and smart waste management, and optimal traffic management to lower operational and utility costs, enhance efficiency and resilience, improve safety, and reduce global carbon emissions by up to 20\% by 2050 \cite{WEF}.

While the number of IoT connections will exceed 30 billion already by 2030 according to recent estimations \cite{iot-analytics}, this figure is expected to explode in the 6G era, as the demand for wireless connectivity is escalating and new applications are introduced. Indeed, the current trends in the development of 6G communication networks suggest that, in the very near future, IoT networks will be widely utilized by wearable devices for continuous health monitoring, for wireless on-chip communication, and in industrial applications, where the IoT devices are expected to support extremely high data rates exceeding $\SI{100}{\giga\bit\per\second}$ \cite{Akyildiz2015, Yi2021}. Other applications include the smart agriculture sector (e.g., for monitoring the health and location of livestock or for monitoring environmental parameters to increase crop yields by adapting accordingly the use of water, nutrients, fertilizers, and pesticides), transportation and logistics (e.g., for monitoring the quality of goods in transit or the inventory stock in warehouses and maximizing the efficiency of supply chains), smart homes (e.g., for monitoring temperature and energy usage) \cite{Vaezi2022}. As a result, the effective realization of future networks heavily depends on technological advancements that enable such extreme connectivity. Common examples of relevant technologies include intelligent reflecting surfaces (IRSs) (also known as reconfigurable intelligent surfaces), terahertz (THz) communications, and holographic multiple-input multiple-output (H-MIMO) antenna systems \cite{Tataria2021}. 

Furthermore, the communication terminals in the considered IoT setups are typically energy-constrained devices, since they are powered by batteries with finite energy storage capacity which support their energy demands in terms of sensing, processing, and transmission. The limited energy resources of the IoT terminals and the requirement for ensuring uninterrupted service provision (and, therefore, continuous device operation) dictate frequent recharging of these nodes. However, their excessive number and the fact that they are often embedded in infrastructure (e.g., walls, vehicles), placed in remote locations (e.g., forests), or installed in areas with restricted/limited access (e.g., industrial facilities) render their wired charging a costly and cumbersome task. On the other hand, ``dead'' nodes can significantly degrade the quality-of-service (QoS). Hence, the development of techniques that prolong the lifetime of these devices is of utmost importance.

The utilization of renewable energy sources, such as solar energy, and the harvesting of ambient radio frequency (RF) energy represent prevailing strategies for extending the lifetime of these devices \cite{WIPTBook}. Nonetheless, these methods present also some drawbacks. Specifically, solar panels require a non-negligible area, which in many cases does not suit the small form factor of compact IoT terminals, such as wireless sensors, or the size limitations imposed by the operational environment (e.g., industrial plants). In addition, they significantly increase the cost. This might be particularly problematic for massive IoT deployments. Perhaps most importantly, both aforementioned approaches are characterized by intermittent and uncontrollable power supply. This opportunistic operation reduces the amount of harvested power in practice, compared to what could be achieved in theory, and raises obstacles for QoS provisioning, which is critical in 6G. Therefore, the notion of harvesting energy from controllable RF sources, that is, the so called wireless energy transfer (WET) paradigm, is becoming an increasingly attractive solution \cite{lu2016}. The main principle of this promising technology is rather simple. Specifically, RF sources transmit energy signals to target devices which are equipped, in turn, with a rectifying antenna (rectenna) circuit that converts the received RF energy to direct current (DC) energy. As such, RF-EH under WET is suitable for replenishing the energy stored at the battery of wireless sensors and other low-power IoT devices in a controllable manner, thereby providing QoS guarantees without substantially increasing the cost and form factor. On the other hand, renewable energy sources, such as solar panels, are typically capable of harvesting much more energy than RF-WET, despite the unavoidable ``no harvesting'' periods mentioned above. Hence, while WET is considered indeed a key component of future sustainable IoT networks, there is a consensus among the industrial and academic communities that the achievement of energy sustainability requires a holistic approach. In particular, WET, along with energy efficiency maximization techniques, is expected to \textit{complement} renewable energy sources and ambient RF-EH \cite{10285066} and provides an alternative to them in extreme scenarios, e.g., when stringent cost or size limitations prohibit the extensive deployment of solar panels or/and in rural/remote areas, where ambient RF signals originating, for example, from nearby cell towers or TV broadcasting stations, are scarce.

\begin{figure*}[t]\centering
\includegraphics[width=0.91\linewidth]{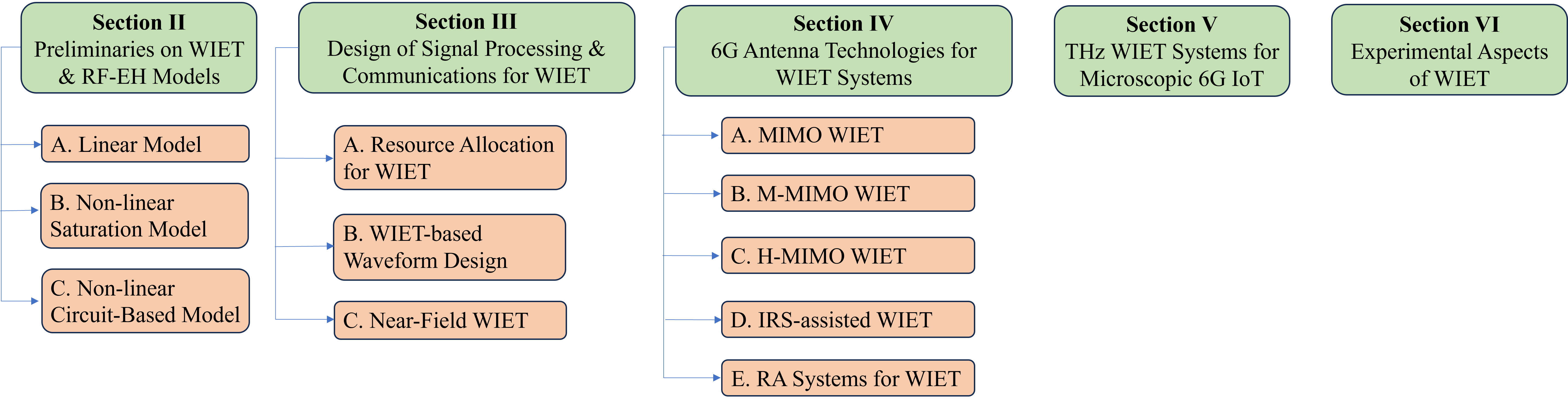}
\caption{Structure and organization of this paper.}\label{fig:2}
\end{figure*}

The integration of WET in radio communication systems gives rise to the wireless information and energy transfer (WIET) paradigm, wherein information and energy signals coexist \cite{krikidis2023}. Through proper co-design of these signals, together with the utilization of emerging 6G technologies \cite{Tataria2021}, the stringent QoS demands of WIET for both data transfer and power delivery can be satisfied, therefore paving the way for the energy-sustainable networks of the future. Specifically, the use of compact high-directivity THz antennas and the exploitation of the huge frequency spectrum available in the THz frequency band allow for both rapid wireless charging of the batteries of small IoT devices \cite{Kuscu2021, akyildiz2019, Akyildiz2015} and ultra-high data rate communication over short distances, whereas the extreme density of antenna elements in H-MIMO implementations enhances beam directionality and data multiplexing, thus improving the efficiency of WET and increasing network capacity. In addition, the combination of large antenna arrays with high carrier frequencies renders the radiative near-field region relevant in 6G networks, e.g., in indoor applications, and gives rise to beam focusing, which concentrates the radiated energy at a specific point instead of a given direction \cite{NFBF, 8957130}. Therefore, near-field beam focusing significantly enhances the efficiency of energy transfer. IRSs, on the other hand, reflect incoming RF signals in a controllable manner to redirect them towards target devices even when the direct path is blocked or to suppress interference in other directions, thus enabling more flexible and efficient energy transfer. Likewise, reconfigurable fluid (or movable) antenna arrays can adjust their topology and electrical characteristics to flexibly optimize data communication and energy transfer \cite{kit2023a, kit2023b, psomas2023}.

\subsection{Contributions and Comparison with Related Works}\label{subsec:1.1}
WIET is a key enabler of 6G IoT. This communication paradigm provides seamless energy transfer alongside data connectivity to IoT nodes in remote areas and enables the realization of energy-sustainable smart environments (e.g., smart homes/offices, cities, and industries), thereby revolutionizing a wide range of industrial sectors and assisting in addressing core sustainability issues and societal challenges. On the other hand, an efficient implementation of 6G-WIET systems not only requires a rethinking of certain communication notions such as resource allocation and waveform design, but also depends on the underlying technologies. Therefore, understanding the unique characteristics, advantages, opportunities and challenges associated with these technologies is critical for the design and performance evaluation of WIET in 6G networks.

There are several surveys and tutorials on WIET, dealing with a variety of topics. These include, for instance, the different WIET variants and protocols, EH models, and EH circuit implementations, as well as waveform design, beamforming optimization, and resource allocation in different network setups (e.g., cooperative networks, cognitive radio networks, cellular networks, etc.) \cite{Clerckx2022, lu2016, KuEHSurv2016, AlsabaBFsurv2018, PereraSWIPTsurv2018, Clerckx2019}. Nevertheless, although WIET has been extensively studied in the literature, to the best of our knowledge, a comprehensive overview of integrated 6G-WIET networks is not available. This article aspires to fill in this gap, presenting the main principles, design aspects, research directions, and challenges associated with the employment of WIET in 6G networks. In particular, the contributions of this work are as follows:
\begin{itemize}
	\item Foundational elements of communication systems employing WIET, including resource allocation, EH circuits, and near-field propagation characteristics, are discussed and efficient WIET designs for 6G networks are provided.
	\item The benefits of new 6G antenna technologies for WIET systems, such as H-MIMO, IRSs, and reconfigurable antennas (RAs) (specifically, fluid antennas) are investigated by exploiting their large size and reconfigurability.
	\item WIET over THz frequency bands is examined, where the potential gains and expected challenges associated with its application for supporting microscopic 6G IoT devices are unveiled.
	\item Finally, a testbed is presented and experimental WIET results are reported based on the implementation of two algorithms. The first algorithm facilitates near-field energy beamforming to support WIET between a phased-array-equipped transmitter (TX) and a receiver (RX), whereas the second one targets power-based IRS beamforming.
\end{itemize}
The 6G technologies studied in this paper enable the co-existence and the management of information and energy signals towards the realization of WIET-based 6G networks.

\subsection{Structure}\label{subsec:1.2}
The rest of this article is organized as follows. Section \ref{sec:models} provides some preliminaries on WIET systems together with the main RF-EH models that exist in the literature. Section \ref{sec:techniques} presents signal processing and communication aspects of WIET, with focus on resource allocation, waveform design, and near-field WIET. Section \ref{sec:antenna} introduces advanced 6G antenna technologies for WIET systems, namely, H-MIMO, IRS, and RAs, and describes their potential gains. Section \ref{sec:thz} discusses the impact of THz communication in WIET systems. Finally, Section \ref{sec:experimental} provides experimental studies on WIET and Section \ref{sec:conc} concludes the article with a discussion on several interesting future research directions. Fig. \ref{fig:2} schematically depicts the paper's structure.

\textit{Notation and Abbreviations:} Boldface lower-case and upper-case letters denote vectors and matrices, respectively. $\PP\{X\}$ and $\E\{X\}$ refer to the probability and the expected value of $X$, respectively; $(\cdot)^\ast$, $(\cdot)^\top$, and $(\cdot)^\text{H}$ denotes the conjugate, the transpose, and the conjugate transpose of a vector or a matrix, respectively; $\mathbf{I}_N$ denotes an $N\times N$ identity matrix; $\mathrm{Tr}(\cdot)$ is the trace of a matrix; $\mathbb{C}^{M\times N}$ represents the set of $M \times N$ complex matrices; $\mathbb{H}^{N}$ denotes the space of $N\times N$ Hermitian matrices; $\| \cdot\|$ is the Euclidean norm; $\mathcal{CN}(0,\sigma^2)$ refers to the distribution of a circularly symmetric complex Gaussian random variable with zero mean and variance $\sigma^2$; $j$ denotes the imaginary unit, $|\cdot|$ is the absolute value of a complex scalar, and the real part of a complex value is denoted by $\Re \{\cdot\}$; $W_0(\cdot)$ and $I_0(\cdot)$ are the principal branch of the LambertW function and the zeroth-order modified Bessel function of the first kind, respectively. A list of abbreviations that are frequently used in this paper is provided in Table \ref{table_abr}.

\renewcommand{\arraystretch}{1.1}
\begin{table}\caption{List of Frequently Used Abbreviations}\label{table_abr}\centering
	\begin{tabular}{|l|l|}\hline
		\textbf{Abbreviation} & \textbf{Description}\\\hline\hline
		CSI & Channel State Information\\\hline
		DC & Direct Current\\\hline
		EH & Energy Harvesting\\\hline
		H-MIMO & Holographic Multiple-Input Multiple-Output\\\hline
		ID & Information Decoding\\\hline
		IoT & Internet-of-Things\\\hline
		IRS & Intelligent Reflecting Surface\\\hline
		MIMO & Multiple-Input Multiple-Output\\\hline
		PAPR & Peak-to-Average-Power Ratio\\\hline
		PS & Power Splitting\\\hline
		QoS & Quality of Service\\\hline
		RA & Reconfigurable Antenna\\\hline
		RF & Radio Frequency\\\hline
		RX & Receiver\\\hline
		SNR & Signal-to-Noise Ratio\\\hline
		SINR & Signal-to-Interference-plus-Noise Ratio\\\hline
		SWIPT & Simultaneous Wireless Information and Power Transfer\\\hline
		TS & Time Switching\\\hline
		TX & Transmitter\\\hline
		WET & Wireless Energy Transfer\\\hline
		WIET & Wireless Information and Energy Transfer\\\hline
		WIT & Wireless Information Transfer\\\hline
	\end{tabular}
\end{table}

\section{Preliminaries on WIET \& RF-EH Models}\label{sec:models}
\graphicspath{{models/}}
There are two types of WIET communication systems, namely, wireless powered communication networks (WPCNs) and simultaneous wireless information and power transfer (SWIPT). A WPCN is a WIET system where the RXs first harvest energy from a TX's RF signals and then use it to power their uplink transmissions \cite{Clerckx2022}. In contrast, in the SWIPT architecture, a TX conveys both information and energy simultaneously to a single RX or multiple RXs \cite{krikidis2023}. In this paper, we focus on SWIPT systems but the main principles that are discussed can also be applied to WPCNs.

Now, in order to decode information and harvest energy from the same RF signal, the SWIPT RXs can employ either a separate or an integrated RX architecture \cite{Zhou2013}. The separate RX architecture splits the received signal into two streams in the RF domain, where one stream is directed to an information decoding (ID) circuit and the other one to an EH circuit. The separation can be realized by time switching (TS), power splitting (PS), or antenna switching (AS) \cite{Zhang2013}. On the other hand, an integrated RX rectifies the entire RF signal and uses the output DC for both ID and EH \cite{Zhou2013, Smida2021}. In this way, the integrated architecture can maximize the RF-EH compared to the separate RX. However, the output signal after the rectification process differs from the original input signal, which negatively affects the ID performance. Therefore, the integrated RX is more suitable for low-power and low-cost applications, as a traditional RF chain for ID is avoided, whereas the separate RX is more suitable for higher data rate applications \cite{Clerckx2022}.

\begin{figure}[t]\centering
	\includegraphics[width=\linewidth]{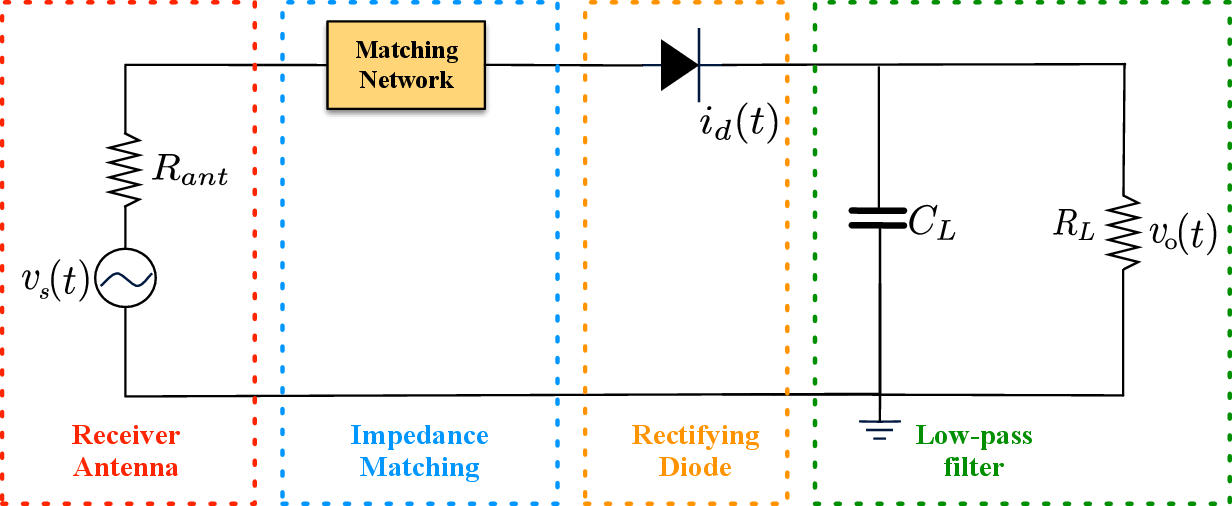}
	\caption{A simple rectenna circuit for RF-EH.}
	\label{fig:rectenna}
\end{figure}

For the implementation of RF-EH, the RX is equipped with a rectenna. In its simplest form, a rectenna consists of an antenna, for receiving the transmitted RF signal, together with a matching network, a rectifying diode and a passive low-pass filter (LPF), for converting RF to DC, as depicted in Fig. \ref{fig:rectenna}. The antenna can be modeled as a signal source $v_s(t)$ followed by a series resistance $R_{\rm ant}$ (the antenna impedance). The received signal is then forwarded to a matching network, which attempts to match the rectifier's input impedance, say $Z_m$, to the antenna impedance $R_{\rm ant}$. Here, we will assume perfect matching, i.e., $R_{\rm ant} = Z_m$, to maximize the power transfer \cite{Morsi2020}. The rectifier contains a single non-linear diode, which outputs current $i_d(t)$. Then, the LPF, consisting of a capacitance $C_L$ and a load resistance $R_L$, removes the high frequency harmonics to produce a DC signal $v_o(t)$ across the load. Finally, the load resistance at the RX models a load which could be a battery or a signal processing unit. Note that more advanced rectenna circuits exist \cite{Durgin2014} and circuit designs with multiple diodes may be beneficial for EH if, e.g., the received signal power at the RX is high \cite{Shanin2020}. Nevertheless, throughout this paper, we mainly consider a simple rectenna circuit with a single diode, as the one in Fig. \ref{fig:rectenna}.

In order to study and design WIET systems, a mathematical model for the RF-EH process of a rectenna is required. The main challenge here is to develop a simple and mathematically tractable model, which, at the same time, is accurate. There are numerous EH models in the literature that capture various aspects of the EH circuit characteristics. There exists a trade-off between tractability and accuracy, depending on the considered model. The selection of a suitable model is determined by the considered WIET problem and the approach used to solve it. In what follows, we provide three main categories of models that will be adopted throughout this article.

\subsection{Linear Model}\label{linear}
In the linear model, the average harvested energy $P_L$ at an EH RX is assumed to be linearly proportional to the average received RF signal power $P_{\rm RF}$ \cite{Zhang2013} and is given by
\begin{equation}
P_{\rm L} = \eta P_{\rm RF},\label{linear_model}
\end{equation}
where $0 < \eta \leq 1$ is a constant parameter that characterizes the energy conversion efficiency. Note that the linear model has been broadly employed in the early works on WIET due to its simplicity as it facilitates theoretical analysis and optimization \cite{Zhang2013, Psomas2016}. However, this model lacks accuracy since it cannot capture the non-linear behavior of practical EH circuits, especially for large RF input powers. More accurate variations of the linear model exist, such as the constant-linear and the constant-linear-constant models \cite{8315145}. The former considers a sensitivity threshold, i.e., the harvested energy is assumed to be zero below a threshold, whereas the latter takes into account both a sensitivity threshold and a saturation level. Furthermore, a general piecewise linear model with multiple segments can provide higher accuracy \cite{8315145}.

\subsection{Non-linear Saturation Model}\label{nl_sat}
To capture the non-linearity and saturation effect of the rectenna, the so called non-linear saturation model was proposed. In this case, by using a non-linear parametric function, the output of the rectenna can be approximated through curve fitting. A popular model that uses a sigmoid function is \cite{7264986}
\begin{equation}
P_{\mathrm{S}} = \frac{\Upsilon - P_{\rm sat}\Psi}{1-\Psi},\label{nl_sat_model}
\end{equation}
where $\Upsilon = P_{\rm sat}/\big(1+\exp\big(-a(P_{\rm RF}-b)\big)\big)$, $\Psi=1/(1+\exp(ab))$, $P_{\rm sat}$ is a positive constant denoting the rectenna's maximum output power (the saturation level), $a$ indicates the charging rate with respect to the average received RF signal power $P_{\rm RF}$ (the slope), and $b$ is a measure for the sensitivity of the EH circuit. Note that $P_S$ is the average harvested energy for a given waveform and so the parameters of \eqref{nl_sat_model} depend on the considered waveform. Therefore, this EH model does not facilitate the design of the transmit signal waveform for WIET. Nevertheless, since the model in \eqref{nl_sat_model} is a continuous quasi-convex function of the average received RF signal power $P_{\rm RF}$, it is widely utilized for theoretical analysis and optimization, e.g., for optimizing resource allocation \cite{Xu2022}.

\subsection{Non-linear Circuit-based Model}\label{nl_circ}
Another category of non-linear EH models that appears in the literature is based on the physical characteristics of the rectifier's diode. These circuit-based models, in contrast to the saturation model, are ideal for waveform design since they are not derived for a specific waveform. Specifically, by using circuit analysis, it can be deduced that the instantaneous harvested energy can be approximated by \cite{Morsi2020}
\begin{equation}
P_{\rm C} = \mathrm{min}\left( \left[\frac{1}{\alpha} W_0(\alpha e^\alpha I_0(\xi))-1 \right]^2 I_s^2 R_\text{L}, \frac{B_v^2}{4R_L} \right),\label{nl_circ_model}
\end{equation}
where $\xi = \sqrt{2 P_{\rm in}}B$, $P_{\rm in}$ is the instantaneous received signal power, constants $\alpha$ and $B$ are parameters that depend on the adopted circuit, and variables $I_s$, $R_\text{L}$, and $B_v$ denote the reverse bias saturation current, the load resistance, and the reverse breakdown voltage of the adopted circuit, respectively \cite{9592187}. The model in \eqref{nl_circ_model} is convex and is a monotonically increasing function of the input power until the saturation level is reached, at which point the harvested energy remains constant. Now, even though this model is suitable for investigating various types of problems, e.g., the optimal input distribution \cite{Morsi2020}, its complicated analytical form makes its application often challenging. Therefore, a simplified and more tractable model was proposed in \cite{Clerckx16} based on the Taylor series expansion of the diode characteristic equation. This results in the following expression for the current $i_d(t)$ flowing through an ideal diode \cite{Clerckx16}
\begin{align}
i_d(t) = \sum_{n=0}^\infty k_n R_{\rm ant}^{n/2} y(t)^n,\label{nl_circ_model2}
\end{align}
where $k_n = \frac{I_s}{n!(\delta V_T)^n}$, $V_T$ is the thermal voltage, $\delta$ is the ideality factor (emission coefficient), $y(t)$ is the received RF signal, and $R_{\rm ant}$ is the antenna impedance. Then, the average harvested energy $P_{\rm E}$ is a monotonic increasing function of \eqref{nl_circ_model2} and is approximated by $P_{\rm E} = \E\{i_d(t)\}$ \cite{Clerckx16}. It should be pointed out that \eqref{nl_circ_model2} does not capture the rectenna's saturation effect and thus it is most suitable for the low-power regime.

\section{Design of Signal Processing \& Communications for WIET}\label{sec:techniques}
In this section, we focus on signal processing and communication aspects that are important in conventional wireless networks. A detailed discussion and investigation on how their design is influenced by WIET is provided. We first present the optimization of the available resources in WIET systems (Subsection \ref{sec:resource}). A resource allocation optimization framework is established and different solution methodologies for addressing the specific type of problems arising in WIET systems are provided. Then, we study the design of the waveform of the transmitted signal, which is a fundamental issue for the efficient operation of WIET systems (Subsection \ref{sec:waveform}). We review the literature on WIET waveform design, discussing its significance for the performance of WIET systems, and then present a recent study on the benefits of chirp waveforms for WIET systems \cite{Roy2023}. Finally, we turn our attention to the design and analysis of WIET systems in the near-field (Fresnel) region (Subsection \ref{sec:nearfield}), where devices in the 6G era may be located in \cite{Zhang22}. Consequently, this has to be accounted for when designing WIET systems.\vspace{-1mm}

\subsection{Resource Allocation for WIET}\label{sec:resource}
\graphicspath{{resource/}}
Over the past decade, how to allocate resources to achieve desired performance objectives has been intensively investigated in the context of wireless communications \cite{bjornson2013optimal, 5200968, yu2020irs, 7372448, 9723093}. In particular, via resource allocation design, we can intelligently distribute the limited wireless resources, e.g., frequency, power, and time, across different users to facilitate efficient information transmission. On the one hand, by accounting for the available system information such as channel state information (CSI), the total wireless resources available, and the adopted performance metrics in the resource allocation optimization framework, we can ensure the QoS requirements for a given radio propagation environment. On the other hand, for different wireless transmission scenarios, different system design objectives are relevant, e.g., sum rate maximization, total power consumption minimization, and energy efficiency maximization. However, the resource allocation optimization frameworks developed for traditional communication-centric wireless networks may be incapable or inefficient for satisfying any WET requirements due to the following reasons. First, compared to information RXs, the mathematical models that accurately capture the EH input–output relationship are more complex, imposing an arduous design challenge for the energy RXs. Second, the non-trivial EH-based performance metrics, e.g., the harvested energy and WET efficiency, introduce new obstacles for resource allocation design. Third, when information RXs and energy RXs coexist in the wireless network, both an information-carrying and an energy-carrying signal may be needed to facilitate energy-efficient WIET, leading to a more complex resource allocation optimization problem. Finally, while in conventional information transmission systems, co-channel interference is in general harmful and has to be carefully controlled or mitigated, in WET systems, strong interference is beneficial to EH as it provides a pivotal energy source for EH RXs, and thus, can be exploited to boost RF-EH \cite{Psomas2016,8959114,7990046}.

Next, we briefly overview seminal works on resource allocation design for WIET systems. WIET system architectures for MIMO broadcasting systems were first proposed in \cite{Zhang2013}, where the achievable rate–energy region was revealed by solving resource allocation design problems. By utilizing the architectures in \cite{Zhang2013}, optimal resource allocation schemes based on TS and PS were, respectively, developed in \cite{6373669} and \cite{Zhou2013} to realize various trade-offs between information transmission and energy transfer. Subsequently, the resource allocation design frameworks were extended to the more general case for multiple EH RXs. For instance, the authors of \cite{6860253} developed an optimization framework for maximizing the total power transferred to EH RXs, while satisfying the minimum signal-to-interference-plus-noise ratio (SINR) requirements for ID RXs. In \cite{6805330}, a resource allocation design framework for minimizing the total transmit power of a multi-user WIET system while satisfying both SINR and EH constraints was proposed. Moreover, a multi-objective optimization (MOO) framework to achieve the conflicting system design goals of simultaneously providing satisfactory service for ID and EH RXs was developed in \cite{7111366}. Note that to facilitate the resource allocation design, the above works adopted the linear EH model (see Subsection \ref{linear}), which may be oversimplified for practical systems. Later works on resource allocation design for WIET systems considered a more practical non-linear saturation model (see Subsection \ref{nl_sat}) and developed the corresponding resource allocation optimization algorithms to satisfy the QoS requirements of WIET \cite{7264986}. On the other hand, few works adopted the non-linear circuit-based model for resource allocation design due to its challenging form (see Subsection \ref{nl_circ}).

\subsubsection{WIET-based Resource Allocation}
A multi-user multiple-input single-output (MISO) WIET system is considered, where a TX equipped with $N$ antenna elements serves $K$ RXs. Without loss of generality, we assume that each RX is simultaneously an ID RX and an EH RX. In each scheduled time slot, the transmitted baseband signal is given by
\begin{equation}
\mathbf{x} = \underset{k\in\mathcal{K} }{\sum} \mathbf{w}_kd_k+\mathbf{v},
\end{equation}
where scalar variable $d_k$ is the information-bearing symbol for RX $k$ and $\mathcal{K} = \{1,\dots, K\}$. Without loss of generality, we assume $\E\{|d_k|^2\} = 1$, $\forall\mathit{k} \in \mathcal{K}$. Also, vectors $\mathbf{w}_k\in \mathbb{C}^{\mathit{N}\times 1}$ and $\mathbf{v}\in \mathbb{C}^{\mathit{N}\times 1}$ denote the information beamforming vector for RX $k$ and the energy signal used for energy transfer. Moreover, vector $\mathbf{v}$ is modeled as a complex random variable, whose covariance matrix is given by $\mathbf{V} = \E\{ \mathbf{v}\mathbf{v}^\text{H} \}$. Then, the baseband signal received by RX $k$ can be expressed as
\begin{equation}
y_k=\mathbf{g}_k^\text{H}\Big(\underset{r\in\mathcal{K} }{\sum} \mathbf{w}_rd_r+\mathbf{v}\Big)+n_k,
\end{equation}
where $n_k\sim \mathcal{CN}\left ( 0,\sigma _{n_k}^2 \right )$ denotes the additive white Gaussian noise (AWGN) at RX $k$ with variance $\sigma _{n_k}^2$. Moreover, vector $\mathbf{g}_k\in \mathbb{C}^{N\times 1}$ characterizes the channel from the TX to RX $k$. Note that we consider the same channel at both RX antennas. Hence, the received RF power at RX $k$ can be calculated as
\begin{equation}
P_k^{\mathrm{EH}} = \E\{|y_k|^2\} = \mathrm{Tr}(\mathbf{V}\mathbf{G}_k)+\underset{r\in\mathcal{K} }{\sum}\mathrm{Tr}(\mathbf{W}_r\mathbf{G}_k),
\end{equation}
where $\mathbf{W}_r\in \mathbb{H}^N$ and $\mathbf{G}_k\in \mathbb{H}^N$ are defined as $\mathbf{W}_r=\mathbf{w}_r\mathbf{w}_r^\text{H}$ and $\mathbf{G}_k=\mathbf{g}_k\mathbf{g}_k^\text{H}$, respectively.

We consider the linear and the non-linear saturation EH models, to highlight the impact that the considered model has on the performance and the insights that can be obtained from the resource allocation design. For the linear EH model, given by \eqref{linear_model}, the harvested energy at RX $k$ is
\begin{equation}
P_{\mathrm{out},k}=\eta P_k^{\mathrm{EH}},
\end{equation}
whereas for the non-linear saturation EH model, given by \eqref{nl_sat_model}, the harvested energy at RX $k$ is
\begin{equation}
P_{\mathrm{out},k}=\frac{\Upsilon_k-P_{\rm sat}\Psi}{1-\Psi},
\end{equation}
where $\Upsilon_k=P_{\rm sat}/(1+\exp(-a(P_k^{\mathrm{EH}}-b)))$.

On the other hand, the achievable rate of RX $k$ is given by
\begin{equation}
R_k = \log_2 \bigg(1+ \frac{\mathrm{Tr}(\mathbf{W}_k\mathbf{G}_k)}{\mathrm{Tr}\big(\sum_{r\in\mathcal{K} \backslash \{k\}}\mathbf{W}_r\mathbf{G}_k+\mathbf{V}\mathbf{G}_k\big)+\sigma _{n_k}^2} \bigg).
\end{equation}

Subsequently, to facilitate the presentation of the resource allocation problem, we introduce two commonly-adopted utility functions for WIET systems, namely, the wireless information transfer (WIT) and WET efficiencies, which are, respectively, given by
\begin{eqnarray}
U^{\mathrm{WIT}}(\mathbf{W}_k,\mathbf{V})&\hspace{-2mm}=\hspace{-2mm}&\frac{R^{\mathrm{WS}}(\mathbf{W}_k,\mathbf{V})}{P^{\rm D}(\mathbf{W}_k,\mathbf{V})-P^{\mathrm{EH}}(\mathbf{W}_k,\mathbf{V})},\\
U^{\mathrm{WET}}(\mathbf{W}_k,\mathbf{V})&\hspace{-2mm}=\hspace{-2mm}&\frac{P^{\mathrm{EH}}(\mathbf{W}_k,\mathbf{V})}{P^{\rm D}(\mathbf{W}_k,\mathbf{V})},
\end{eqnarray}
where $R^{\mathrm{WS}}(\mathbf{W}_k,\mathbf{V})= \sum_{k\in\mathcal{K}}\alpha_k R_k$ is the weighted sum rate with $\alpha_k\geq 0$ being the weight factor for RX $k$, which can be used to adjust the priorities of different RXs and ensure fairness for the resource allocation by setting $\sum_{k\in\mathcal{K}} \alpha_k = 1$.

Moreover, $P^{\rm D}(\mathbf{W}_k, \mathbf{V}) = P_c + \zeta \mathrm{Tr}(\sum_{k\in\mathcal{K}} \mathbf{W}_k+\mathbf{V})$ represents the required power dissipation, where $P_c$ is the circuit power of the system, $\zeta \geq 1$ is the power amplifier efficiency, and $P^{\mathrm{EH}}(\mathbf{W}_k,\mathbf{V}) = \sum_{k\in\mathcal{K}} P_{\mathrm{out},k}$ is the total power harvested by all RXs. Note that the WIT efficiency $U^{\mathrm{WIT}}$ is usually adopted for communication-centric resource allocation design as it characterizes how many bits per joule of energy can be delivered. Finally, the WET efficiency $U^{\mathrm{WET}}$ accounts for the ratio of the harvested power and the power consumed by the TX, which is a key performance indicator for EH-centric designs.

\subsubsection{Problem Formulation}
We now propose a typical resource allocation optimization problem formulation for WIET systems. In particular, one can maximize the WIT efficiency, while guaranteeing a minimum WET efficiency, by solving the following optimization problem:
\begin{eqnarray}\label{prob1}
&&\hspace*{-10mm}\underset{\mathbf{W}_k,\mathbf{V}\in\mathbb{H}^N}{\mathrm{maximize}}\hspace*{4mm}U^{\mathrm{WIT}}(\mathbf{W}_k,\mathbf{V})\notag
\\
&&\hspace*{-10mm}\mathrm{subject}\hspace*{1.6mm}\mathrm{to}\hspace*{2mm}
\mbox{C1:}\hspace*{1mm} \sum_{k\in\mathcal{K}} \mathrm{Tr}(\mathbf{W}_k)+\mathrm{Tr}(\mathbf{V})\leq P^{\mathrm{max}},\notag\\
&&\hspace*{8mm}\mbox{C2:}\hspace*{1mm}R_k\geq R_{\mathrm{req},k},\hspace*{1mm}\forall k,\notag\\
&&\hspace*{8mm}\mbox{C3:}\hspace*{1mm}P_{\mathrm{out},k}\geq P_{\mathrm{req},k},\hspace*{1mm}\forall k,\notag\\
&&\hspace*{8mm}\mbox{C4:}\hspace*{1mm}U^{\mathrm{WET}}(\mathbf{W}_k,\mathbf{V})\geq U^{\mathrm{WET}}_{\mathrm{req}},\notag\\
&&\hspace*{8mm}\mbox{C5:}\hspace*{1mm}\mathbf{W}_k\succeq \mathbf{0},\mathbf{V}\succeq \mathbf{0},\hspace*{1mm}\forall k,\notag\\
&&\hspace*{8mm}\mbox{C6:}\hspace*{1mm}\mathrm{Rank}(\mathbf{V})\leq 1, \mathrm{Rank}(\mathbf{W}_k)\leq 1,\hspace*{1mm}\forall k.
\end{eqnarray}
Here, constant $P^{\mathrm{max}}$ in constraint C1 limits the maximum transmit power of the TX. Constraint C2 specifies the WIT QoS constraint for each RX, and constant $R_{\mathrm{req},k}$ is the corresponding minimum required achievable rate. Constraint C3 is the EH constraint for RX $k$, and $P_{\mathrm{req},k}$ is the corresponding minimum required harvested DC power. Constraint C4 guarantees that the WET efficiency is always above a given threshold $U^{\mathrm{WET}}_{\mathrm{req}}$ to facilitate energy-efficient WET. Constraint C5 ensures that the optimization variables $\mathbf{W}_k$ and $\mathbf{V}$ are valid covariance matrices, while the rank-one constraints in C6 ensures that $\mathbf{V} =  \mathbf{v}\mathbf{v}^\text{H}$ and $\mathbf{W}_k = \mathbf{w}_k \mathbf{w}_k^\text{H}, \forall k$ hold after optimization. 

The problem in \eqref{prob1} is a general problem formulation for WIET systems which includes several typical resource allocation problems as special cases. In fact, by slightly adjusting the objective function in \eqref{prob1}, we can obtain other typical resource allocation optimization problems such as:
\begin{itemize}
\item \textit{Weighted sum rate maximization problem:}
\begin{equation}\label{prob2}
\underset{\mathbf{W}_k,\mathbf{V}\in\mathbb{H}^N}{\mathrm{maximize}}\hspace*{4mm}R^{\mathrm{WS}}(\mathbf{W}_k,\mathbf{V})\hspace*{2mm}\mathrm{subject}\hspace*{1.6mm}\mathrm{to}\hspace*{2mm}\mbox{C1-C6}.
\end{equation}
This optimization problem is suitable for designing conventional communication-centric systems such as cellular network-enabled WIET systems.
\item \textit{Total power consumption minimization problem:}
\begin{equation}
\label{prob3}\underset{\mathbf{W}_k,\mathbf{V}\in\mathbb{H}^N}{\mathrm{minimize}}\hspace*{4mm}P^{\rm D}(\mathbf{W}_k,\mathbf{V})\hspace*{2mm}\mathrm{subject}\hspace*{1.6mm}\mathrm{to}\hspace*{2mm}\mbox{C1-C6}.
\end{equation}
This resource allocation optimization problem can be adopted when the power consumption of the TX is strictly constrained, e.g., in on-board battery-powered unmanned aerial vehicles.
\item \textit{Total harvested power maximization problem:}
\begin{equation}
\label{prob4}\underset{\mathbf{W}_k,\mathbf{V}\in\mathbb{H}^N}{\mathrm{maximize}}\hspace*{4mm}P^{\mathrm{EH}}(\mathbf{W}_k,\mathbf{V})\hspace*{2mm}\mathrm{subject}\hspace*{1.6mm}\mathrm{to}\hspace*{2mm}\mbox{C1-C6}.
\end{equation}
When wireless charging is the key objective of a WIET system, the above resource allocation optimization problem is usually considered to prolong the duration of power-limited RXs.
\end{itemize}

All the aforementioned optimization problems are non-convex and are NP-hard to solve. Nevertheless, there are powerful advanced algorithms that find globally optimal solutions, including monotonic optimization algorithms \cite{zhang2013monotonic,8648498} and branch-and-bound (BnB) algorithms \cite{lawler1966branch,9154337}. In particular, although the aforementioned optimization problems are non-convex in nature, they do preserve monotonicity, i.e., their objective functions are monotonically increasing or decreasing with respect to the optimization variables over the feasible set \cite{8974403}. Hence, we can exploit this property to iteratively reduce an upper bound on the objective function until the optimum is obtained. On the other hand, by successively dividing the feasible set into different subsets and continuously removing non-optimal subsets by checking lower and upper bounds, the globally optimal solution for all the aforementioned optimization problems can be found with the help of BnB theory \cite{Xu2022}.

However, these globally optimal algorithms entail in general prohibitive computational complexity, which is not affordable for practical WIET systems. As a compromise, we can also employ several computationally efficient suboptimal algorithms to strike a balance between optimality and complexity. In fact, for the above optimization problems, the non-convexity generally originates from the non-convex objective function or a few non-convex constraints, while the remaining parts of the optimization problem are still convex. This observation inspires the application of successive convex approximation (SCA) \cite{dinh2010local,9183907}. Specifically, SCA methods allow us to approximate the involved non-convex functions by tractable convex functions and iteratively solve a sequence of approximated versions of the original problem until convergence \cite{yu2020irs}. In addition, the aforementioned optimization problems involve the rank-one constraints in C6, which are highly non-convex. As such, in the literature, semidefinite relaxation (SDR) is commonly adopted to tackle this obstacle. In particular, the SDR method allows us to omit the rank-one constraint and focus on the relaxed version of the optimization problem \cite{xu2022robust}. Moreover, the tightness of such a relaxation can be verified by examining the Lagrangian function of the original optimization problem with respect to the beamforming matrices $\mathbf{V}$ and $\mathbf{W}_k, \forall k$. More details on the proof of optimality of the SDR method can be found in \cite{6860253} and \cite{6781609}.

\begin{figure}\centering
\includegraphics[width=0.95\linewidth]{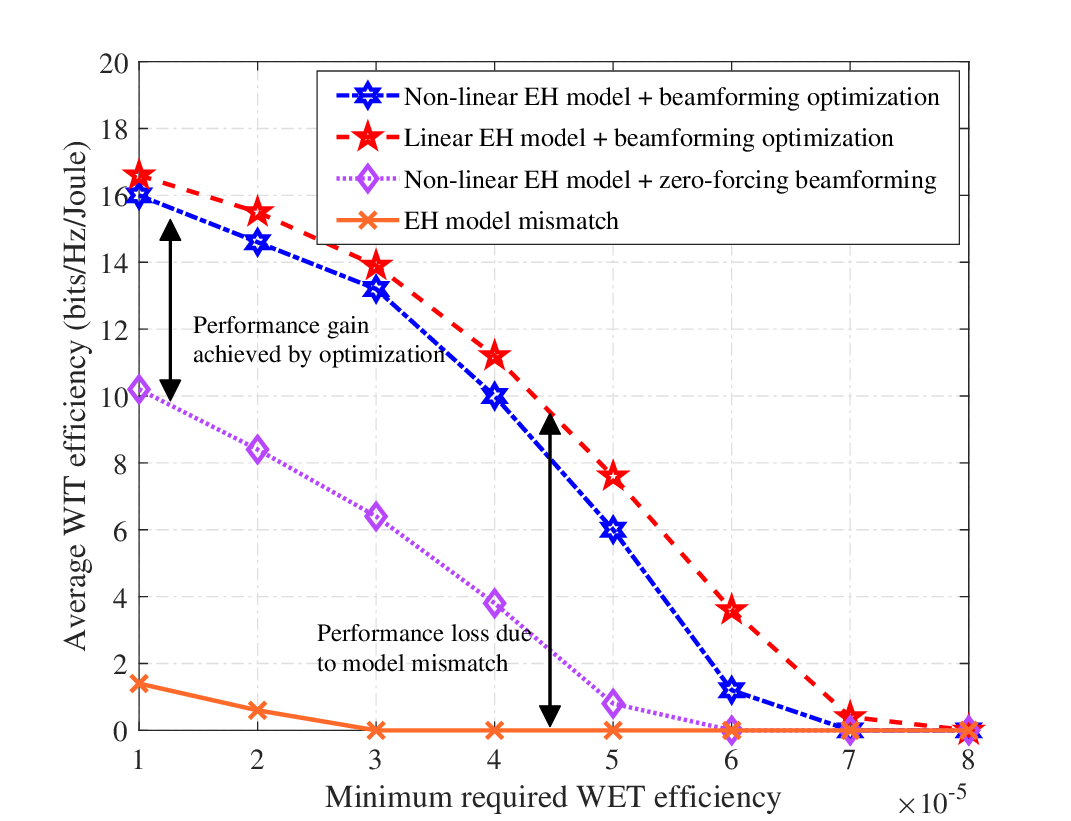}
\caption{Average system WIT efficiency (bit/Hz/Joule) versus minimum required WET efficiency.}\label{fig:RA_simulation}
\end{figure}

\subsubsection{Simulation Results}
Next, the performance of the considered WIET system is evaluated for $K=3$ RXs. The TX is equipped with $N=10$ antenna elements and has a maximum transmit power of $46$ dBm. The RXs are uniformly distributed in the system. The channels between the TX and the RXs are modeled as Rician fading channels with a Rician factor of $2$ dB. The path-loss exponent is $2.5$ and the variance of the AWGN is $-100$ dBm. The minimum required EH power is $2$ $\mu$W, while the minimum required spectral efficiency is $1$ bits/s/Hz. Moreover, we consider the following parameters for the EH models: $\eta = 1$, $a = 0.003$, $b = 6400$, and $P_{\rm sat} = 20$ mW. The resource allocation design is formulated as a WIT efficiency maximization problem, cf. optimization problem \eqref{prob1}. Fig. \ref{fig:RA_simulation} illustrates the average WIT efficiency (bits/Hz/Joule) versus the minimum required WET efficiency for different schemes. As can be observed from the figure, the average WIT efficiency of all considered schemes decreases as the required WET efficiency increases. Moreover, we can observe that the WIT efficiency for the scheme with linear EH model and beamforming optimization is higher than that of the scheme with non-linear saturation EH model and beamforming optimization since the latter model usually introduces more constraints on the optimization problem compared to the former one. However, the linear power conversion, assumed for the linear EH model, is not achievable in practice due to the inevitable non-linearity of EH circuits.

We further consider a scheme where the WIT maximization problem is solved for the linear EH model, but the performance is then evaluated for the non-linear EH model. This illustrates the impact of the resulting resource allocation mismatch by comparing the designs that are optimized based on the same model, i.e., the linear EH model. Note that the WIT efficiency is set to zero if a communication or EH QoS constraint is violated. We observe that the resource allocation policy optimized based on the linear EH model inevitably leads to a significant performance loss, for all considered WET efficiencies, when applied in a practical WIET system with a non-linear EH circuit. For comparison, we also consider a scheme where the information beamforming is implemented based on the zero-forcing beamforming policy. It can be seen that compared to the schemes with optimized beamforming, the one employing the zero-forcing beamforming policy sacrifices performance in favor of a simplified implementation. This underlines the importance of jointly optimizing the information beamforming and energy signal.\vspace{-1mm}

\subsection{WIET-based Waveform Design}\label{sec:waveform}
\graphicspath{{waveform/}}
The design of efficient rectenna circuits has been one of the main research directions of the community working on the WET technology \cite{Durgin2014}. Despite the significance of the actual rectenna circuit design, its overall efficiency also depends on the waveform of the received RF signal. Indeed, it has been shown experimentally that signals with high peak-to-average-power ratio (PAPR), such as white noise signals, orthogonal frequency division multiplexing (OFDM) signals, and chaotic signals, can achieve high WET efficiency as they have periodic high power picks that can overcome the build in potential of diodes \cite{collado2014,boaventura2015}. Moreover, the work in \cite{boaventura2011}, reveals through theory and with measurements that multisine signals provide a higher DC output compared to a single tone excitation.

An analytical methodology for the optimization and design of multisine waveforms is presented in \cite{Clerckx16}. This work highlights the necessity of considering the non-linear effects of the rectenna in waveform design. A similar approach is used in \cite{Clerckx23}, where the non-linear effects of the high power amplifier (HPA) at the TX are also taken into account. In this case, the non-linearity of the HPA causes the optimal waveforms to have low PAPRs. A transmission scheme based on the properties of chirp waveforms is proposed in \cite{Roy2023}, where it is shown that chirp signals are ideal for boosting EH compared to multisine waveforms. Moreover, in \cite{Mukherjee2021}, the authors consider a RX employing an analog correlator and present a signal design based on differential chaos shift keying, which enhances the RF-EH. The waveform design for WET in a multi-user setting has also been studied \cite{Abeywickrama21,Kim2019}. Specifically, in \cite{Abeywickrama21}, a non-linear current-voltage rectenna model is proposed and, based on this model, the optimal waveforms are derived to maximize the harvested energy at the RXs. The authors of \cite{Kim2019} focus on the fairness aspect of waveform design. They develop a low-complexity algorithm that regulates the level of fairness between the RXs, thus mitigating the near-far problem.

The aforementioned works focus on WET and do not take into account information transfer. Nevertheless, waveform design for SWIPT systems has also been studied in several works \cite{Morsi2020, heath2021, Clerckx2017, Zawawi2019, Ayir21}. In particular, the authors of \cite{Morsi2020} investigate the optimal input distribution that maximizes the information transfer conditioned on a minimum RF-EH constraint for multiple RXs. It is proven that the optimal distribution, which maximizes the rate-energy region subject to a peak transmit power constraint, is discrete with a finite number of mass points. An RX architecture for OFDM SWIPT systems is proposed in \cite{heath2021}, where the cyclic prefix of the modulated OFDM signal and a portion of the received information-carrying signal are exploited for EH. In \cite{Clerckx2017}, the signal-to-noise ratio (SNR)-energy region of wireless powered backscatter communications is investigated. The region is maximized by optimizing the amplitudes and phases of a multisine waveform. This work is extended in \cite{Zawawi2019} to multi-user systems with a focus on the SINR-energy region. Moreover, a practical approach is pursued in \cite{Ayir21}, where based on software-defined radios, the effects of clipping and non-linear amplification on multisine waveforms are studied. It is shown that the end-to-end efficiency (i.e., the ratio of the DC power harvested at the RX to the total DC power needed at the TX) highly depends on the average radiated power of the waveform, which is inversely proportional to the PAPR.

In what follows, we present the gains achieved with chirp waveforms in WIET systems \cite{Roy2023}. Chirp is a frequency modulated signal, whose duration can be varied independent of its bandwidth (unlike fixed-frequency waveforms) and whose instantaneous frequency varies as a function of time. By exploiting the unique properties of chirp, the superposition of multiple chirps within a given subband is proposed and the benefits for WIET are presented. Specifically, in contrast to conventional multisine signals, the superposition of chirp waveforms achieves frequency diversity as they are transmitted over a selected subset of subbands with higher channel gains. Moreover, despite not utilizing all available frequency subbands, the total number of waveforms is maintained, thus achieving non-linearity gains due to high PAPR, as with conventional multisine signals. Applications that make use of chirp waveforms include radar systems as well as LoRa-based wide area networks \cite{Cook}. However, it is important to note that these systems employ basic chirp signals and are not concerned with superimposed chirp waveforms.

\begin{figure}[t]\centering
\includegraphics[width=0.6\linewidth]{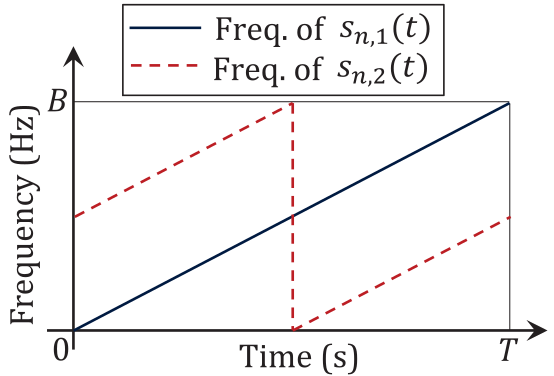}
\caption{Instantaneous frequency of chirp waveforms with $\xi = 2$ \cite{Roy2023}.}\label{fig:dual_chirp_sc_selec}
\end{figure}

\subsubsection{Superimposed Chirp Waveform for WIET}
A chirp is a waveform in which the instantaneous frequency increases (an up-chirp) or decreases (a down-chirp) with time. Specifically, an up-chirp waveform of bandwidth $B$ and duration $T$ over a subband $n$ can be expressed as \cite{Roy2020}
\begin{equation}
s_n(t) = \cos \left[2\pi t \left(f_n + \frac{1}{2} \mu t \right) \right], 0 \leq t \leq T,\label{eq:signal_chirp}
\end{equation}
where $f_n = f_0 + (n-1)B$, $f_0$ is the initial frequency, $n\in \{1, 2, \dots, N\}$, $N$ is the total number of subbands, and $\mu = B/T$ is the rate of change of the instantaneous frequency with $BT \geq 1$ \cite{Cook}. In contrast to a fixed-frequency waveform, by selecting an appropriate value for $\mu$, one can vary the duration of a chirp independently of its bandwidth.

Now, consider the superposition of multiple chirp waveforms in a given subband. In this case, the chirp waveform over subband $n$ can be expressed as \cite{Roy2023}
\begin{align}
s_{n,l}(t) = \begin{cases}
\cos\left[2\pi t \left(f_n + \frac{l-1}{\xi}B +\frac{1}{2} \mu t \right) \right], 0 \leq t \leq T_l,\\
\cos\left[2\pi t\left(f_n + \frac{l-1-\xi}{\xi}B +\frac{1}{2} \mu t \right) \right], T_l < t \leq T,
\end{cases}
\label{eq:signal_xi-chirp}
\end{align}
where $\xi \in \mathbb{Z}^{+}$ is the number of chirps superimposed within the same time and signal bandwidth $B$, $l \in \{1,2,\dots,\xi\}$, $T_l = T (1-\frac{l-1}{\xi})$, $T_\text{o}=\frac{1}{B}$, and $T = \xi T_\text{o}$. The instantaneous frequency corresponding to these superimposed waveforms appears at distinct time instants. Specifically, the separation in instantaneous frequency between any two adjacent chirps in the set of $\xi$ waveforms is $B/\xi$, while each maintains a bandwidth of $B$. Moreover, note that the above waveform preserves orthogonality across subbands. The instantaneous frequency variation for $\xi=2$ is shown in Fig. \ref{fig:dual_chirp_sc_selec}.

In what follows, we consider the proposed superposition of chirp waveforms in a SWIPT setup and show how it benefits WIET. For simplicity, we focus on the case $\xi = 2$, illustrated in Fig. \ref{fig:dual_chirp_sc_selec} and referred to as chirp-$2$; the general case, for any $\xi$, can be found in \cite{Roy2023}. Specifically, consider a SWIPT system consisting of a TX with $M$ antennas and a single-antenna WIET RX. The bandwidth of the system is divided into $N$ subbands of equal bandwidth $B$. The channel over each subband experiences frequency-flat fading and varies independently across subbands \cite{Zeng15}. The complex baseband channel vector between the TX and the RX over the $n$-th subband is
\begin{align}
\mathbf{g}_n = \sqrt{\beta} \mathbf{h}_n,	
\end{align}
where $\beta$ is the large-scale fading, $\mathbf{h}_n = [h_n^1 ~ h_n^2 ~ \cdots ~ h_n^M]^\top$ is the small-scale fading vector with $h_n^m \sim \mathcal{CN}(0,1)$, $m=1,2,\dots,M$ and $n=1,2,\dots,N$. We assume that the TX is subject to an average transmit power constraint of $\Pt$.

\begin{figure}[t]\centering
	\includegraphics[width=0.8\linewidth]{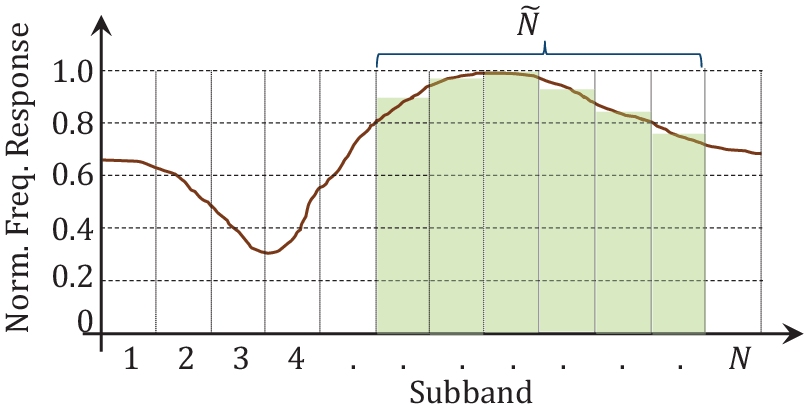}
	\caption{Selection of $\Nt$ strongest subbands out of $N$ available subbands \cite{Roy2023}.}\label{fig:subands}
\end{figure}

The channel remains time-invariant over a coherence block and varies independently across blocks. The RX transmits a pilot of length $\tau_\text{p}$ to the TX for estimating the channel impulse response. Then, the TX transmits information and energy to the RX for a duration of $\tau_\text{DL}$ using a beamforming vector whose design is based on the estimated channel. Note that $\tau_\text{p}+\tau_\text{DL}$ is the length of the coherence block. Thus, the minimum mean square error (MMSE) estimate $\widehat{\mathbf{g}}_n$ of channel vector $\mathbf{g}_n$ is \cite{Marzetta}
\begin{equation}
\widehat{\mathbf{g}}_n = \mathbf{g}_n + \widetilde{\mathbf{g}}_n,
\label{eq:estimated_channel_vector}
\end{equation}
where $\widetilde{\mathbf{g}}_n \sim \mathcal{CN}(\mathbf{0},(\beta - \gamma) \mathbf{I}_M)$ denotes the channel estimation error with
\begin{align}
\gamma = \frac{\frac{P_\text{p} \tau_\text{p}}{N} \beta^2}{\sigma^2 +\frac{P_\text{p} \tau_\text{p}}{N}\beta},
\end{align}
where $P_\text{p}$ is the pilot power and $\sigma^2$ denotes the noise power. Then, the estimated channel vectors over all subbands can be arranged in descending order of their channel gains, i.e.,
\begin{align}
\|\widehat{\mathbf{g}}_{[1]}\|^2 \geq \|\widehat{\mathbf{g}}_{[2]}\|^2 \geq \cdots \geq \|\widehat{\mathbf{g}}_{[n]}\|^2 \geq \ldots \geq \|\widehat{\mathbf{g}}_{[N]}\|^2,
\end{align}
where
\begin{equation}
\|\widehat{\mathbf{g}}_{[n]}\|^2 = \big|{\widehat{g}_{[n]}^1}\big|^2 + \big|{\widehat{g}_{[n]}^2}\big|^2 + \cdots + \big|{\widehat{g}_{[n]}^M}\big|^2,
\end{equation}
denotes the estimated channel vector between the TX and the RX over the $n$-th strongest subband (in terms of its channel gain). The TX selects the $\Nt = N/\xi = N/2$ (since we consider $\xi=2$) strongest subbands out of the available $N$ subbands, as illustrated in Fig. \ref{fig:subands}, for the superposition of the chirps. It then performs power allocation proportional to the channel gains of the selected subbands such that the transmit power constraint $\Pt$ is satisfied with equality. Note that even though $\Nt < N$ subbands are selected for the superposition, the number of transmitted chirp waveforms remains equal to $N$. Therefore, through this process, frequency diversity is achieved while, at the same time, the non-linearity gains at the rectenna due to the high PAPR of the multisine are maintained.

For data transmission, we consider binary phase-shift keying (BPSK) modulation, without loss of generality. Then, the transmitted signal over the $n$-th subband is
\begin{equation}\label{eq:Tx_chirp_Downlink_ecsi}
\mathbf{x}_n(t) = \sqrt{\Pt} \sqrt{\eta_n} {\pmb{\varphi}_n^\ast} S_n(t), 0\leq t\leq T,
\end{equation}
where $T = \xi T_\text{o} = 2T_\text{o}$, $\eta_n$ is the power control coefficient such that $\sum_{n=1}^N \E\{\left\|\mathbf{x}_n(t) \right\|^2\} \leq \Pt$, $\pmb{\varphi}_n$ is the beamforming precoder given by \cite{Roy2023}
\begin{equation}
\pmb{\varphi}_n = \frac{\widehat{\mathbf{g}}_n}{\big\|{\widehat{\mathbf{g}}_n }\big\|},\label{eq:beamform_Downlink_ecsi}
\end{equation}
and
\begin{equation}
S_n(t) = \sqrt{2} (d_{n,1} s_{n,1}(t) + d_{n,2} s_{n,2}(t)),
\end{equation}
where $s_{n,l}(t)$ is given by \eqref{eq:signal_xi-chirp} and $d_{n,l}$ are the data symbols; the scaling by $\sqrt{2}$ ensures that the average power of the chirp waveform over the $n$-th subband is unity.

\subsubsection{Energy and Information Transfer}
\begin{figure}[t]\centering
	\includegraphics[width=0.8\linewidth]{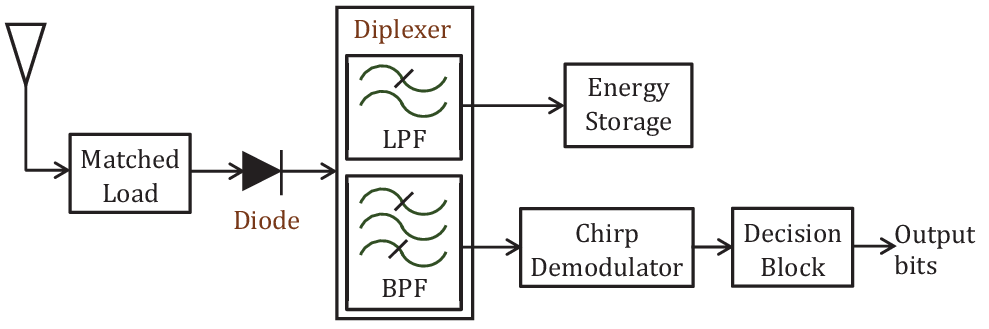}
	\caption{The block diagram of a diplexer-based integrated RX \cite{Roy2023, Smida2021}.}\label{fig:diplexer}
\end{figure}

We consider an integrated RX model, based on the design in \cite{Smida2021}, where an RF diplexer replaces the LPF (see Fig. \ref{fig:diplexer}). The diplexer allows frequency-domain multiplexing by separating the low and high frequency signals from the same input signal without splitting. As such, low-pass band-pass diplexers can be used to separate the low frequency signals for EH from the higher frequency signals for ID \cite{Smida2021}. The diode output current is processed by the LPF that removes the high-frequency components and the DC signal component at the output of the LPF is harvested. The output of the band-pass filter (BPF) is fed to the chirp demodulator that performs multiplication of the input signal with the respective reference chirp followed by integration. Then, the output of the demodulator is fed to a decision block for the estimation of the transmitted bits through the chirps over the considered subbands.

Therefore, for energy transfer, the received signal $y(t)$ is given by
\begin{align}
y(t) = \sqrt{\Pt} \sum_{n=1}^{\Nt} \sqrt{\eta_{[n]}} {\mathbf{g}_{[n]}^\top} {\pmb{\varphi}_{[n]}^\ast} S_{[n]}(t),\label{eq:Rx_chirp_sensork_DLEBF2}
\end{align}
where the antenna noise is ignored as its contribution to the average harvested energy is considered negligible. By applying \eqref{nl_circ_model2}, the average energy harvested over the $N$ subbands is \cite{Roy2023}
\begin{align}
P_{\rm E} \approx \E \bigg\{\int_0^T \bigg[k_2 R_{\rm ant} y(t)^2 + k_4 R_{\rm ant}^2 y(t)^4\bigg] dt\bigg\},\label{eq:harveat_energy_chirp_sensork_ecsi}
\end{align}
which is obtained by considering only the first two even terms at the output of the LPF. On the other hand, for information transfer, the BPF output of the diplexer includes signal-frequency components starting at $f_n$ with bandwidth $B$, whereas other frequency components are removed \cite{Smida2021}. Hence, the equivalent signal at the input of the chirp demodulator over the $n$-th strongest subband is given by \cite{Roy2023}
\begin{align}
y_{[n]}(t) &= \frac{I_s}{\delta V_T} \sqrt{\Pt\eta_{{[n]}} } \widehat{\mathbf{g}}_{[n]}^\top {\pmb{\varphi}_{[n]}^\ast} S_{{[n]}}(t) \nonumber\\
&\qquad- \frac{I_s}{\delta V_T}  \sqrt{\Pt\eta_{{[n]}} } \widetilde{\mathbf{g}}_{[n]}^\top {\pmb{\varphi}_{[n]}^\ast}  S_{[n]}(t) + w_{[n]} (t),\label{eq:received_info_signal_chirp}
\end{align}
where $w_{[n]} (t)$ denotes the antenna noise corresponding to the $n$-th strongest subband and the last line is related to the fact that ${\mathbf{g}}_n = \widehat{\mathbf{g}}_n-\widetilde{\mathbf{g}}_n$.

At the chirp demodulator block, $y_{[n]}(t)$ is multiplied with the associated reference chirp $s_{[n],1}(t)$ followed by integration. Thus, we obtain the demodulated signal $r_{[n],1}$ corresponding to symbol $d_{[n],1}$ in the $n$-th strongest subband, given by \cite{Roy2023}
\begin{equation}
r_{[n],1} = \frac{1}{\sqrt{T}} \int_0^T y_{[n]}(t) s_{[n],1}(t) dt.
\label{eq:received_info_signal_chirp2}
\end{equation}

\begin{figure}[t]\centering
\includegraphics[width=0.85\linewidth]{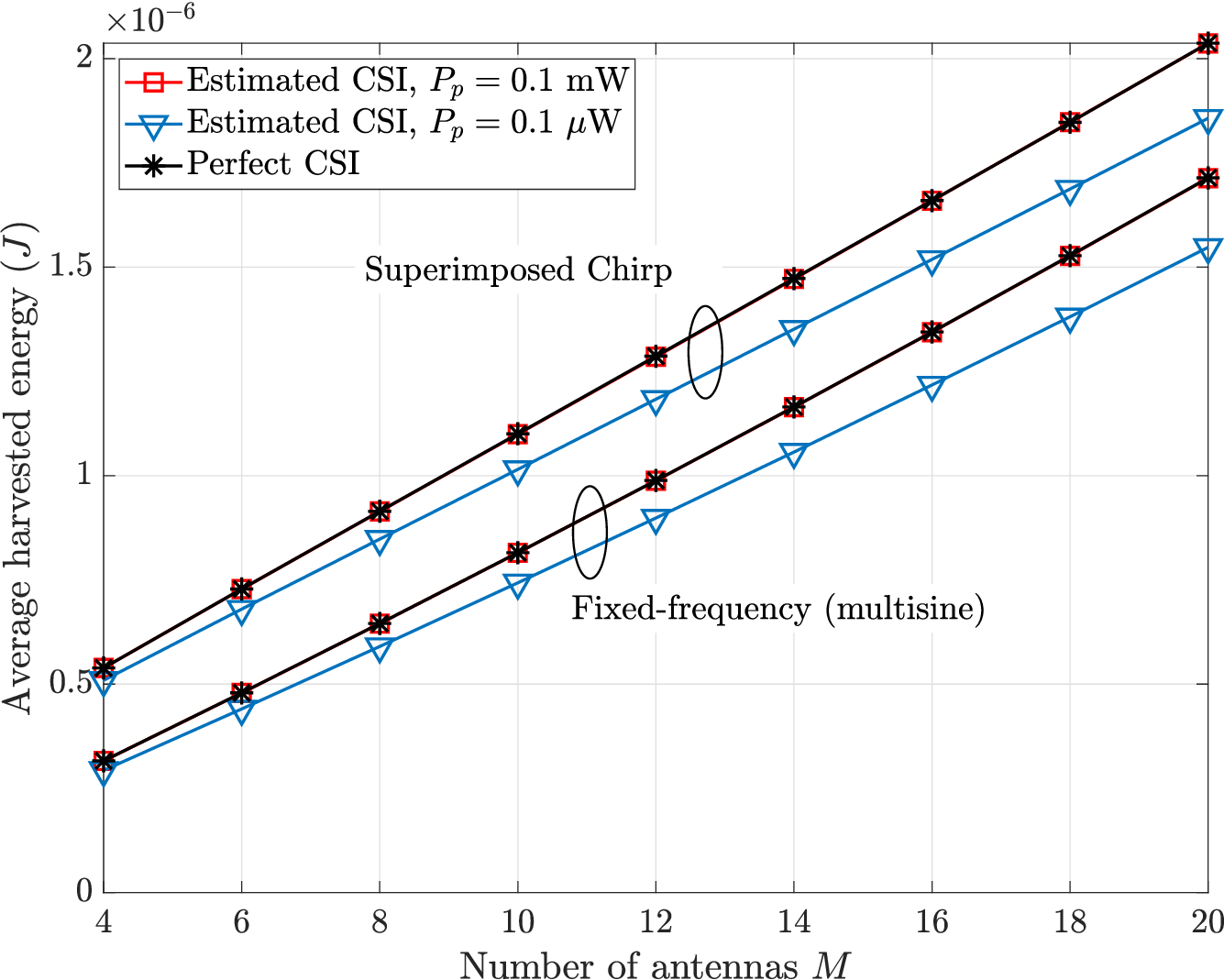}
\caption{Average harvested energy with respect to the number of antennas $M$ ($N=16$) \cite{Roy2023}.}\label{fig:chn_err_HE_chirp}
\end{figure}

\begin{figure}[t]\centering
	\includegraphics[width=0.85\linewidth]{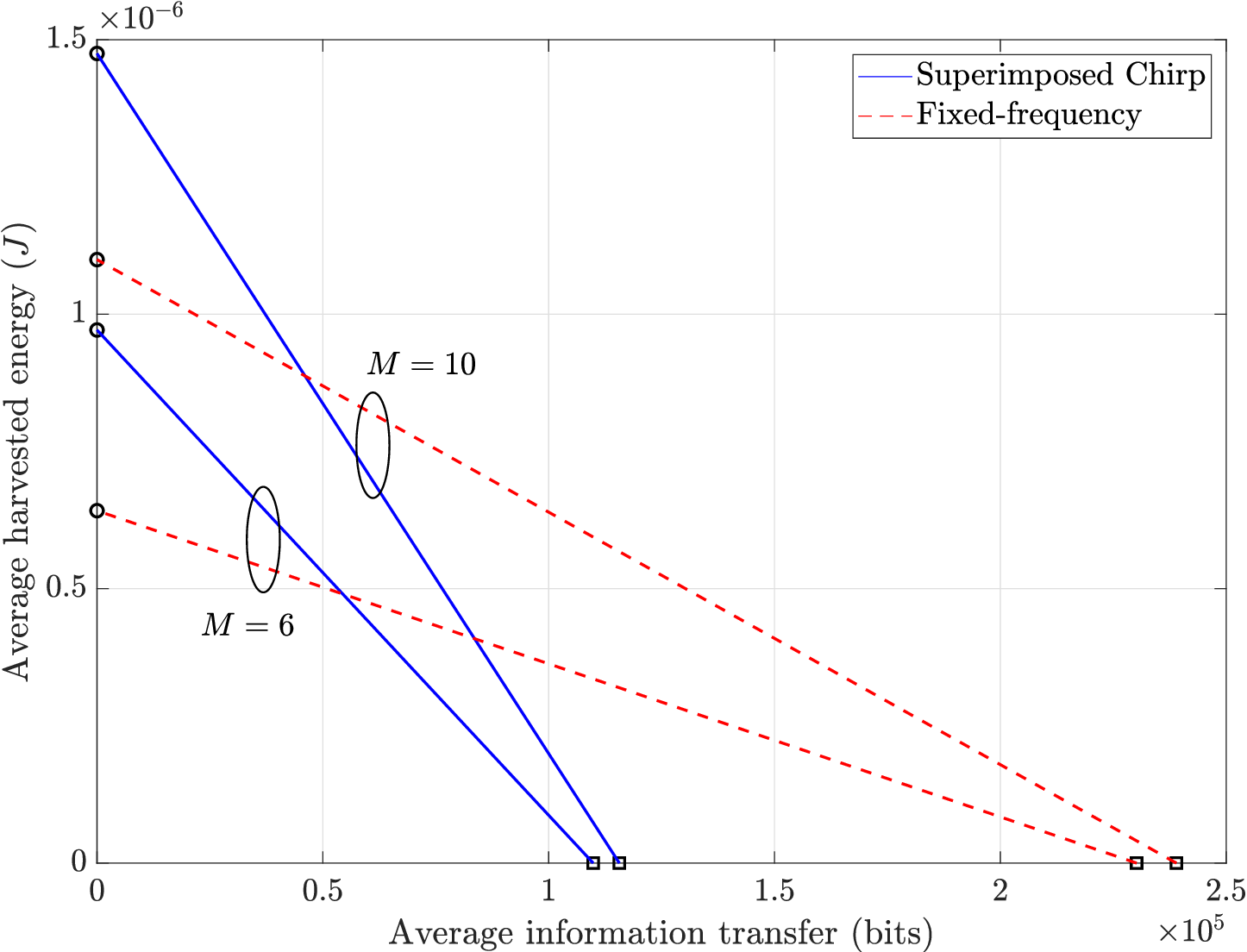}
	\caption{Energy-information transfer region ($M=6,10$ and $N=16$) \cite{Roy2023}.}
	\label{fig:avg_HE_info_chirp}
\end{figure}

Then, $r_{[n],1}$ is fed to a decision block and the associated spectral efficiency corresponding to the data symbol $d_{[n],1}$ over symbol duration $T$ is \cite{Yang2017}
\begin{equation}
R_1 \approx \sum_{n=1}^{\Nt} \sqrt{\frac{\rho_{[n],1}}{\rho_{[n],1} + 2}} \sum_{u=1}^\infty \frac{2(-1)^u}{\sqrt{1+\frac{2}{\rho_{[n],1}}} + 2u + 1},\label{eq:avg_received_info_chirp}
\end{equation}
where $\rho_{[n],1}$ is the average SINR given by \cite{Roy2023}
\begin{equation}
\rho_{[n],1} = \E\Bigg\{\frac{\varepsilon^2_1 \Pt T \eta_{[n]} \big|\widehat{\mathbf{g}}_{[n]}^\top {\pmb{\varphi}_{[n]}^\ast} \big|^2 d^2_{[n],1}}{\varepsilon^2_1 \Pt T \eta_{[n]} \big|\widetilde{\mathbf{g}}_{[n]}^\top {\pmb{\varphi}_{[n]}^\ast} \big|^2 d^2_{[n],1} + |w_{[n]}|^2}\Bigg\},\label{eq:avg_received_sinr_chirp}
\end{equation}
 and
\begin{align}
w_{[n]} = \frac{1}{\sqrt{T}} \int_0^T w_{[n]} (t) s_{[n]}(t) dt.
\end{align}

In a similar manner, we can obtain $\rho_{[n],2}$ and $R_{2}$. Thus, assuming imperfect CSI estimation, the amount of information $R$ (in bits) that is conveyed via the superimposed chirp based transmission to the RX over $N$ subbands for a duration of $\tau_\text{DL}$ is given by \cite{Roy2023}
\begin{equation}
R \approx \frac{\tau_\text{DL}}{T} (R_1 + R_2).
\label{eq:prop_infor_chirp_sensork}
\end{equation}

\subsubsection{Simulation Results}
Without loss of generality, we consider the following parameters: $N=16$, $\Nt = 8$, $B = 200$ kHz, $R_{\rm ant} = 1 ~\Omega$, $V_T = 25$ mV, $\delta \approx 1$, $I_s = 0.6$ mA \cite{Smida2021}, $P_{\rm tx} = 1$ W, and $\sigma^2 = 10^{-20}$. Fig. \ref{fig:chn_err_HE_chirp} illustrates the average harvested energy with respect to the number of antennas $M$ at the TX and the RX at a distance of $10$ m. The results clearly show the significant gains in harvested energy that can be achieved with superimposed chirp waveforms compared to multisine waveforms \cite{Clerckx16}. The multisine waveforms consist of a set of fixed-frequency cosine signals and the transmitted signal power in a subband is proportional to the subband gain. It is important to note that the proposed scheme with superimposed chirps allocates the power proportionally to the channel gains over the set of selected subbands. As a result, by exploiting frequency diversity, more energy is conveyed over the selected set of subbands having higher channel gains. For instance, for the case $M = 12$, the signal design based on superimposed chirps achieves a gain of around $30$\% in average harvested energy compared to fixed-frequency (multisine) waveforms. Furthermore, Fig. \ref{fig:chn_err_HE_chirp} illustrates the impact of pilot power on the average harvested energy for different values of $P_\text{p}$. We observe that as the pilot power increases, the quality of channel estimation improves (i.e., the channel estimation error decreases) and the average harvested energy increases. The results also indicate that for imperfect CSI with pilot power $P_\text{p} = 0.1$ mW, the harvested energy is very close to that obtained for perfect CSI.

Fig. \ref{fig:avg_HE_info_chirp} shows the energy-information transfer region for WIET with superimposed chirp and fixed-frequency waveforms for $M=6$ and $M=10$. The RX is placed at a distance of $9.1$ m from the TX. Since the superimposed chirp operates over a set of subbands $\Nt<N$ with high channel gains, more energy can be harvested compared to fixed-frequency transmission but the achieved information rate is lower (since the system does not operate over all $N$ bands). The case $\xi = 1$, $N = \Nt$, corresponds to the transmission of a single chirp waveform over all $N$ bands which attains the same WIET performance as the fixed-frequency waveforms. Thus, by varying $\xi$, the superimposed chirp provides flexibility and extra degrees of freedom that facilitate a trade-off in terms of harvested energy and information transfer.

\subsection{Near-field WIET}\label{sec:nearfield}
\graphicspath{{nearfield/}}
In order to facilitate the futuristic applications intended for the era of 6G, the performance requirements placed on communication and WET systems significantly surpass the capabilities of today's systems \cite{Saad20,Zhang19,Dang20}. The enabling features for meeting the performance targets impact the design of WIET and include operating at extremely high frequencies such as in the THz frequency band (Section \ref{sec:thz}), and employing antenna arrays with a very large aperture (Section \ref{sec:antenna}). The combination of these two enabling features, i.e., an extremely high operating frequency and very large antenna arrays, impacts the propagation characteristics of the EM wavefronts in the region where RXs are typically located compared to conventional systems.

Normally, planar EM wavefronts are assumed for the wireless channel, which is known as the far-field approximation. However, this assumption, on which the design of conventional WIET systems is based, may no longer be applicable for some WIET applications in the 6G era. The propagation environment of EM waves comprises the following three regions, which are distinguished by their prevalent EM wavefront type \cite{Yaghjian86}:
\begin{itemize}
\item In the \textbf{reactive near-field region}, the wireless channel can be modeled effectively by evanescent EM wavefronts. Typically, this region is limited to the space close to the transmit antenna, i.e., less than a few wavelengths. Within this region the EM field is primarily reactive.
\item In the \textbf{radiating near-field (Fresnel) region}, the wireless channel is modeled using spherical EM wavefronts. This region lies in between the reactive near-field region and the far-field region.
\item In the \textbf{far-field region}, the wireless channel can be described through planar wavefronts, which approximate the spherical wavefronts. Typically, the angular component of the EM field is approximated to be independent of the TX-RX distance within this region.
\end{itemize}
The reactive near-field region is generally small in size and so it is usually ignored in the literature \cite{NFBF}. Therefore, in what follows, the focus is placed on the WIET system design for the radiating near-field. Moreover, the term near-field pertains to the radiating near-field region. 

The Fraunhofer distance (or Rayleigh distance) is commonly utilized to define the boundary between the near-field and far-field regions. The size of the Fresnel region is given by $d_F = 2L_\mathrm{A}^2 / \lambda$ and depends on the wavelength $\lambda$ and the diameter of the antenna array's aperture $L_\mathrm{A}$ \cite{NFBF}. The reactive near-field extends to the Fresnel distance  $d_{\rm Fresnel} = \sqrt[3]{\frac{L_{AE}^4}{8\lambda}}$, where $L_{AE}$ denotes the aperture of an array element \cite{7942128}. While the near-field could often be deemed negligibly small in conventional wireless systems, when operating in the mmWave band or at even higher frequencies, the near-field may span a region which is relevant, e.g., for indoor applications, when employing a large antenna array.
To illustrate this point, consider two wireless systems with $L_\mathrm{A}=20$ cm operating at $2.4$ GHz and $60$ GHz, respectively. 
The Fresnel region of the first system is limited to a distance of $d_F = 64$ cm from the transmit antenna array. 
In contrast, the near-field of the second system spans $d_F = 16$ m around the antenna array.
Moreover, since the size of the Fresnel region increases with decreasing wavelength, accounting for the spherical nature of the EM wavefronts will be necessary for a large range of applications, especially when operating in the THz band.

In summary, the far-field approximation may indeed no longer be appropriate for certain applications when designing WIET systems in the 6G era that employ a large antenna array in conjunction with a high operating frequency.
Naturally, this requires a suitable channel model for capturing the effects of the spherical nature of the EM wavefronts.

\begin{figure}[t]\centering
	\includegraphics[width=0.8\linewidth]{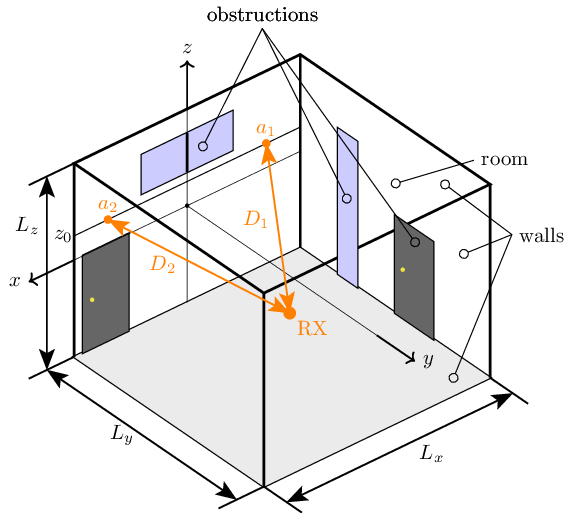}
	\caption{Illustration of the system model with two transmit antennas \cite{Mayer23}.}\label{fig: System Model Overview}
\end{figure}

\subsubsection{Near-Field Channel Model}
Modeling the wireless channel in the near-field requires a spherical wavefront model (SWM), whereas in the far-field, a uniform planar wavefront model (PWM) approximates the spherical nature of the EM wavefronts sufficiently well.

Moreover, the line-of-sight (LoS) component is predominant in wireless channels at high frequencies, such as mmWave and beyond.
This is due to the severe reflection and scattering losses in these bands \cite{Zhang22}, \cite{Zhang23}. 
Consequently, an LoS channel is typically assumed for wireless systems operating in the Fresnel region \cite{NFBF}, \cite{Zhang23}.
When considering LoS conditions, the equivalent complex baseband link between an antenna array element at the TX and the RX is defined as 
\begin{align}\label{eq: LOS Channel}
g_{mn} = \frac{\sqrt{c}}{D_{mn}} \exp\left(-j\frac{2\pi}{\lambda} D_{mn}\right),
\end{align}
where $c$ is the channel power gain at the reference distance of $1$ meter and $D_{mn}$ is the distance between the $m$-th array element at the TX and the $n$-th element at the RX. Both the amplitude and the phase of $g_{mn}$ depend on distance $D_{mn}$ since the wireless channel is modeled using the SWM. In contrast, for the PWM, the path loss between all elements of the transmit antenna array and the receive antenna array is approximately constant, i.e., $D_{mn} \approx D, \forall m,n$. Note that the phase of $g_{mn}$ depends on $D_{mn}$ for both the SWM and the PWM.

Therefore, the spherical nature of the EM wavefronts is reflected by the amplitude variations in the wireless channel. The far-field channel model can be considered as a special case of the near-field channel model when applying the far-field approximation. The accuracy of this approximation depends on whether the system is operating in the near-field region or not. However, when operating in the near-field, neglecting this information results in an inadequate channel description regarding the WIET system.

\subsubsection{Beam Focusing in Near-Field WIET Systems}
The design of wireless systems for the Fresnel region is of interest for communications \cite{NFBF} and WET \cite{Zhang22}. By exploiting the spherical nature of the EM wavefronts, beams can not only be steered in a certain direction, but instead one may focus them precisely on the desired location \cite{Zhang23}. Thus, beamforming in the Fresnel region is often referred to as beam focusing \cite{Zhang23}. Beam focusing yields a larger received power compared to conventional beam steering, which is relevant to communications as well as WET \cite{Zhang22, Zhang23}. Besides increasing the energy transfer efficiency towards the RXs, pollution of the remaining environment by energy beams is reduced compared to conventional beam steering \cite{Zhang23}. Consequently, this decreases the exposure to radiation at undesired locations in the environment and it reduces interference in multi-user communications \cite{NFBF, Zhang23}.

By capturing the spherical nature of the EM wavefronts in the channel model \eqref{eq: LOS Channel}, novel design criteria are established, which were not considered in the far-field. The following system design provides an example where the distance information contained in the amplitude variations of the wireless channel is exploited. Moreover, a comparison to a far-field system, which neglects this distance information, is drawn. Note that the presented concept may be extended to a WIET system, e.g., through a time sharing protocol.

In particular, we investigate the optimal transmit antenna placement for a MISO WET system in the Fresnel region \cite{Mayer23}. Specifically, a three-dimensional cuboid room is considered under LoS conditions, where the environment is defined through its width $L_x$, depth $L_y$, and height $L_z$, as depicted in Fig. \ref{fig: System Model Overview}. The RX is located in the room, whereby portions of the wall are obstructed. There are two transmit antennas placed on a horizontal line of height $z_0$ at one of the room's walls. The antennas are located at $a_1$ and $a_2$ on the horizontal line and their distances to the RX are given by $D_1$ and $D_2$, respectively.

\begin{figure}[t]\centering
	\includegraphics[width=0.85\linewidth]{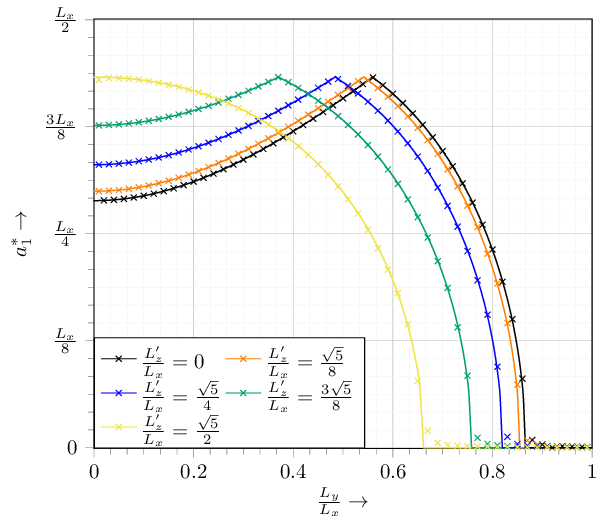}
	\caption{Optimal placement of antenna $a_1^*$ \cite{Mayer23}.}
	\label{fig: Optimal a_1 position vs. L_y/L_x}
\end{figure}

The antenna placement is optimal if it maximizes the received power $\gamma$ at the worst possible RX location by considering the spherical nature of the EM wavefronts in the Fresnel region while assuming perfect knowledge of the channel at the TX. The received power $\gamma$ is maximized by matching the transmit signal $\boldsymbol{s} \in \mathbb{C}^{N_t \times 1}$ to the channel $\boldsymbol{g} \in \mathbb{C}^{N_t \times 1}$, i.e., $\boldsymbol{s} = \sqrt{P_\mathrm{tx}} \, \boldsymbol{g} / \| \boldsymbol{g} \|$, where the total transmit power $P_\mathrm{tx}$ is shared among the $N_t$ antennas. Every component of the MISO channel is given by \eqref{eq: LOS Channel}, whereby the subscript of the RX is dropped for better legibility, i.e., $\boldsymbol{g} = [g_{1},\dots,g_{N_t} ]^T$. Therefore, the received power $\gamma$ is given by
\begin{align}\label{eq: RX Signal Power}
\gamma = \lVert \boldsymbol{g}^H \boldsymbol{s} \rVert_2^2 = P_\mathrm{tx} \,\lVert \boldsymbol{g} \rVert_2^2 = P_\mathrm{tx} c\sum_{i=1}^{N_t} \frac{1}{D_{i}^2}.
\end{align}
Thus, the received power is directly proportional to the sum of the squared, inverse distances $D_i$, $\forall i=1,\dots,N_t$, between the RX and the transmit antennas. We emphasize that the optimal transmit placement is specific to the indoor environment under consideration and is designed for a WET system with a single RX. The proposed methodology can be easily extended to multiple RXs through time division multiple access (TDMA). Indeed, in \cite{9729101}, TDMA and space division multiple access (SDMA) protocols were considered for an indoor WET system with radio stripes at the TX and multiple RXs. Hereby, TDMA was found preferable when supplying power to a small number of RXs and SDMA was deemed beneficial for supplying a large number of users. An extension of our proposed methodology to support a concurrent supply of power to multiple RXs is an interesting topic for future work. However, the resulting optimal transmit antenna placement may no longer be representable via a closed-form analytical solution. When considering $N_t=2$, the maximum received power at the locations yielding the worst performance is quantified by \cite{Mayer23}
\begin{align}\label{eq: Optimal received power}\hspace{-2mm}
\gamma^* = P_\mathrm{tx}c
\begin{cases}
\frac{2}{\frac{4}{3}  L_y^2 + \frac{1}{12}  L_x^2 + \frac{1}{3}{L_z^\prime}^2} &\!\text{if $4 \frac{L_y^2}{L_x^2} + \frac{{L_z^\prime}^2}{L_x^2} \leq \frac{5}{4}$}, \\[2mm]
\frac{2\sqrt{4\frac{L_y^2}{L_x^2} + \frac{{L_z^\prime}^2}{L_x^2}+1}+2}{4L_y^2 + {L_z^\prime}^2} &\!\text{if $\frac{5}{4} \leq 4 \frac{L_y^2}{L_x^2} + \frac{{L_z^\prime}^2}{L_x^2} \leq 3$}, \\
\frac{2}{L_y^2 + \frac{1}{4}  L_x^2 + \frac{1}{4}  {L_z^\prime}^2 } &\!\text{if $4 \frac{L_y^2}{L_x^2} + \frac{{L_z^\prime}^2}{L_x^2} \geq 3$},
\end{cases}
\end{align}
where $L_z^\prime = L_z + 2 \vert z_0 \vert$. The optimal transmit antenna positions $a_1^*$ and $a_2^*$ for achieving $\gamma^*$ satisfy $a_2^* = -a_1^*$ and they are determined by the geometry of the environment.
Specifically, $a_1^*$ is defined as a closed-form expression depending on the parameters $L_x$, $L_y$, and $L_z^\prime$, which is visualized in Fig. \ref{fig: Optimal a_1 position vs. L_y/L_x} and found in \cite{Mayer23}.

\begin{figure}[t]\centering
	\includegraphics[width=0.85\linewidth]{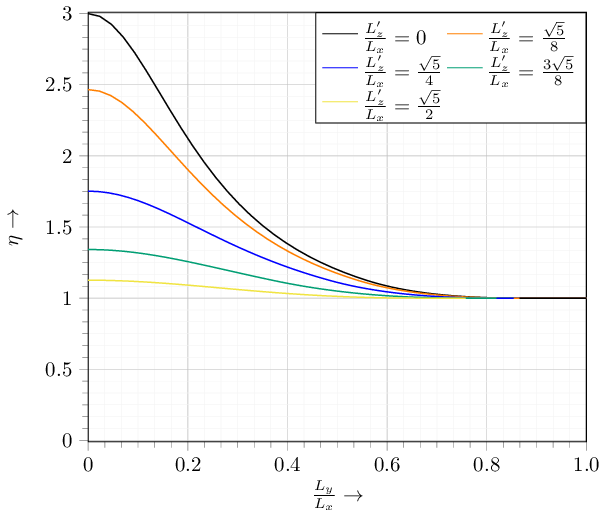}
	\caption{Power gain $\eta$ compared to the optimal far-field solution \cite{Mayer23}.}\label{fig: Gain}
\end{figure}

The optimal antenna placement for the far-field is $a_1^*=a_2^*=0$, which is valid if the environment contains locations where the far-field approximation holds. As shown in Fig. \ref{fig: Optimal a_1 position vs. L_y/L_x}, for certain environment geometries the far-field solution is optimal in general, and thus, the WET system does not have to be designed for operation in the Fresnel region. However, when the solution deviates from the far-field placement, operating in the Fresnel region is necessary for the solution to be optimal, i.e., the environment must be contained within $d_\mathrm{F}$.  By considering the spherical nature of the EM waves and utilizing this information for optimizing the transmit antenna placement for $N_t=2$, the received power can be increased by up to a factor of $3$. This power gain compared to the optimal far-field solution is presented in Fig. \ref{fig: Gain}. An extension of the methodology to determine the optimal antenna placement for systems with $N_t > 2$ is briefly discussed in \cite{Mayer23}. Hereby, an asymmetrical placement is deemed suboptimal in general.

\section{6G Antenna Technologies for WIET Systems}\label{sec:antenna}
This section investigates the crucial role that advanced antenna technologies, namely H-MIMO, IRSs, and fluid RAs, play in 6G WIET systems. The benefits provided by these technologies for WIET are complementary. Specifically, H-MIMO improves WIET with the appropriate design at the TX (or RX) side, IRSs increase the overall WIET efficiency by controlling the propagation environment, and fluid RAs enhance the WIET performance at the RX (or TX) through their reconfigurability. H-MIMO is expected to succeed classical MIMO and massive MIMO (M-MIMO) in future 6G communication networks due to its ability to offer unprecedented high energy efficiency, low latency, and massive connectivity \cite{Hu2018}. It is best suited for high-demand scenarios, where ultra-high data throughput and spectral efficiency are required. The IRS technology can increase the data rates, enhance the capacity, extend the coverage, and improve the reliability of wireless communication systems \cite{IRSSurvey}. Due to the passive nature of its elements, it is more suitable for scenarios, where cost and power efficiency are critical. Finally, RAs are flexible antennas with reconfigurable properties, which can provide new degrees of freedom for the design of 6G wireless communication systems \cite{martinez2022}. They are particularly useful in scenarios involving dynamic environments or multiple spectrum bands and signal waveforms that are tailored to different applications (e.g., 5G/6G services) as well as in use cases that require enhanced flexibility. In terms of complexity, accurate CSI acquisition for all three technologies is challenging, especially for H-MIMO, due to the large number of densely deployed antenna elements.

In the following, we review the historical evolution of MIMO communication networks and discuss the recent progress in the development of MIMO WIET systems (Sub-sections \ref{sec:mimo} and \ref{sec:mmimo}). Then, we focus on each of the above mentioned antenna technologies separately and study their performance in the context of WIET (Subsections \ref{sec:hmimo} to \ref{sec:ras}).

\graphicspath{{mimo/}}
\subsection{MIMO WIET}\label{sec:mimo}
\begin{figure}[!t]\centering
	\includegraphics[width=\linewidth]{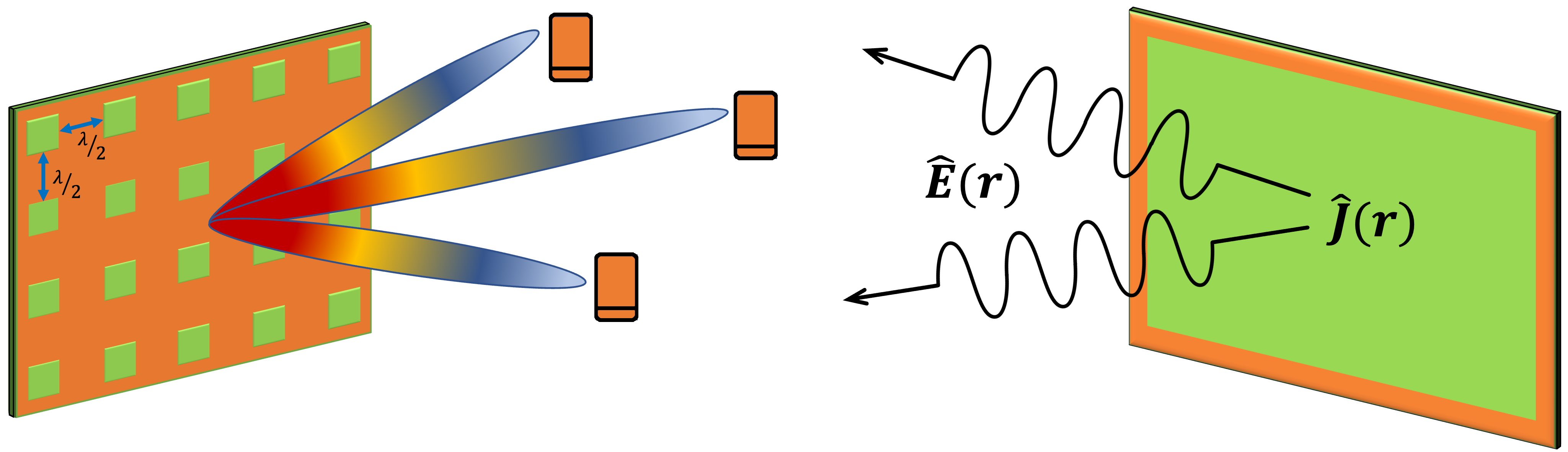}
	\caption{Classical MIMO TX (left) with $\lambda/2$ separation between antenna elements and H-MIMO TX (right) with a nearly-continuous antenna aperture generating a time-harmonic EM wave $\hat{\boldsymbol{E}}(\boldsymbol{r})$.}\label{hmimo}
\end{figure}

Early works on WIET-based communication systems have shown that multi-antenna TXs provide significantly higher data rate and more harvested power than TXs with a single antenna \cite{Zhang2013, Liu2014}.
In classical MIMO communication systems, TXs and RXs employing a finite number of half-wavelength spaced antenna elements, as shown in Fig. \ref{hmimo}, exploit multipath signal propagation to provide diversity and multiplexing gains, and thus, improve the overall system performance \cite{Telatar1999, Chuah2002, Tse2005}.
In a typical multi-antenna TX (RX), each antenna element is connected to a dedicated RF chain composed of an RF low-noise power amplifier, a local oscillator, a mixer, and a digital-to-analog (analog-to-digital) converter followed by a computational unit for digital processing of the transmit (received) signals.
This approach is referred to as digital precoding (digital combining) and enables a flexible control of the transmit (received) beampattern, i.e., transmission (reception) of a certain amount of power in a given spatial direction.
However, for devices equipped with hundreds or even thousands of antennas, such an architecture may be costly and impractical \cite{Ahmed2018}.
Furthermore, for multi-antenna users in WET systems, one can employ channel-adaptive RF domain combining for processing of the signals received by different antenna elements to feed the total received power to a single EH circuit.
In practice, such RF signal combining schemes may be infeasible due to the computational capabilities required for channel estimation and the insertion losses of practical power dividers and phase shifting devices \cite{Shanin2021a}.
Therefore, each antenna of a user is typically connected to a dedicated EH circuit, whose output signals are transferred to a load device or stored in a battery \cite{Shen2020}.

In single-user MIMO WIET systems, maximum ratio transmission beamforming at the TX was shown to maximize the received power at the user device, and thus, is optimal \cite{Clerckx2018}.
However, for multi-user MIMO information transmission, the optimal signal precoding scheme should not only maximize the received signal power at the user devices, but also suppress the interference between the transmit data streams of individual users \cite{Goldsmith2005}.
Therefore, modern 5G communication networks utilize practical linear zero-forcing (ZF) and MMSE precoding schemes, which are able to effectively suppress the inter-user interference \cite{Sun2010}.
Furthermore, it is known that the capacity of Gaussian multi-user MIMO broadcast channels is achieved by dirty-paper precoding at the TX \cite{Caire2003}.
However, this precoding scheme entails a higher computational complexity than the ZF and MMSE approaches.

For multi-user MIMO WET, a single transmit energy beamforming vector, which is collinear with the dominant eigenvector of the channel matrix, is shown in \cite{Zhang2013} to be optimal for maximizing the sum of harvested powers the EH RXs by considering the linear EH model in \eqref{linear_model}. Since EH circuits are in general non-linear, the energy beamforming in \cite{Zhang2013} may not be optimal for practical multi-user MIMO WET systems \cite{Clerckx2019}. Based on the non-linear sigmoidal saturation EH model, given by \eqref{nl_sat_model} \cite{7264986}, the beamforming design for WET-based communication networks has been thoroughly studied in the literature, see, e.g., \cite{Boshkovska2018, Xiong2017, Ma2019}. Furthermore, utilizing the circuit-based non-linear EH model in \eqref{nl_circ_model} to characterize the instantaneous power at the user device, the authors of \cite{Shanin2021a} demonstrate that two beamforming vectors are needed for optimal multi-user MIMO WET. Although the optimal transmission policy for MIMO WET is known, obtaining the optimal performance of multi-user MIMO WIET taking into account the non-linear effects of practical EH circuits is challenging. Moreover, to achieve high spatial efficiencies and mitigate path loss, a large number of transmit antenna elements are needed, which, in turn, requires novel hybrid signal precoding algorithms at the TX for MIMO WIET.

\subsection{M-MIMO WIET}\label{sec:mmimo}
Since signal processing at multi-antenna TXs and RXs may be computationally complicated, the authors of \cite{Larsson2014} propose an M-MIMO system, where an asymptotically infinite number of antenna elements is adopted at the TX to generate extremely narrow beams.
The key for the theoretical analysis of M-MIMO is the asymptotic orthogonality assumption of the channels between the TX and the RXs, which holds, e.g., for Rayleigh fading channels, if the ratio between the number of RX and TX antennas tends to zero and the number of TX antennas tends to infinity \cite{Bjornson2016}. 
In particular, maximum ratio transmission beamforming, which is known to be optimal for single-user MIMO systems, is able to not only maximize the received signal power at the RXs, but also suppresses the inter-user interference, and hence, is optimal for M-MIMO information transmission \cite{Nayebi2017}.
Thus, signal processing schemes for M-MIMO are typically less computationally expensive than those used in classical MIMO systems.

For multi-user M-MIMO WET with the linear EH in \eqref{linear_model}, the optimal energy beamforming vector reduces to a weighted sum of the normalized maximum ratio transmission beamforming vectors of the different users \cite{Yang2015}.
Furthermore, accounting for the non-linear effects of practical EH circuits, by utilizing the EH model in \eqref{nl_circ_model}, the authors of \cite{Shanin2023} prove that similar to \cite{Shanin2021a}, the optimal transmission policy minimizing the average transmit power for multi-user M-MIMO WET also employs multiple beamforming vectors, which are obtained as a weighted sum of the normalized channel vectors.
However, the derivation of the optimal transmit signal waveform and beamforming for M-MIMO WIET systems taking into account the non-linearities of practical EH circuits is still challenging.

\subsection{H-MIMO WIET}\label{sec:hmimo}
H-MIMO is an extension of M-MIMO, where not only the antenna aperture at the TX is assumed large, but also the spacing between the antenna elements is negligibly small \cite{Hu2018, Chafii2023}. Such nearly-continuous antenna surfaces, see Fig. \ref{hmimo}, employ programmable electromagnetic (EM) excitable elements, which are envisioned to be accurately tunable, low cost, and low power and should allow TXs not only to transmit the RF signal in a certain direction, but to also fully control the radiated EM wave \cite{Gong2023}. Furthermore, by facilitating an unprecedented flexibility in controlling the EM radiation with RF-domain signal processing, these holographic antennas can provide extremely high energy efficiency for future WIET systems. Moreover, they can be utilized at high frequencies, e.g., in the THz frequency band ($\SI{300}{\giga\hertz}-\SI{3}{\tera\hertz}$), where the path loss is significantly higher compared to that at lower frequencies \cite{Luo2023, Yi2023, Wan2021}.

In contrast to traditional half-wavelength spaced MIMO antennas, where individual RF feeding lines are used for each antenna element, the RF feeds of H-MIMO TXs are employed to generate reference EM waves, which first interfere inside the holographic surface to compose a desired EM signal, and then, excite the closely spaced elements of the nearly-continuous TX antenna, see Fig. \ref{hmimo} \cite{Huang2020}.
Moreover, each antenna of the holographic metasurface can tune the amplitude of the radiated signal to provide further flexibility in controlling the output EM wave.
From a physical perspective, such holographic surfaces are realized using leaky-wave antennas based on metamaterials, where the antenna elements are LC resonators controlled by PIN-diodes \cite{Araghi2021, Deng2023}.
The experimental results in \cite{Deng2023} demonstrate that H-MIMO metasurfaces are able to efficiently generate a desired transmit far-field beampattern.
Furthermore, since mutual coupling between antenna elements may compromise H-MIMO performance \cite{DAmico2023}, a decoupling procedure increasing the overall H-MIMO antenna gains is proposed in \cite{Kim2022a}.

The propagating EM waves are fully characterized by Maxwell's equations, which determine the connection between the electrical and magnetic fields and can be described in differential form as follows \cite{Bladel2007}:
\begin{align}
\nabla \times \boldsymbol{E}(\boldsymbol{r}, t) &= - \frac{\partial}{\partial t} \boldsymbol{B}(\boldsymbol{r}, t), \label{Eqn:HMIMO_Maxwell1}\\
\nabla \times \boldsymbol{B}(\boldsymbol{r}, t) &= \frac{1}{c_0^2} \frac{\partial}{\partial t} \boldsymbol{E}(\boldsymbol{r}, t) + \mu_0 \boldsymbol{J}(\boldsymbol{r}, t), \label{Eqn:HMIMO_Maxwell2}\\
\nabla \cdot \boldsymbol{E}(\boldsymbol{r}, t) &= \frac{\boldsymbol{\rho}(\boldsymbol{r}, t)}{\epsilon_0}, \label{Eqn:HMIMO_Maxwell3}\\
\nabla \cdot \boldsymbol{B}(\boldsymbol{r}, t) &= 0. \label{Eqn:HMIMO_Maxwell4}
\end{align}
Here, $\boldsymbol{E}(\boldsymbol{r}, t)$, $\boldsymbol{B}(\boldsymbol{r}, t)$, $\boldsymbol{\rho}(\boldsymbol{r}, t)$, and $\boldsymbol{J}(\boldsymbol{r}, t)$ are the electrical field, magnetic field, charge density, and current density, respectively, as functions of location $\boldsymbol{r}$ and time $t$.
Furthermore, $\epsilon_0$, $\mu_0$, and $c_0 = (\epsilon_0 \mu_0)^{-\frac{1}{2}}$ denote the electric permittivity, magnetic permeability, and speed of light of the propagation medium, respectively.
Finally, $\nabla \times$ is the curl operator and $\nabla \cdot$ is the divergence operator.
Since practical EM waves are time-harmonic, the electrical field, magnetic field, and current density can be represented as $\boldsymbol{E}(\boldsymbol{r}, t) = \Re\{\hat{\boldsymbol{E}}(\boldsymbol{r}) \exp(j 2\pi f_c t)\} $, $\boldsymbol{B}(\boldsymbol{r}, t) = \Re \{\hat{\boldsymbol{B}}(\boldsymbol{r}) \exp(j 2\pi f_c t)\}$, and $\boldsymbol{J}(\boldsymbol{r}, t) = \Re\{\hat{\boldsymbol{J}}(\boldsymbol{r}) \exp(j 2\pi f_c t)\} $ with location-dependent envelope fields $\hat{\boldsymbol{E}}(\boldsymbol{r})$, $\hat{\boldsymbol{B}}(\boldsymbol{r})$, and $\hat{\boldsymbol{J}}(\boldsymbol{r})$, respectively, where $f_c$ is the carrier frequency.
Thus, the time-harmonic envelope $\hat{\boldsymbol{E}}(\boldsymbol{r})$ of the electrical field defined in \eqref{Eqn:HMIMO_Maxwell1}-\eqref{Eqn:HMIMO_Maxwell4} can be equivalently represented as \cite{Bladel2007}:
\begin{align}
\nabla \times (\nabla \times \hat{\boldsymbol{E}}(\boldsymbol{r})) - k_0^2 \hat{\boldsymbol{E}}(\boldsymbol{r}) = j \epsilon_0 \mu_0 \hat{\boldsymbol{J}}(\boldsymbol{r}),\label{Eqn:HMIMO_WaveEqn}
\end{align}
where $k_0$ is the wave number. Note that the envelope of the magnetic field in \eqref{Eqn:HMIMO_Maxwell1}-\eqref{Eqn:HMIMO_Maxwell4} can be derived similarly.

The wave equation \eqref{Eqn:HMIMO_WaveEqn} reveals the dependence between the current density $\hat{\boldsymbol{J}}(\cdot)$ at the TX surface and the electrical field $\hat{\boldsymbol{E}}(\cdot)$ of the generated EM wave (see Fig. \ref{hmimo}).
Thus, by calculating the electrical field at an RX antenna, one can define the H-MIMO system from the EM wave perspective.
We note that the model in \eqref{Eqn:HMIMO_WaveEqn} differs substantially from that of traditional linear MIMO/M-MIMO systems.
Since holographic TX surfaces with vanishing antenna spacing are able to generate arbitrary transmit EM waves satisfying \eqref{Eqn:HMIMO_WaveEqn}, it may be possible to derive the optimal continuous transmit wave maximizing the H-MIMO data rate, and hence, defining the capacity of the continuous space \cite{An2023P3}. 
Blending the frameworks of Shannon's information theory and EM wave theory is an interesting direction for future research on H-MIMO communication networks.
In preliminary works \cite{Gong2023a} and \cite{Zhang2023}, the authors combine Shannon's results with Maxwell's equations to characterize the capacity limits of H-MIMO.
Furthermore, the statistical characteristics of H-MIMO channels are studied and estimated based on EM wave theory in \cite{Pizzo2022, Ghermezcheshmeh2023, Demir2022, Wei2023}.
Moreover, exploiting OFDM and hybrid digital-analog MIMO precoding, a wavenumber-division multiplexing scheme and hybrid digital-EM beamforming are proposed in \cite{Sanguinetti2023} and \cite{Deng2022}, respectively.
Finally, since holographic surfaces enable high spatial resolution, integrated sensing and communication systems utilizing H-MIMO are considered in \cite{Zhang2022a, Zhang2023a, Adhikary2023}.

For WET, the flexibility and near-continuity of holographic TX surfaces enable extremely high spatial energy efficiencies and an accurate tuning of the power distribution among EH RXs \cite{Zhou2022}.
In particular, the authors in \cite{Yurduseven2017} numerically demonstrated that holographic surfaces based on PIN-diodes are able to efficiently focus a high amount of power at an EH RX located in the near-field Fresnel zone.
Similar to \cite{Shanin2021a, Clerckx2018, Zhang2013, Boshkovska2017a, Liu2014}, and other related works, one can derive the EM signal at the H-MIMO TX maximizing the average harvested power at multiple user devices.
Finally, although transmit EM wave design for integrated information and energy transfer is challenging, the flexibility offered by holographic matasurfaces may enable a qualitative theoretical analysis of H-MIMO WIET.

\graphicspath{{reconfigurable/}}
\subsection{IRS-assisted WIET}\label{sec:IRS}
Next, we discuss the emerging frontier research field of wireless communication technologies with reconfigurable and flexible characteristics, which can accommodate the rising demands of future networks. Towards this end, IRSs have been recently proposed to control the propagation environment via software-controlled	metasurfaces \cite{liaskos2018}. An IRS is a programmable planar surface equipped with a large number of passive elements that reflect the incident radio signals whilst controlling their amplitude or/and phase \cite{liu2021b}. In this way, IRSs can reconfigure the wireless channel between a TX and a RX. The extensive interest in IRSs and the vast number of research results reported in the last five years, e.g., \cite{liaskos2018, liu2021b, basharat2021}, are indicators for the benefits arising from reconfigurability, flexibility, and intelligence in wireless networks.

This emerging paradigm dynamically reconfigures the wireless channel via passive beamforming (PB) \cite{IRS2}. An IRS can strike a desirable balance between the conflicting information and energy transfer requirements, e.g., by effectively managing the interference, which deteriorates radio communication, yet constitutes a green RF energy source, or/and by leveraging PB gains to form local wireless charging hotspots. Consequently, several studies have investigated the joint optimization of the transmit and reflect beamformers in a variety of IRS-aided WIET setups and scenarios. In \cite{WCL2}, the authors formulate an MOO problem for maximizing the sum rate and the sum harvested energy in a multi-user MISO downlink system with separate information and energy RXs, while the authors of \cite{IRSSWIPT0, IRSSWIPT1}, and \cite{IRSSWIPT2} focus on maximizing the minimum power, maximizing the weighted sum-power, and minimizing the transmit sum-power, respectively. The latter problem is revisited for an equivalent setup with PS RXs in \cite{IRSSWIPTImpCSI} and \cite{NtougiasProb}, assuming imperfect CSI and interference constraints under a spectrum underlay regime, respectively, whereas the authors of \cite{Proposed1} explore the weighted sum rate maximization problem for a MIMO broadcasting setup. Also, the authors of \cite{WCL1} propose a scheme that maximizes a utility function which balances the efficiency of information and energy transmission and \cite{Provable2} tackles the max-min energy efficiency fairness problem.

In what follows, we focus on two IRS-assisted SWIPT use cases that are of utmost importance for the envisioned 6G paradigm. In particular, we consider $(i)$ an analog/digital (A/D) TX that applies hybrid precoding \cite{NtougiasHPIRS}, and $(ii)$ a battery-powered IRS that utilizes TS or PS to harvest RF energy.

\subsubsection{Joint Hybrid Precoding/Reflect Precoding for SWIPT}
MIMO technology is often utilized in SWIPT systems to enhance both the data rate and transfer of energy via transmit precoding. Nevertheless, the stringent cost and energy consumption constraints of the TXs in IoT setups limit the number of antennas and, therefore, the corresponding performance gains, under a fully-digital (FD) implementation. The adoption of an A/D architecture combining a digital baseband precoder with an analog RF precoder (RFP), which is commonly realized via a network of phase shifters, has been considered as a workaround to this problem \cite{Masouros,NtougiasHP}. This implementation utilizes fewer RF chains than antennas. A hybrid precoder (HP) can be combined with IRS-based PB to further boost the performance. Joint HP/PB optimization has been considered for IRS-assisted MISO broadcasting systems in scenarios that involve millimeter-wave (mmWave) communication \cite{HPIRS, HPIRS2} or massive MISO setups \cite{HPIRS3}. Both of these setups limit the applicability of the derived schemes. Likewise, the work in \cite{IRSHyb}, considers secure mmWave SWIPT with separate information and energy RXs and focuses on a decoupled HP/PB design that utilizes a codebook-based RFP.

We focus on joint HP/PB designs that do not require an excessive number of TX antennas or sparse mmWave channels \cite{NtougiasHPIRS}. Furthermore, we consider a more general PS setting, which encompasses separate information and energy RXs setups as a special case. 

\begin{figure}[!t]\centering
	\includegraphics[width=\linewidth]{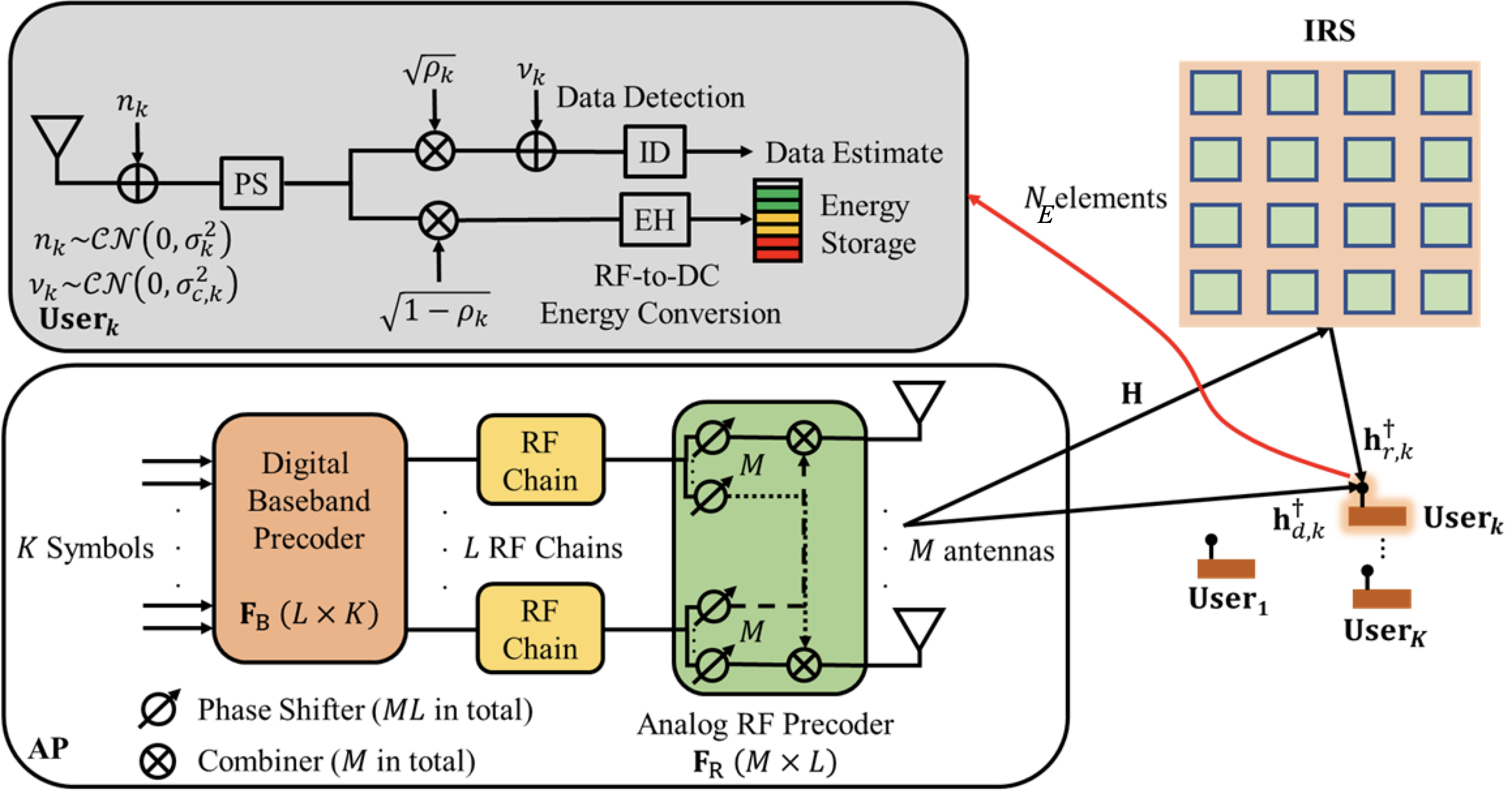}
	\caption{A fully-connected A/D TX, an IRS, and a RX using PS.}
	\label{fig:KN4}
\end{figure}

\paragraph{System Model}
The considered system consists of $K$ single-antenna WIET RXs using PS and a TX with $M$ antennas and $L$ RF chains ($K \leq L < M$). A fully-connected A/D structure and an IRS with $N_E$ passive reflecting elements are employed, as shown in Fig. \ref{fig:KN4}. We assume a quasi-static, flat fading channel model. The direct TX--RX $k$, reflecting IRS--RX $k$, and incident TX--IRS baseband equivalent channels are denoted by $\mathbf{h}_{d,k}^\text{H} \in \mathbb{C}^M$, $\mathbf{h}_{r,k}^\text{H} \in \mathbb{C}^{N_E}$, and $\mathbf{H} \in \mathbb{C}^{N_E\times M}$, respectively, $\forall k \in \mathcal{K}$, where $\mathcal{K}\triangleq \{1,\dots,K\}$. Then, the IRS-assisted and the effective TX--RX $k$ channel, $\mathbf{H}_k \in \mathbb{C}^{N_E\times M}$ and $\mathbf{h}_k^\text{H} \in \mathbb{C}^{M}$, are defined as $\mathbf{H}_k \triangleq \operatorname{diag}(\mathbf{h}_{r,k}^\text{H})\mathbf{H}$ and $\mathbf{h}_k^\text{H} \triangleq \mathbf{h}_{d,k}^\text{H} + \mathbf{h}_{r,k}^\text{H}\bm{\Theta}\mathbf{H} = \mathbf{h}_{d,k}^\text{H} + \bm{\theta}^\text{H}\mathbf{H}_k$, respectively, where $\bm{\Theta} = \operatorname{diag}\left(e^{j\theta_{1}},\dots,e^{j\theta_{N_E}}\right)$ and $\bm{\theta}=\operatorname{diag}\left(\bm{\Theta}^{*}\right).$

Let $\mathbf{F}_{\mathrm{B}} \triangleq\left[\mathbf{f}_1, \ldots, \mathbf{f}_K\right] \in \mathbb{C}^{L \times K}$ denote the digital baseband precoder, where $\mathbf{f}_k \in \mathbb{C}^L$ is the digital precoding vector assigned to RX $k$. Then, the transmitted signal $\mathbf{x}\in\mathbb{C}^{M}$ is written as $\mathbf{x} = \sum_{k\in\mathcal{K}}\mathbf{F}_{\text{R}}\mathbf{f}_k s_k$, where $\mathbf{F}_{\text{R}} \in \mathbb{C}^{M\times L}$ represents the RFP and $s_k \sim \mathcal{CN}(0,1)$ is the data symbol intended for RX $k$. For a fully-connected structure, $\mathbf{F}_{\text{R}}(m,l) = e^{j\varphi_{m,l}}$ holds, where $\mathbf{F}_{\text{R}}(m,l)$ refers to the element in the $m$-th row and $l$-th column of matrix $\mathbf{F}_{\text{R}}$, and $\varphi_{m,l}\in[0,2\pi)$, $\forall m\in \{1,\dots,M\}$, $\forall l\in \{1,\dots,L\}$. The transmit sum-power can be expressed as $P_{\rm sum} = \sum_{k\in\mathcal{K}}\left\|\mathbf{F}_{\text{R}}\mathbf{f}_{k}\right\|^{2}$.

The SINR at the ID branch of RX $k$ is given by
\begin{equation}\label{eq:2}
\operatorname{SINR}_k = \frac{\rho_k\left|\mathbf{h}_k^\text{H}\mathbf{F}_{\text{R}}\mathbf{f}_k\right|^2}{\rho_k\big(\sum_{j\in\mathcal{K}\setminus\{k\}}\left|\mathbf{h}_k^\text{H}\mathbf{F}_{\text{R}}\mathbf{f}_{j}\right|^{2} + \sigma_{k}^{2}\big) + \sigma_{C,k}^2},
\end{equation}
where $\sigma_{k}^{2}$ and $\sigma_{C,k}^{2}$ denote the power of RX noise $n_k$ and baseband conversion noise $\nu_k$, respectively (see Fig. \ref{fig:KN4}), whereas $\rho_k \in [0,1]$ is the PS ratio. The received RF power at the EH branch of the $k$-th RX, ignoring the negligible noise power, can be expressed as $E_k = (1-\rho_k) \sum_{j\in\mathcal{K}} \big|\mathbf{h}_k^\text{H}\mathbf{F}_{\text{R}}\mathbf{f}_{j}\big|^{2}$. The harvested energy at the output of the EH unit is given by $E_{H,k} = F(E_k)$, where $F(\cdot)$ is a monotonically increasing, non-linear EH function, e.g., the sigmoid function in \eqref{nl_sat_model}.

The SINR and EH constraints of RX $k$ are $\operatorname{SINR}_k \geq \gamma_{k}$ and $E_{H,k} \geq \tilde{Q}_{k} \Leftrightarrow E_k \geq F^{-1}(\tilde{Q}_k) \triangleq Q_{k}$, where $\gamma_{k}>0$, $\tilde{Q}_{k}>0$, and $Q_{k}>0$ denote the minimum SINR, minimum harvested power, and minimum received RF power target, respectively. Then, the optimization problem of interest is formulated as:
\begin{subequations}\label{eq:4}
\begin{alignat}{2}
&&&\underset{\mathbf{F}_{\text{R}},\left\{\mathbf{f}_{k}\right\},\bm{\Theta},\left\{\rho_{k}\right\}}{\mathrm{minimize}} \ P_{\rm sum} \label{eq:4a} \\
&&&\text{subject to} \ \ \text{C1:} \ \operatorname{SINR}_{k} \geq \gamma_{k}, \ \text{C2:} \ E_{k} \geq Q_{k}, \label{eq:4b} \\
&&&\hspace{16.5mm} \text{C3:} \ \left|\bm{\Theta}(n,n)\right|=1, \ \text{C4:} \ \left|\mathbf{F}_{\text{R}}(m,l)\right|=1, \label{eq:4c} \\
&&&\hspace{16.5mm} \text{C5:} \ 0 \leq \rho_{k} \leq 1, \label{eq:4d}
\end{alignat}
\end{subequations}
where the use of non-strict inequalities in \eqref{eq:4d} is dictated by the employed numerical solver, which expects as input an optimization problem written in standard form. Since the focus is on SWIPT, we omit solutions that correspond to the cases of $\rho_k=0$ or $\rho_k=1$, which represent EH-only or ID-only scenarios, respectively.

\begin{figure}[!t]\centering
	\includegraphics[height=7cm,width=\linewidth]{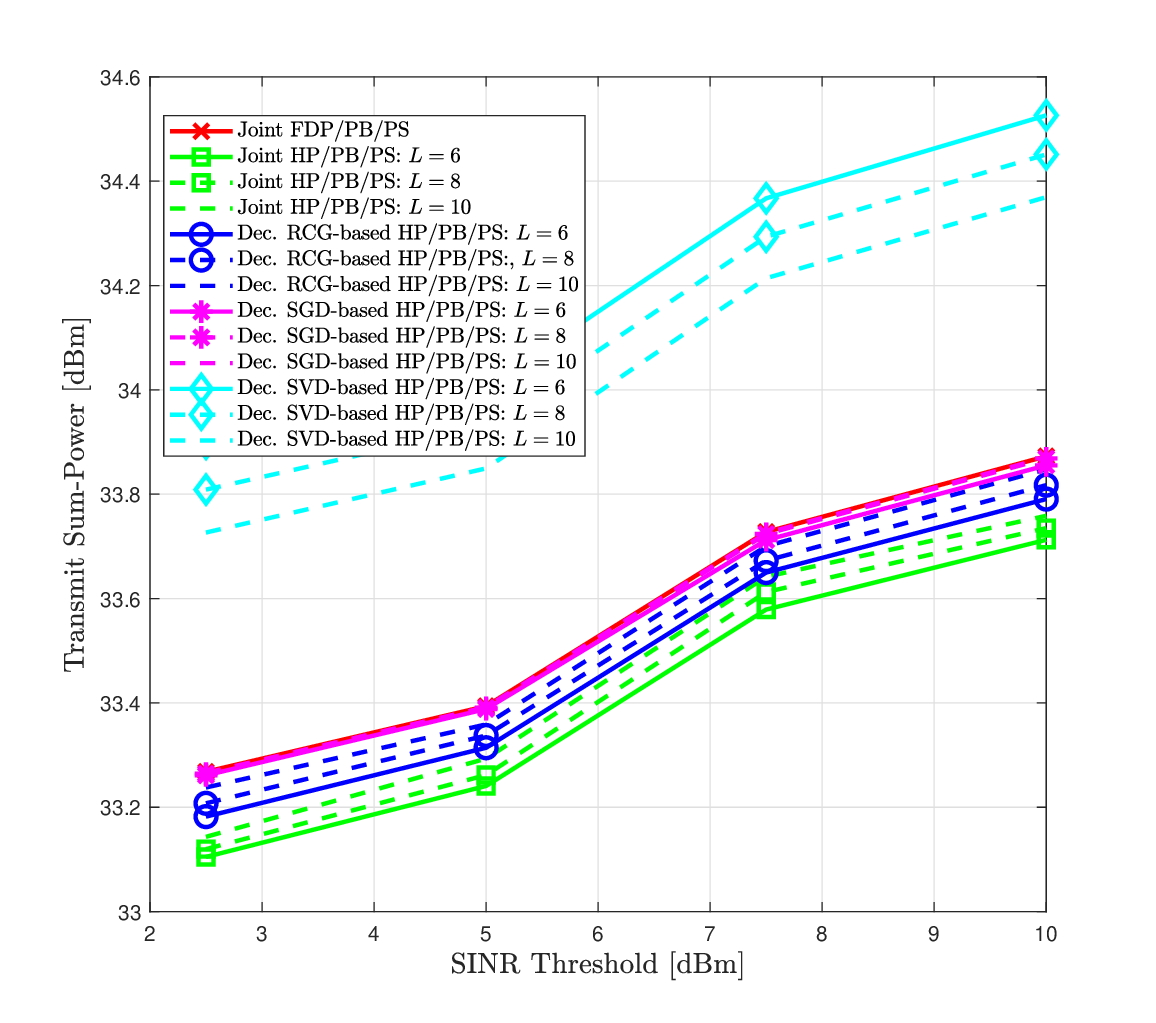}
	\caption{Transmit sum-power versus the SINR threshold.}
	\label{subfig:KN5a}
\end{figure}

The authors in \cite{NtougiasHPIRS} developed a two-layer, penalty-based block coordinate descent algorithm that involves the Riemannian conjugate gradient (RCG) method as well as a low-complexity, decoupled iterative design, that makes use of the stochastic gradient descent (SGD) and SCA methods under the alternating minimization framework, to tackle \eqref{eq:4}.

\paragraph{Simulation Results}
We now evaluate the performance of the proposed designs and compare it against benchmarks via numerical simulations. The following parameters are considered: $K=2$, $M=24$, $\sigma_k^2 = -70$ dBm, $\sigma_{C,k}^{2} = -50$ dBm, and $\tilde{Q}_{k} =-15$ dBm, $\forall k\in\mathcal{K}$. We adopt the EH model in \eqref{nl_sat_model} with $P_{\text{sat}}=13.8$ dBm, $a=150$ and $b=0.024$. We also apply the correlated Rayleigh fading channel model \cite{IRSRice2} with path loss exponent set to $2.5$. The results are averaged over $100$ independent channel realizations. The considered schemes are the joint FD precoder (FDP)/PB/PS design \cite{IRSSWIPT2}, the proposed joint and decoupled HP/PB/PS designs (with RFP matrix updates in the latter case based on either the SGD or the RCG algorithm), and a heuristic decoupled HP/PB/PS design where the RFP is computed based on the singular value decomposition (SVD) of the composite matrix of the effective channels \cite{NtougiasHP}. We also consider scenarios where random IRS phase shifts are utilized or no IRS is deployed.

\begin{figure}[!t]\centering
\includegraphics[height=7cm,width=\linewidth]{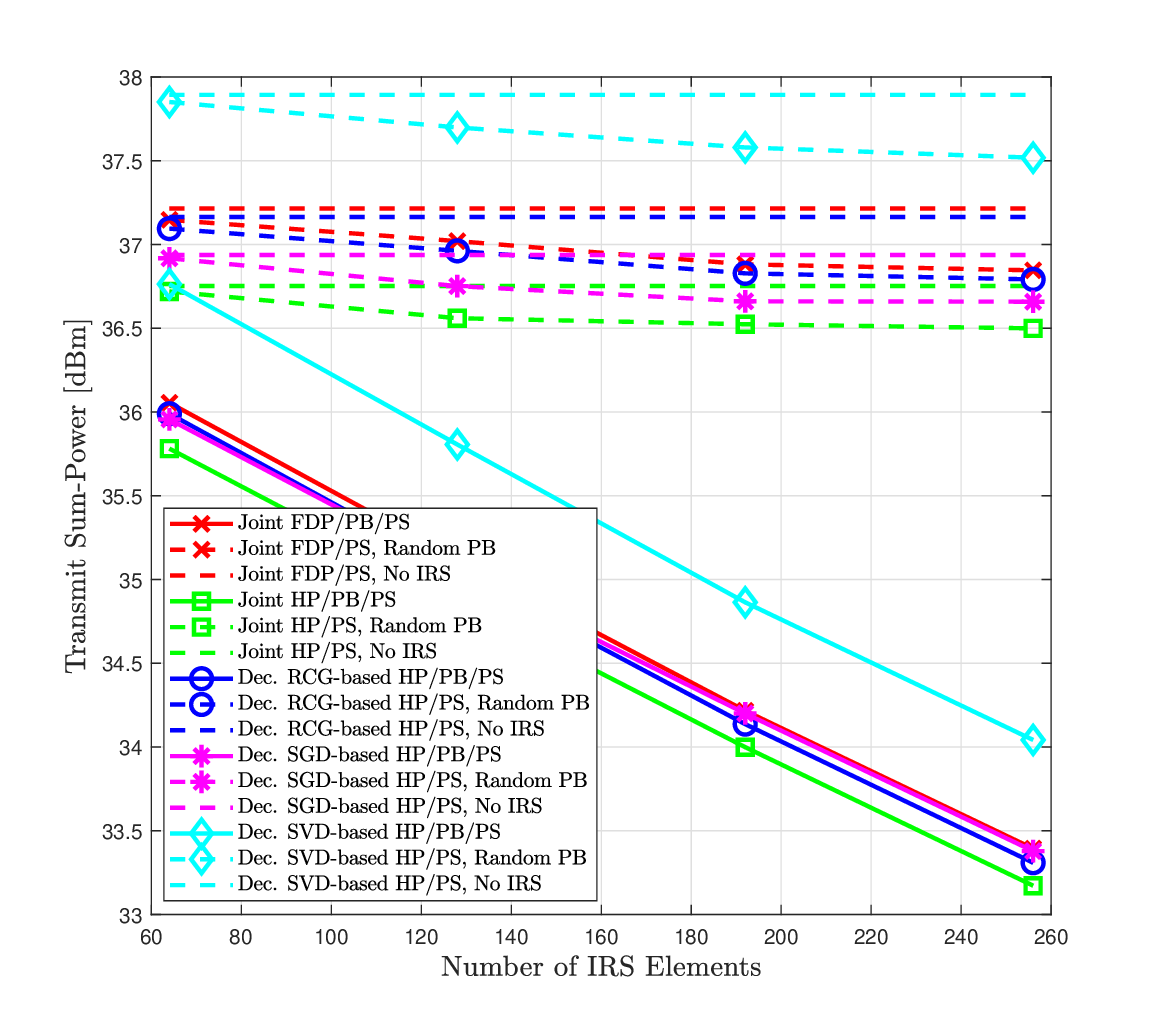}
\caption{Transmit sum-power versus the number of IRS elements.}
\label{subfig:KN5b}
\end{figure}

Fig. \ref{subfig:KN5a} illustrates the transmit sum-power in terms of the SINR threshold for different number of RF chains $L \in \{6,8,10\}$, assuming $N_E=256$. Fig. \ref{subfig:KN5b} depicts the transmit sum-power with respect to the number of IRS elements, assuming $\gamma_k = 0$ dB, $\forall k$, and $L=6$. We observe that the proposed designs perform better than the joint FDP/PB/PS design and the joint HP/PB/PS design outperforms the respective decoupled ones. Among these decoupled design variants, the RCG-based scheme performs better than the SGD-based one. The SVD-based decoupled design presents the worst performance, since the RFP computation is not optimization-based, but instead it relies on a heuristic. Moreover, higher SINR thresholds and larger number of RF chains result in higher transmit sum-power. The use of an IRS substantially improves the performance and the transmit sum-power decreases as the number of the IRS elements increases. Even when random phase shifts are utilized, there is some performance gain in comparison to the case where no IRS is deployed. As expected, the curves for the no IRS benchmarks are constant lines in Fig. \ref{subfig:KN5b} and do not depend on the number of IRS elements. However, in this scenario, the system is less sensitive to the increase of the number of IRS elements. Also, there is a large performance gap with respect to the case where the PB is optimized, since when random phase shifts are employed, the IRS does not exploit the channel knowledge to achieve coherent PB gains.

\subsubsection{Self-Sustainable IRS-aided SWIPT}
IRSs are equipped with a large number of low-power passive reflection elements to achieve high PB gains and compensate for the path loss caused by the cascaded TX-IRS-RX channel. Hence, although the overall power consumption of the reflection elements is much smaller, in general, than that of active relays, it is non-negligible \cite{IRSEE}. This is particularly problematic for battery-powered IRS implementations, which have been recently proposed to enhance the flexibility of network deployment, e.g., in remote areas where it is difficult to attach the IRS to the power grid. The use of RF-EH at the IRS via SWIPT has been proposed in response to this issue. Specifically, one of the most widely adopted practical implementations of IRS is based on the use of PIN diodes integrated with the reflecting elements \cite{IRSSurvey}. These same diodes can be used to realize an EH circuit, such that an EH state, instead of a simple no-phase-shift (OFF) state, complements the reflect (ON) state. We note that the power of the noise introduced by the PIN diodes is negligible in comparison to the harvested power (IRSs are generally considered to be noiseless components) and, therefore, it is typically omitted in system modeling. Corresponding works considered sum rate maximization in a wireless powered communication network with single-antenna nodes \cite{IRSSelfSustain} or in a secure MISO broadcasting setup \cite{IRSSelfSecure} and capacity maximization in a single-user MISO communication system \cite{IRSSelfMISO}. In this section, we investigate self-sustainable-IRS-aided SWIPT. Specifically, we assume that the IRS applies SWIPT (i.e., either the TS or the PS scheme), similar to the RX, and investigate the interplay and trade-offs between the EH requirements of the self-sustainable IRS and the RX under these fundamental SWIPT protocols. 

\begin{figure}[t]\centering
	\includegraphics[width=\linewidth]{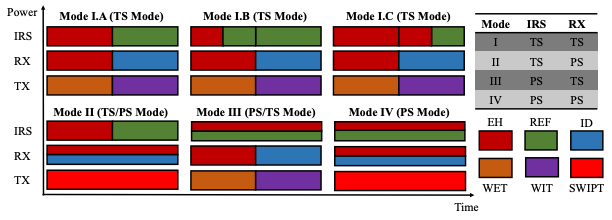}
	\caption{Operation modes of the self-sustainable IRS-aided MISO SWIPT system.}
	\label{fig:KN6}
\end{figure}

\paragraph{System Model}
Consider a TX with $M$ antennas and a self-sustainable IRS with $N_E$ passive reflection elements that synergistically serve a single-antenna WIET RX using PS or TS. We assume a time-slotted transmission frame structure, i.e., a sequence of $T_{f}$ time slots with duration of $T$ seconds each.

The system's operation modes are depicted in Fig. \ref{fig:KN6}. When the RX adopts the TS protocol, the TX divides time slot $t$ into a WET and a WIT sub-slot of duration $(1-\tau(t))T$ and $\tau(t)T$, $\tau(t)\in(0,1)$, where the RX operates in EH and ID state, respectively. When the RX applies the PS scheme, the TX transmits both information and power and the RX operates in both EH and ID state for the whole time slot duration, $T$. Likewise, when the IRS adopts the PS protocol, it operates in both EH and reflection states for the whole time slot duration. In Modes I.A and II, the IRS divides time slot $t$ into an EH and a reflection sub-slot of duration $(1-\tau(t))T$ and $\tau(t)T$, respectively. In Mode I.A., these sub-slots coincide with the WET and WIT sub-slots, respectively. In Mode I.B, the IRS divides the WET sub-slot into EH and reflection mini-slots with respective duration $(1-\tau(t))(1-\delta(t))T$ and $(1-\tau(t))\delta(t) T$, where $\delta\in(0,1)$, while it operates in reflection state in the WIT sub-slot. Likewise, in Mode I.C, the IRS divides the WIT sub-slot into EH and reflection mini-slots with respective duration $(1-\delta(t))\tau(t) T$ and $\delta(t)\tau(t) T$, while it operates in EH state in the WET sub-slot. Hence, while in Mode I.A the IRS treats the EH and reflection (REF) operations equally, in Modes I.B and I.C favors the latter or the former, respectively. That is, in Mode I.B the IRS is more altruistic than in Mode I.A, since it does not assist the TX only in WIT but it also supports WET, whereas in Mode I.C it is more selfish, since it harvests RF energy for more time.

\begin{figure}[t]\centering
	\includegraphics[height=7cm,width=\linewidth]{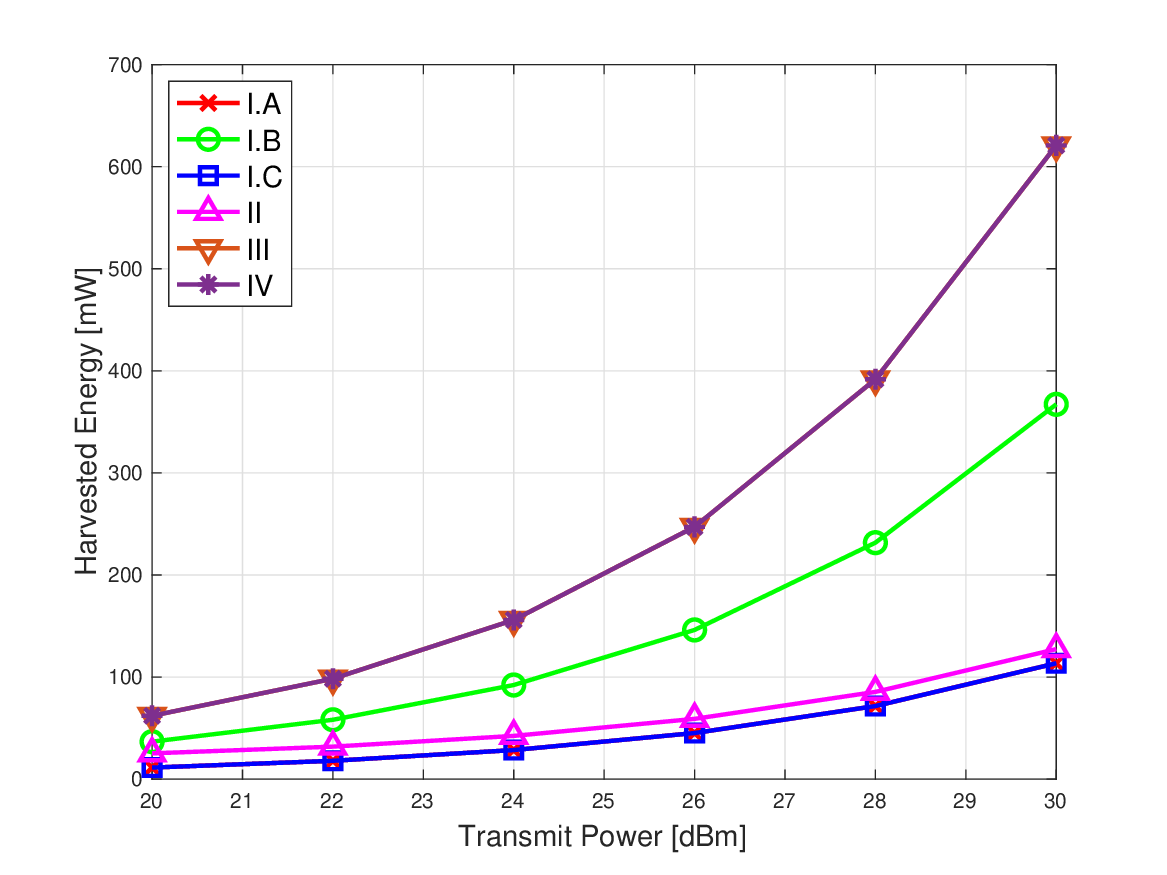}
	\caption{Transmit sum-power versus the harvested energy at the RX for different operating modes.}\label{subfig:KN7a}
\end{figure}

\paragraph{Simulation Results}
We are interested in jointly optimizing the transmit precoding vectors, the PB matrix, and the TS and PS ratios in each scenario, in order to maximize the SNR in each time slot and, therefore, optimize both information and energy transfer, subject to the transmit sum-power constraint, the EH constraints of the IRS and the RX, and the unit-modulus constraints of the former. We omit here the problem formulation for each operation mode and present instead the respective numerical simulation results that provide insights regarding the behavior of the derived designs in the different operation modes. For the simulation setup, we have $M=8$, $N_E=64$, $\sigma^2=-70$ dBm, and $\sigma_C^{2}=-50$ dBm. The power consumption of each reflecting element and the RX is $1.5$ mW and $6$ dBm, respectively. The EH model in \eqref{nl_sat_model} is adopted, with parameters $P_{\rm sat}=0.02$ W, $a = 6400$, and $b=0.003$. We assume a normalized unit frame duration and consider Rician fading channels with a mean path loss of $30$ dB, a Rician factor of $5$ dB, and a path loss exponents $2.2$ for the TX--IRS/IRS--RX channels and $3.6$ for the TX--RX channel.

\begin{figure}[t]\centering
	\includegraphics[height=7cm,width=\linewidth]{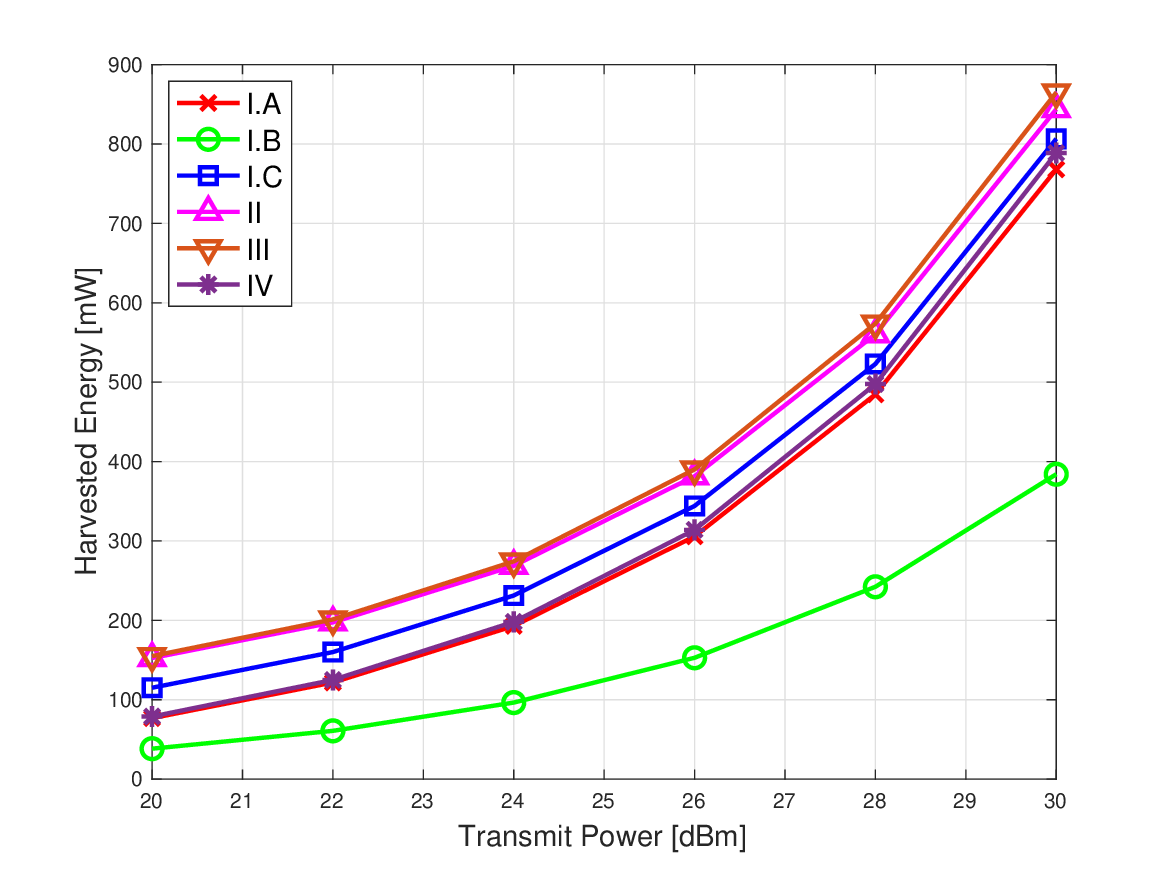}
	\caption{Transmit sum-power versus the harvested energy at the IRS for different operating modes.}\label{subfig:KN7b}
\end{figure}

In Figs. \ref{subfig:KN7a} and \ref{subfig:KN7b}, we plot the harvested energy of the RX and the IRS, respectively, versus the transmit power. We note in Fig. \ref{subfig:KN7a} that in Modes III and IV, where the IRS adopts the PS protocol and, therefore, assists the transmissions of the TX for the whole duration of each slot, the RX harvests the largest amount of energy. Next comes Mode I.B, where the IRS adopts the TS mode and harvests energy for the least amount of time, in comparison with the I.A and I.C Modes. Then, we have Mode II, where the RX employs the PS protocol. Finally, Modes I.A and I.C yield the worst performance. Their curves, as the ones for Modes III and IV, coincide due to the fact that the TS or PS ratios coincide. For the IRS, in turn, the situation is different. The best performance is observed for Mode III, followed by Mode II, where the RX harvests energy and decodes data for the whole duration of the slot under the PS protocol, hence essentially freeing resources that can be utilized by the IRS for EH purposes. Next, Mode I.C follows, where the IRS employs TS and harvests energy for the larger portion of the slot, followed last by the PS Mode IV and TS Modes I.A and I.B, where the IRS harvests energy for half of the time or less, respectively, of the slot.

\subsection{RA Systems for WIET}\label{sec:ras}
A different approach towards reconfigurability is from the transceiver’s point-of-view, where so called RAs can be employed to add both flexibility and reconfigurability to the RF front-end \cite{costantine2015}. This type of antenna has tuneable characteristics, which provide multiple switchable configurations that are acquired by modifying their geometry (i.e., their physical structure) and/or electrical behavior (i.e., the current distribution over the antenna's surface) \cite{piazza2008, huang2021, rodrigo2016, song2020}. In this way, a single RA can support different network requirements (e.g., changes in operating frequency or radiation pattern) and can readjust in dynamic environments (e.g., due to mobility or obstacles). Therefore, RAs can provide more cost-efficient and energy-efficient antenna array designs (using fewer RF elements) as well as more flexibility and intelligence in wireless networks \cite{kit2023a}. Even though the concept is not new, there has been revived interest in RAs due to recent advances in fluid/movable antenna systems \cite{martinez2022, wu2023}.

Conventional antennas adhere to a static and stationary approach as they are usually made of metal, making them rigid and inflexible and, in some cases, costly. In view of this, the RA technology provides new degrees of freedom for the design of wireless networks, including WIET systems, and has the potential to push further their performance limits \cite{martinez2022}. Since an RA can be small and energy efficient, it is a suitable technology for low-power devices such as sensors, IoT devices, and nano-machines \cite{costantine2015}, and thus, ideal for WIET. Due to their physical characteristics, RAs can modify their shape and adapt their configuration to change the operating frequency, the radiation pattern as well as the gain and polarization \cite{piazza2008,huang2021,rodrigo2016}. This reconfiguration can be beneficial in multi-user networks for interference mitigation; can assist in mobility scenarios to ensure good quality of service; can reduce the human exposure to EM fields through the selection of an appropriate configuration and can provide significant WIET performance gains via the dynamic switching between configurations \cite{kit2023b}.

The reconfigurability of RAs can be achieved in a programmable and controllable manner using either non-movable or movable components. In the former case, the RAs employ RF-switches or variable capacitors to modify the load at specific locations of the antenna structure using RF micro-electromechanical systems (RF-MEMS), PIN diodes, or varactors \cite{piazza2008}. The latter case refers to an antenna that alters its structure by modifying its position, orientation and/or shape. A popular approach for this type of RAs is through the use of metallic liquids (e.g., Mercury, Galinstan) with microfluidic and electrowetting techniques, which displace the liquid inside a dielectric holder \cite{huang2021}; these are known as fluid antennas. In both cases, the RA can switch to a pre-defined configuration either by modifying the state (ON/OFF) of the RF-switches or by displacing the liquid to a location inside the holder.

The research community working on this technology has focused mostly on the design and development of circuits and antennas for RAs, e.g., \cite{piazza2008, huang2021, rodrigo2016, song2020, ali2007, shen2020a, ahmed2015, rodrigo2012, xing2016}. On the other hand, the investigation of RAs from a wireless communication theory point of view is still in its infancy. Specifically, the benefits of RAs towards the achieved degrees of freedom have been shown in \cite{ke2012} for a MIMO setup and in \cite{yang2017b} for full-duplex cellular networks. In \cite{fazel2009}, a switching scheme with a three-dimensional block code was proposed for RAs and was proven to achieve maximum diversity gains. Some initial results on the performance analysis of RAs are given in \cite{senanayake2017, smith2012}, which show the benefits of RAs in wireless networks in terms of rate and diversity. A channel estimation method for RAs with low-overhead is developed in \cite{bahceci2017}. Moreover, the authors of \cite{piazza2009} propose a selection scheme for reconfigurable circular patch arrays and demonstrate that it outperforms the conventional MIMO approach.

Recently, there has been interest in movable RAs for wireless communications, where a flexible single-element RA adapts its position over a small space. In \cite{wong2021}, it was shown that a fluid RA system with sufficiently large number of configuration can outperform a maximum ratio combining system with conventional antennas. Similar observations were obtained in \cite{Zhu2023}, where it was demonstrated that a movable RA achieves significant performance gains compared to a conventional fixed-position antenna. The work of \cite{wong2021} was extended in \cite{wong2022} for the multiple access, illustrating that the network multiplexing gain grows linearly with the number of configurations while being ultimately limited by the number of RXs. In \cite{psomas2023}, a linear prediction scheme is proposed to overcome the delays due to the liquid displacement and space-time coded modulation schemes are designed that exploit the end-to-end displacement of the liquid. Finally, the work in \cite{wu2023} investigated the optimal resource allocation in a multi-user MISO system. By jointly optimizing the beamforming and the positions of the movable RAs at the TX, the transmit power was minimized while ensuring a minimum SINR at each RX.

Still, the performance gains from RAs in WIET systems have not been demonstrated, yet. Therefore, in what follows, we study fluid RAs, in the context of WIET, and show how the extra degrees of freedom provided by RAs can be exploited to enhance both the ID and EH.

\subsubsection{SWIPT-enabled RXs with fluid RAs}
Consider a simple point-to-point SWIPT topology consisting of a conventional single-antenna TX and an RX with a single fluid RA \cite{Psomas2023b}. The TX has continuous power supply and transmits with power $P_{\rm tx}$, aiming to deliver both information and energy to the RX. It is assumed that the RX also experiences $N_I$ interfering signals. All wireless links are considered to be Rayleigh fading. The RX can displace the liquid to a position, say $v$, inside a dielectric holder of linear dimension and length $V$, $0\leq v\leq V$; we assume that the displacement can occur instantly and without delays \cite{wong2020, wong2022}. So the position of the liquid corresponds to a specific configuration of the RA.

Let $g(v) \sim \mathcal{CN}(0,\beta_0)$ be the desired signal's channel coefficient at the $v$-th position and $h_i(v) \sim \mathcal{CN}(0,\beta_I)$, $i\in\{1,\dots,N_I\}$, the interfering channels, assumed to be independent and identically distributed (i.i.d.). Then, the received signal at the $v$-th position is given by
\begin{align}
y(v) &= \sqrt{P_{\rm tx}} g(v) x_0\nonumber\\
&\qquad + \sqrt{P_{\rm tx}} \sum_{i=1}^{N_I} h_i(v) x_i + n(v), ~0\leq v \leq V,
\end{align}
where $x_0$ is the transmitted data symbol with $\E\{|x_0|^2\} = 1$, $x_i$ is the data symbol of the $i$-th interferer with $\E\{|x_i|^2\} = 1$, and $n(v)$ is an AWGN with zero mean and variance $\sigma^2$; for simplicity, we assume the interferers also transmit with power $P_{\rm tx}$. Moreover, it is assumed that the RX has perfect CSI knowledge \cite{wong2020,Zhu2023,wu2023}.

\begin{figure}[t]\centering
	\includegraphics[width=0.75\linewidth]{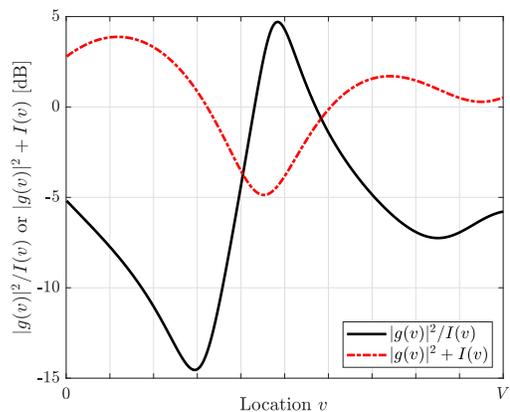}
	\caption{Arbitrary instances of $|g(v)|^2/I(v)$ and $|g(v)|^2+I(v)$; the liquid can be displaced over the length of the fluid RA.}\label{fig1}
\end{figure}

\subsubsection{Information and Energy Transfer} The RA RX employs SWIPT based on PS, where the aggregate received signal is split into two parts: one is converted to baseband for ID and the other is directed to the rectenna for EH (similar to Fig. \ref{fig:KN4}). Let $\rho$ denote the PS parameter and so $100\rho\%$, $0 < \rho \leq 1$, of the received power is used for ID and the remaining portion is send for EH. During the RF-to-baseband conversion phase, additional circuit noise is generated, which is modeled as AWGN with zero mean and variance $\s^2_C$.

Therefore, the SINR at the $v$-th position of the fluid RA is
\begin{align}
S(v) = \frac{\rho P_{\rm tx}|g(v)|^2}{\rho (\s^2 + I(v))+\s_C^2},
\end{align}
where
\begin{align}
I(v) = P_{\rm tx}\sum_{i=1}^{N_I} |h_i(v)|^2,
\end{align}
is the aggregate interference power at position $v$. Then, the achieved rate at position $v$ is
\begin{align}\label{rate}
R(v) = \log_2(1+S(v)).
\end{align}
Therefore, to maximize its performance in terms of ID, the RA RX will displace the liquid to some position that maximizes \eqref{rate}. In other words, by assuming full channel knowledge, the RX displaces the liquid to the location that achieves
\begin{align}\label{sup}
R^* = \sup_{0\leq v\leq V} \{R(v)\},
\end{align}
where $\sup_{0\leq v\leq V} \{ R(v) \}$ denotes the supremum of $R(v)$.

On the other hand, we study the EH performance with respect to the average harvested energy. Since $100(1 - \rho)\%$ of the received power is used for rectification, the average harvested energy from position $v$, considering the first two even terms of \eqref{nl_circ_model2}, is
\begin{align}
P_{\rm E}(v) &= k_2 R_{\rm ant} (1-\rho) \E\{|y(v)|^2\}\nonumber\\
&\qquad + k_4 R_{\rm ant}^2 (1-\rho)^2 \E\{|y(v)|^4\}.\label{avg_en}
\end{align}
In this case, the maximum performance is obtained at the location that achieves
\begin{align}\label{sup2}
y^* &= \sup_{0\leq v\leq V} \{|y(v)|\},
\end{align}
that is, the supremum of $|y(v)|$. Fig. \ref{fig1} depicts arbitrary instances of the ratio $|g(v)|^2/I(v)$ and the sum $|g(v)|^2 + I(v)$ for $P_{\rm tx} = 0$ dBm. The former affects the information transfer, whereas the latter affects the energy transfer. It is clear that the decision rules \eqref{sup} and \eqref{sup2} correspond to different positions of the liquid in the RA. Therefore, an RX chooses a decision rule for the RA according to which metric needs to be maximized.

\begin{figure}[t]\centering
	\includegraphics[width=0.9\linewidth]{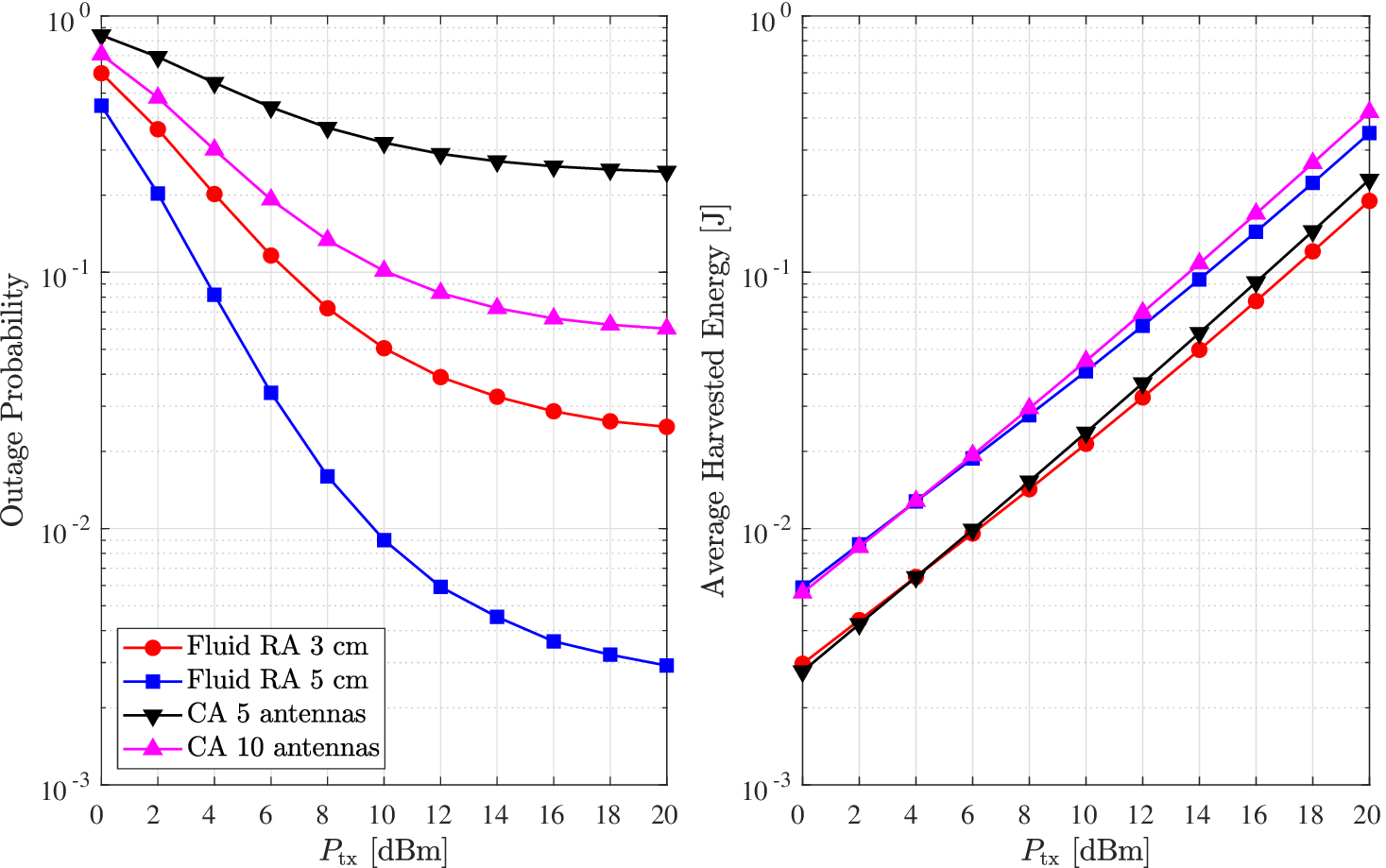}
	\caption{Prioritizing ID; outage probability (left) versus the average harvested energy (right).}\label{fig2}
\end{figure}

Now, since the liquid can take any position inside the holder, the signals $y(v)$, $0\leq v \leq V$, are assumed to be spatially correlated. By considering two-dimensional isotropic scattering, the correlation between the received signals at two positions spaced by $\tau$ can be modeled by the well-known Jake's model as \cite{heath}
\begin{align}\label{correlation}
\mu(\tau) = J_0\Big(2\pi \frac{\tau}{\lambda}\Big),
\end{align}
where $\lambda$ is the signal's wavelength and $J_0(\cdot)$ is the zeroth order Bessel function of the first kind.

\subsubsection{Simulation Results}
For the sake of illustrating the benefits of fluid RAs in WIET systems, we focus on the performance of the two extreme scenarios, given by \eqref{sup} and \eqref{sup2}. In particular, we consider: ($i$) prioritizing ID, based on the outage performance $\PP\{R^* < \theta\}$, where $\theta$ is the target SINR and $R^*$ is obtained by \eqref{sup}; and ($ii$) prioritizing EH, based on the average harvested energy $P_{\rm E}$, with the received signal given by \eqref{sup2}. We consider the following parameters: TX-RX distance $10$ m, $\beta_0 = 1$, $\beta_I = 1$, $N_I=2$, $\theta = 0$ dB, $\sigma^2 = 10^{-20}$, $\sigma^2_C = 10^{-5}$, $k_2 = 0.0034$, $k_4 = 0.3829$, $R_{\rm ant} = 50~\Omega$ \cite{Clerckx16}, $\rho = 0.5$, $\lambda = 1$ cm and length for the fluid RA at $3$ cm and $5$ cm. The comparison is made with a selection combiner of $5$ and $10$ conventional antennas (CAs); we assume that the channels at the CAs are mutually independent.

\begin{figure}[t]\centering
	\includegraphics[width=0.9\linewidth]{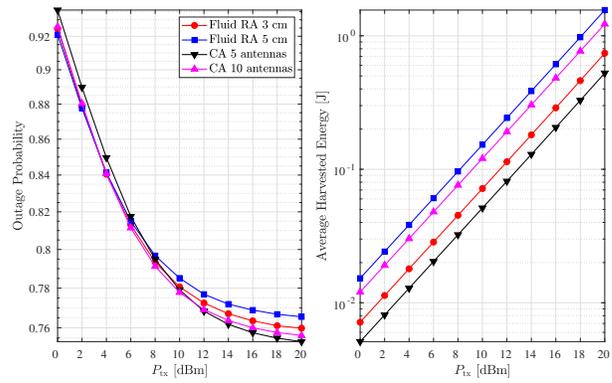}
	\caption{Prioritizing EH; outage probability (left) versus the average harvested energy (right).}\label{fig3}
\end{figure}

In Fig. \ref{fig2}, when ID is prioritized, the gains in terms of outage probability can be clearly seen, where both considered RA configurations outperform the CA configurations, despite the high correlation between the positions of the RA. In terms of EH, at low values of $P_{\rm tx}$, the larger RA performs similar to the CA with $10$ antennas, whereas the smaller RA performs similar to the CA with $5$ antennas. However, as $P_{\rm tx}$ increases, both CAs achieve greater harvesting performance than the RAs. Due to the extra degrees of freedom provided by the RAs, when prioritizing information transfer, the RX can find a position where the interference is in deep fade \cite{wong2022}, thus outperforming CAs. This however does not assist with EH since it results in lower aggregate received signal strength than the CA counterpart.

On the other hand, when EH is prioritized, the opposite behavior can be observed, as shown in Fig. \ref{fig3}. Specifically, the small and large RA achieves a better performance, with respect to harvesting, compared to the CA with $5$ and $10$ antennas, respectively. For the outage probability, the RAs have slightly better performance at low values of $P_{\rm tx}$ but are outperformed as $P_{\rm tx}$ increases. Indeed, when prioritizing harvesting, the RA finds a position with high interference but this achieves a lower SINR. Despite these extreme scenarios, the flexibility provided by the RA can be exploited to find a good balance between information and energy transfer, as depicted in Fig. \ref{fig1}. Firstly, the RX can choose a different position that achieves specific QoS thresholds for both metrics. Moreover, an adaptive PS method can be employed, which will adapt the splitting between information and energy accordingly. In this way, the RA can sacrifice some of the gains achieved in one metric to boost the gains in the other.
	
\section{THz WIET Systems for Microscopic 6G IoT}\label{sec:thz}
\graphicspath{{thz/}}
In this section, we consider THz band WIET for microscopic 6G IoT devices. Specifically, we first provide a discussion on THz EH circuits for micro-scale IoT devices. Then, we propose a general framework for the analysis of micro-scale THz WIET systems and the optimization of their transmit signal distributions.

Future 6G IoT devices will become microscopic in size and consume low amounts of power \cite{Tataria2021}. These high-speed 6G IoT networks will be enabled by THz band communication, where the available frequency spectrum is extremely large and can be effectively utilized to support the required high data rates \cite{Lemic2021, Chen2019}. Even though THz communications suffer from high propagation and molecular absortion losses, the application of IRSs, H-MIMO, and distance-aware resource allocation schemes can help to overcome these limitations \cite{Sarieddeen2021}. However, the need to supply these devices with power is a challenging limiting factor that hinders the design of truly microscopic THz IoT devices. Moreover, for micro-scale IoT, the regular replacement of bulky batteries may be impractical or even impossible. A promising solution to this difficult problem are 6G THz WIET networks, in which not only information but also energy is wirelessly transferred using THz band signals \cite{Shanin2023d, Zhang22, Rong2017}. In the following, we discuss microscopic THz EH circuits and present a general approach to the design of THz WIET systems.

\subsection{Microscopic THz EH Circuits}
When operating at frequencies below $\SI{6}{\giga\hertz}$, electrical EH RXs are typically equipped with Schottky diodes to rectify the received RF signal \cite{Tietze2012}.
However, Schottky diodes may not be the most suitable choice for EH at THz frequencies, where resonant tunneling diodes (RTDs) and metal–insulator–metal (MIM) diodes are more efficient.
In the following, we discuss different THz diodes for EH.

\begin{figure}[!t]\centering
	\includegraphics[width=0.9\linewidth]{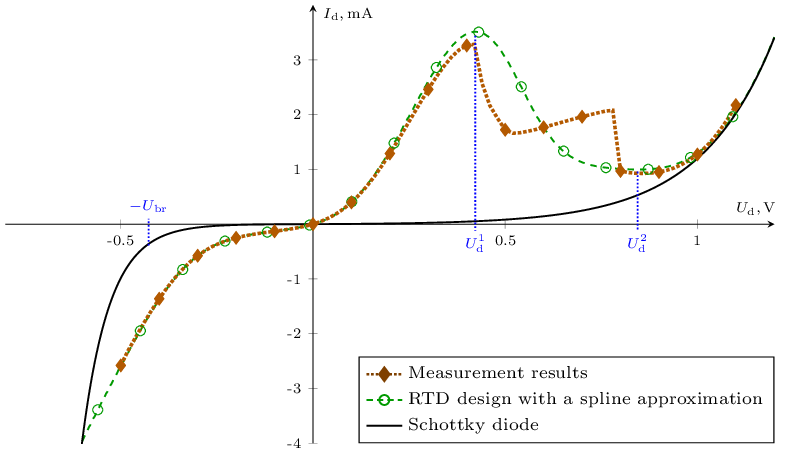}
	\caption {I-V characteristic of the Keysight ADS RTD design in \cite{Clochiatti2022} matched to the measurements in \cite{Clochiatti2022} and I-V characteristic of a Schottky diode \cite{Tietze2012}.}\label{Fig:THz_RtdIV}
\end{figure}

\subsubsection{THz Schottky Diodes} GaAs Schottky diodes comprising a single p-n junction are widely utilized for the design of THz electronic devices, such as mixers, multipliers, and signal generators \cite{Mehdi2017}.
Similar to GHz Schottky diodes, the current-voltage (I-V) characteristic of a THz Schottky diode can be modeled by the Shockley diode equation \cite{Tietze2012}.
In particular, when a positive voltage $U_\text{d} > 0$ is applied to a Schottky diode, the forward-bias current flow $I_\text{d}$ through the diode can be expressed as a monotonic increasing exponential function of $U_\text{d}$, as shown in Fig. \ref{Fig:THz_RtdIV}.
Furthermore, when the voltage $U_\text{d} < 0$ is negative and small, the leakage reverse-bias current is negligibly low \cite{Morsi2020}.
However, when the negative applied voltage exceeds a threshold $U_\text{d} \leq -U_\text{br}$, typically referred to as the breakdown voltage, the current flow increases exponentially and can even cause damage to the device, see Fig. \ref{Fig:THz_RtdIV}.

Due to the diode's non-linear I-V characteristic, the dependence between the received and harvested powers at EH RXs equipped with Schottky diodes also exhibits a non-linear behavior \cite{Morsi2020, Clerckx2019, Shanin2020}.
In particular, when the power of the received signal at the RX is low, the non-linearity of the rectifier circuit is determined by the non-linear forward-bias I-V characteristic of the diode \cite{Morsi2020}.
However, for high received signal powers, the harvested power at EH RXs saturates due to the breakdown of the employed Schottky diodes \cite{Shanin2020}.
The non-linear effects of EH circuits have to be accurately taken into account for an efficient design of communication systems based on EH, regardless of the operating frequency.

Although Schottky diodes have been widely used for the design of signal rectifiers operating at low frequencies, they may not be suitable for THz band EH.
First, at THz frequencies, rectifying diodes have to have a very short response time, i.e., their RC time constant must be much smaller than one period of the received sinusoidal signal.
Due to their large response time, GaAs Schottky diodes are not able to efficiently rectify signals above $2-\SI{3}{\tera\hertz}$ \cite{Citroni2022}.
Next, as shown in Fig. \ref{Fig:THz_RtdIV}, the exponential I-V characteristics of Schottky diodes have a low curvature at the zero-bias point $U_\text{d} = 0$, and hence, may not be efficient for EH when the received signal power is small \cite{Shanin2023d}.
Therefore, in the following, we discuss MIM diodes and RTDs, which are able to operate efficiently at THz frequencies. 

\subsubsection{THz MIM Diodes} A THz MIM diode is an ultra-fast switching device, which consists of a thin dielectric layer within two metal electrodes \cite{Citroni2022}.
Due to the quantum tunneling effect, MIM diodes are able to rectify signals at frequencies up to $\SI{100}{\tera\hertz}$ \cite{Shriwastava2019}.
Similar to the I-V characteristics of Schottky diodes in Fig. \ref{Fig:THz_RtdIV}, the I-V characteristic of MIM diodes is also monotonic increasing for all values of the applied voltage $U_\text{d}$.
Additionally, the curvature of THz MIM diodes at low bias voltages is similar to that of Schottky diodes, and hence, MIM diodes are also not efficient for EH when the received signal power is low \cite{Citroni2022}.

\subsubsection{THz RTDs} THz RTDs also have a significantly shorter transient time compared to Schottky diodes and can be efficiently utilized at frequencies of up to $\SI{5}{\tera\hertz}$ \cite{Villani2021}.
In contrast to MIM diodes, RTDs feature not only multiple quantum barriers, but also a potential well \cite{Villani2021}.
The experiments in \cite{Clochiatti2022} demonstrated that due to an added tunneling current, the I-V characteristic of a triple-barrier RTD is not only non-linear, but also non-monotonic, as shown in Fig. \ref{Fig:THz_RtdIV}.
In particular, when the applied voltage $U_\text{d}$ is positive and low, the forward-bias current flow $I_\text{d}$ of the triple-barrier RTD increases monotonically and significantly exceeds the current flow of Schottky diodes.
For voltages in the interval $U_\text{d} \in [U_\text{d}^1, U_\text{d}^2]$, the RTD is driven into a region of negative resistance, where the current flow decreases when $U_\text{d}$ grows.
Finally, when the second critical point is reached, i.e., $U_\text{d} \geq U_\text{d}^2$, the I-V characteristic of the RTD starts to increase and follows that of the Schottky diode.
Thus, due to the potential well and multiple potential barriers, RTDs typically have a large curvature at the zero-bias point and efficiently harvest energy also for low received signal powers.
Moreover, due to their small form factor, RTDs can be integrated with on-chip THz antennas for application in microscopic IoT devices \cite{Clochiatti2022}. Furthermore, when driven into the region of negative resistance, i.e., when the applied voltage is in the interval $U_\text{d} \in [U_\text{d}^1, U_\text{d}^2]$, RTDs behave as THz oscillators and generate a THz sinusoidal signal \cite{Villani2021, Clochiatti2022}.
The THz signal generated by oscillating RTDs of microscopic IoT RXs can be utilized, for example, for sensing or information transmission in the uplink.

In contrast to Schottky diodes, for RTDs, a complete and accurate analytical model capturing all non-linear and non-monotonic effects is not available, yet.
However, the authors in \cite{Clochiatti2022} developed a compact Keysight Advanced Design System (ADS) \cite{ADS2017} model for RTDs, which is based on a spline approximation and fits the measurements presented in \cite{Clochiatti2022} for a wide range of operating frequencies, as seen in Fig. \ref{Fig:THz_RtdIV}. Please note that the discrepancy between the I-V characteristic of the ADS RTD design in \cite{Clochiatti2022} and the measurement results in Fig. \ref{Fig:THz_RtdIV} is caused by non-idealities (parasitic oscillations) of the measurement setup used in \cite{Clochiatti2022}.
Furthermore, we note that the reverse-bias leakage current of the RTD in \cite{Clochiatti2022} is high and the development of more efficient RTDs for EH is an open research problem.

Similar to electrical circuits utilized for EH at low frequencies, the dependence between the received and harvested powers at the THz EH RX is also characterized by a non-linear function \cite{Shanin2023d, Citroni2022}.
Furthermore, since an analytic accurate model characterizing the current flow through a MIM diode and an RTD is not available, the analytical derivation of an accurate EH model for THz RXs could also be infeasible.
However, similar to \cite{7264986, Boshkovska2017a, Boshkovska2018}, and other related works, where the authors characterized the average harvested power at the RX utilizing a parameterized sigmoid function (see Subsection \ref{nl_sat}), one can develop a parameterized non-linear non-monotonic function to describe the dependence between the instantaneous input power and instantaneous harvested power at a THz microscopic IoT RX \cite{Shanin2023d}.

\subsection{THz WIET System Design}
In the following, we discuss the design and optimization of transmit signal distributions for THz WIET systems.

In modern GHz-band communications, e.g., 5G mobile and Wi-Fi networks, phase-modulated RF signals are typically utilized for information transmission. However, such signal waveforms have low PAPR and are highly suboptimal for WET. Furthermore, the design of coherent THz information RXs is challenging due to the instability and phase noise of THz local oscillators. Also, the received signal power at THz IoT devices is typically rather low due to high path loss and the low efficiency of THz power amplifiers \cite{Sarieddeen2021, zhu2022, wu2024}. On the other hand, the spectrum available in the THz frequency band is large and high data rates can be achieved with simple modulation schemes, encoding the information only into the amplitude of the THz signal, such as unipolar amplitude shift keying (ASK) \cite{Sarieddeen2021}. Moreover, since THz antennas are very small, extremely high gains exceeding $30$ dBi can be realized with compact THz horn antennas \cite{he2020}.

\begin{figure}[!t]\centering
\includegraphics[width=0.85\linewidth]{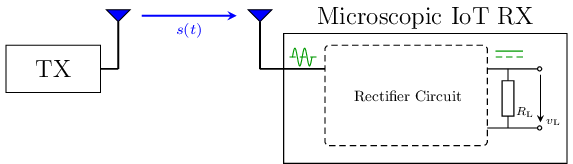}
\caption{THz WIET system model with a single microscopic IoT RX equipped with a non-linear rectifier circuit for both EH and information detection.}\label{Fig:THz_SystemModel}
\end{figure}

Let us consider a THz WIET system employing a single microscopic IoT RX for both EH and information detection, as shown in Fig. \ref{Fig:THz_SystemModel}.
We express the real-valued THz signal received at the IoT RX as follows \cite{Shanin2023d}:
\begin{equation}
r(t) = g s(t) \cos(2\pi f_\text{c} t) + \tilde{n}(t),
\end{equation}
\noindent where $g$ is the real-valued channel gain between TX and IoT RX, $f_c$ is the carrier THz frequency, $\tilde{n}(t)$ is the additive noise at the RX, and $s(t) = \sum_k s[k] \phi(t-kT)$ is the pulse-modulated THz transmit signal.
Here, $\phi(t)$ is a rectangular pulse that takes value $1$ if $t \in [0, T)$ and $0$, otherwise, $T$ is the symbol duration, and $s[k]$ is an i.i.d. realization of a non-negative random variable $s \in [0, +\infty)$ with probability density function (pdf) $f_s$ in time slot $k, k \in \{1,2,\dots\}$.

Since EH RXs are envelope detectors, whose output signals are characterized by the powers of the received signals and are independent of the phase of the input signal, we utilize the output signal $v_\text{L}(t)$ at the THz IoT RX not only for EH, but also for information detection \cite{Tietze2012}.
To this end, we define the real-valued output RX signal generated in time slot $k, k \in \{1,2,\dots\},$ as
\begin{equation}
y[k] \triangleq \frac{v_\text{L}[k]}{\sqrt{R_\text{L}}} = \sqrt{ \psi( |g s[k]|^2 ) } + n[k],\label{Eqn:THz_OutputSignal}
\end{equation}
where $\psi(\cdot)$ is a non-linear non-monotonic function that maps the instantaneous received THz signal power in time slot $k$ to the output power at the load resistance $R_\text{L}$.
For a given rectifier circuit at the THz RX and instantaneous input power $\rho$, the value of function $\psi(\rho)$ can be measured experimentally or obtained using a circuit simulation tool, such as Keysight ADS \cite{ADS2017}.
We note that due to the breakdown of the rectifying diode, function $\psi(\rho)$ is typically bounded, i.e., there exists a value $P_\text{max}$ such that $\psi(\rho) \leq P_\text{max}, \forall \rho \in [0, +\infty)$.
Furthermore, in \eqref{Eqn:THz_OutputSignal}, $n[k]$ is the equivalent noise, which distorts the output signal $y[k]$ in time slot $k$ and can be modeled as an i.i.d. realization of a zero-mean AWGN with variance $\sigma_n^2$.

The ultimate goal of designing the transmit signal distribution for a communication system is to maximize the mutual information between the transmit and received signals, and thus, to achieve the channel capacity.
For the considered THz WIET system, the mutual information between the transmit signal $s$ and output signal $y$ at the IoT RX can be expressed as \cite{Grover2010}
\begin{equation}
I(s, y) = \int_s f_{s} (s)  \int_y f_{y|s} (y|s) \ln \frac{f_{y|s} (y|s)}{f_y(y)} \text{d} y \;\text{d}s, \label{Eqn:THz_MutInf}
\end{equation}
where $f_{y|s}(y|s)$ and $f_y(y)$ are the conditional pdf of random variable $y$ given input signal $s$ and the marginal pdf of random variable $y$, respectively.
Hence, the capacity-achieving input pdf $f_s^*$ can be obtained as solution of the following optimization problem \cite{Grover2010}
\begin{equation}
C = \max_{f_s \in \mathcal{F}_s} \; I(s,y),	\label{Eqn:THz_GeneralCapacity}
\end{equation}
where the set of feasible pdfs $\mathcal{F}_s$ is determined by the considered system model.
In particular, it is shown in \cite{Morsi2020} that the channel capacity of WIET systems with spatially separated EH and information RXs is achieved with a unique and discrete distribution of the transmit signal.
However, in general, determining the optimal distribution as solution of \eqref{Eqn:THz_GeneralCapacity} is a challenging optimization problem, which can not be solved in closed form if, e.g., not only the average power, but also the peak amplitude of the transmit signal $s$ is limited or a constraint on the average harvested power $\mathbb{E}_s \{\psi(|g s[k]|^2)\}$ at the IoT RX is imposed and has to be satisfied.

To optimize the distribution of the transmit signal and characterize the performance of THz WIET systems, we define an achievable information rate $J(f_s)$ \cite{Lapidoth2009, Shanin2023d}
\begin{equation}
J(f_s) = \frac{1}{2} \ln \left( 1 + \frac{\exp\big(2 h_x(f_s)\big)}{2 \pi e \sigma_n^2} \right),\label{Eqn:THz_AIR}
\end{equation}
where $h_x(f_s)$ is the entropy of random variable $x = \sqrt{\psi(g s)}$ for a given pdf $f_s$ of random variable $s$.
The achievable information rate in \eqref{Eqn:THz_AIR} constitutes a lower-bound on the mutual information in \eqref{Eqn:THz_MutInf} \cite{Lapidoth2009}
\begin{equation}
C = \max_{f_s \in \mathcal{F}_s} \; I(s,y) \geq \max_{f_s \in \mathcal{F}_s} J(f_s) = J^*_{\mathcal{F}_s},\label{Eqn:THz_MaxAIR}
\end{equation}
where $J^*_{\mathcal{F}_s}$ is the maximum achievable information rate.
In general, the computation of $J^*_{\mathcal{F}_s}$ in \eqref{Eqn:THz_MaxAIR} is a less challenging problem compared to \eqref{Eqn:THz_GeneralCapacity}.
In particular, for single-user THz WIET systems with constrained peak transmit signal amplitude $A$ and a required average harvested power at the RX of $P_\text{req} \leq P_\text{max}$, where $P_\text{max}$ is the maximum achievable instantaneous harvested power for given $A$, i.e., with $\mathcal{F}_s = \{f_s \vert s \in [0, A], \mathbb{E}\{x^2\} \geq P_\text{req}, \int_s f_s(s) \text{d} s = 1\}$, the maximum achievable information rate is determined by the power ratio $\frac{P_\text{req}}{P_\text{max}}$.
If the ratio satisfies $\frac{P_\text{req}}{P_\text{max}} \leq \frac{1}{3}$, the maximum value $J^*_{\mathcal{F}_s}$ is achieved for a uniform distribution of signal $x = \sqrt{\psi(g s)}$ and is given by $J^*_{\mathcal{F}_s} = \frac{1}{2} \ln \big( 1 + \frac{ P_\text{max} }{2 \pi e \sigma_n^2} \big)$ \cite[Appendix D]{Shanin2023d}.
Furthermore, for ratios exceeding $\frac{P_\text{req}}{P_\text{max}} > \frac{1}{3}$, the maximum achievable rate can be written as \cite[Appendix F]{Shanin2023d}
\begin{equation}
J^*_{\mathcal{F}_s} = \frac{1}{2} \ln \left( 1 + \frac{ 2 \mu_0 - 2 \mu_1^2 \frac{P_\text{req}}{P_\text{max}} }{2 \pi e \sigma_n^2} \right),\label{Eqn:THz_OptimalAchRate}
\end{equation}
where $\mu_0 = \mu_1^2 + \ln \left( \frac{P_\text{max}}{1 + 2 \mu_1^2 \frac{P_\text{req}}{P_\text{max}}} \right)$ and $\mu_1$ is the real non-negative coefficient that is obtained as the unique solution of the following equation
\begin{equation}
\frac{P_\text{req}}{P_\text{max}} = \frac{\exp(\mu_1^2)}{\sqrt{\pi} \mu_1 \text{Ei}(\mu_1) } - \frac{1}{2\mu_1^2},
\end{equation}
with imaginary error function $\text{Ei}(\cdot)$.
Furthermore, the corresponding pdf of the transmit signal that yields \eqref{Eqn:THz_OptimalAchRate} is given by \cite{Shanin2023d}
\begin{equation}
f_s^*(s) = \frac{\partial}{\partial s} F_x^*({\psi} (|g s|^2)),
\label{Eqn:THz_InputDistribution}
\end{equation}
with $F_x^*(x) = \int_{0}^{x} f_x^*(\tilde{x}) \text{d}\tilde{x}$, $f_x^*(x) = \exp(- {\mu}_0 + {\mu}_1^2 \frac{x^2}{\bar{P}_\text{\upshape max}})$, and $x \in [0, \sqrt{P_\text{max}}]$.

\begin{figure}[!t]\centering
\includegraphics[width=\linewidth]{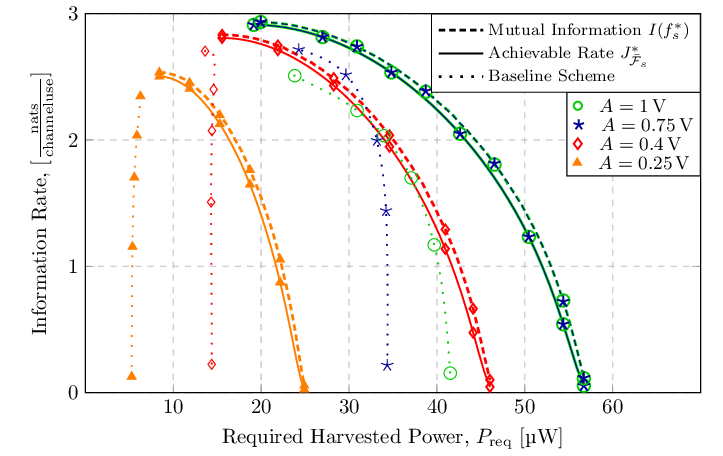}
\caption{Achievable rate-power trade-offs for different values of the maximum transmit signal amplitude $A$ \cite{Shanin2023d}.}\label{Fig:THz_Tradeoffs}
\end{figure}

In Fig. \ref{Fig:THz_Tradeoffs}, we plot the maximum achievable rate $J^*_{\mathcal{F}_s}$ for different values of maximum THz signal amplitude $A$ and required harvested power $P_\text{req}$ \cite{Shanin2023d}.
For our simulations, we model the LoS channel as $g = \frac{c_l}{4 \pi d f_c} \sqrt{G_\text{T} G_\text{R}}$ and adopt carrier frequency $f_\text{c} = \SI{300}{\giga\hertz}$ and noise variance $\sigma_n^2 = \SI{-50}{\dBm}$.
We set the TX and RX antenna gains and the distance between the TX and IoT RX to $G_\text{T} = \SI{30}{\deci\belisotropic}$, $G_\text{R} = \SI{10}{\deci\belisotropic}$, and $d = \SI{0.3}{\meter}$, respectively.
To validate the tightness of the lower-bound \eqref{Eqn:THz_MaxAIR}, for the input distributions $f_s^*$ of the transmit signal $s$ in \eqref{Eqn:THz_InputDistribution}, we additionally calculate and plot in Fig. \ref{Fig:THz_Tradeoffs} the corresponding mutual information $I(s, y)$.
Finally, as a baseline scheme, we adopt transmit symbols $s$ following a Gaussian distribution truncated to the interval $[0, A]$ with mean $\frac{A}{2}$ and variance $\sigma_s^2$.
For the baseline scheme, we determine and plot in Fig. \ref{Fig:THz_Tradeoffs} the corresponding values of mutual information $I(s, y)$ and average harvested powers by adjusting the variance $\sigma_s^2 \in [0, \infty)$ of the truncated Gaussian distribution $f_s$.
To calculate the average harvested power, we adopt the parameterized non-linear EH model for RTD-based IoT RXs, which is derived and tuned to the measurement results in \cite{Shanin2023d}.

Since, for all considered values of $A$ and $P_\text{req}$, the proposed THz WIET system is able to achieve significantly higher information rates in Fig. \ref{Fig:THz_Tradeoffs} than the baseline scheme, we conclude that Gaussian distributed signals are highly suboptimal and the optimization of the transmit signal distribution is needed for efficient THz WIET.
Furthermore, the observed gap between the achievable rate and mutual information in Fig. \ref{Fig:THz_Tradeoffs} is small for all considered values of $A$ and $P_\text{req}$.
Moreover, since the mutual information and achievable rate decrease when the required harvested power $P_\text{req}$ grows for any $A$, there is a trade-off between information rate and harvested power in THz WIET systems that is determined by the maximum transmit signal amplitude $A$.
Finally, note that for low maximum transmit amplitudes $A < \SI{0.75}{\volt}$, both the information rates and harvested powers grow with $A$, whereas, for large $A \geq \SI{0.75}{\volt}$, the rate-power trade-off does not depend on $A$ since the THz RTD-based EH circuits are driven into saturation when the received signal power is high \cite{Shanin2023d}.
	
\section{Experimental Aspects of WIET}\label{sec:experimental}
\graphicspath{{experimental/}}
The preceding sections have emphasized the essential technological advancements that facilitate WIET. In the following, we will showcase several cutting-edge prototypes for WET and WIET developed by Sungkyunkwan University (SKKU), along with several experimental results. In addition to the implementation at SKKU, we encourage readers to explore other WET and WIET experimental studies conducted by various research groups worldwide, such as blind adaptive beamforming for IoT devices \cite{Yeda:2017} and selective, tracking, and adaptive beamforming solutions based on backscattering feedback \cite{Belo:2019}.

For effective performance, WIET systems must incorporate adaptive beamforming and beam focusing, which necessitate accurate CSI for wireless channel estimation. Nevertheless, most of the existing channel estimation methods rely on the phase information of the received signals, which is not consistently accessible at the RX for the following reasons. Firstly, obtaining precise phase information at the RX is challenging due to frequency drift and phase noise originating from local oscillators. Secondly, measuring the phase incurs additional costs and complexity for EH RXs during implementation. Blind (CSI-free) techniques \cite{lopez2022, clerckx2018twc} can be a good alternative to avoid the acquisition of CSI. However, these techniques are not capable of focusing power beams to several focal points, which greatly degrades the power transfer efficiency (PTE) for RF WET. Therefore, received power-based energy beamforming solutions are indispensable in the context of WIET applications, demanding prompt attention. As a result, we focus primarily on proposing and investigating power-based beam-focusing algorithms to address the practical limitations of WET and WIET. Specifically, we present
\begin{itemize}
\item[1.] a power-based beamforming scheme for near-field phased array-based WET \cite{Aziz:2019,Jehyeon:2021},
\item[2.] a multi-tile IRS beam scanning (MTBS) scheme for IRS-aided WET \cite{Tran:2021,Tran:2022}, and
\item[3.] a beam sharing scheme for IRS-aided WIET \cite{TranICASSP:2023}.
\end{itemize}

\begin{figure}[t]\centering
	\subfigure[]{\label{fig:Txphased_array}
		\includegraphics[trim={0in 0.25in 0.1in 0.25in},clip=true,width=0.3\textwidth]{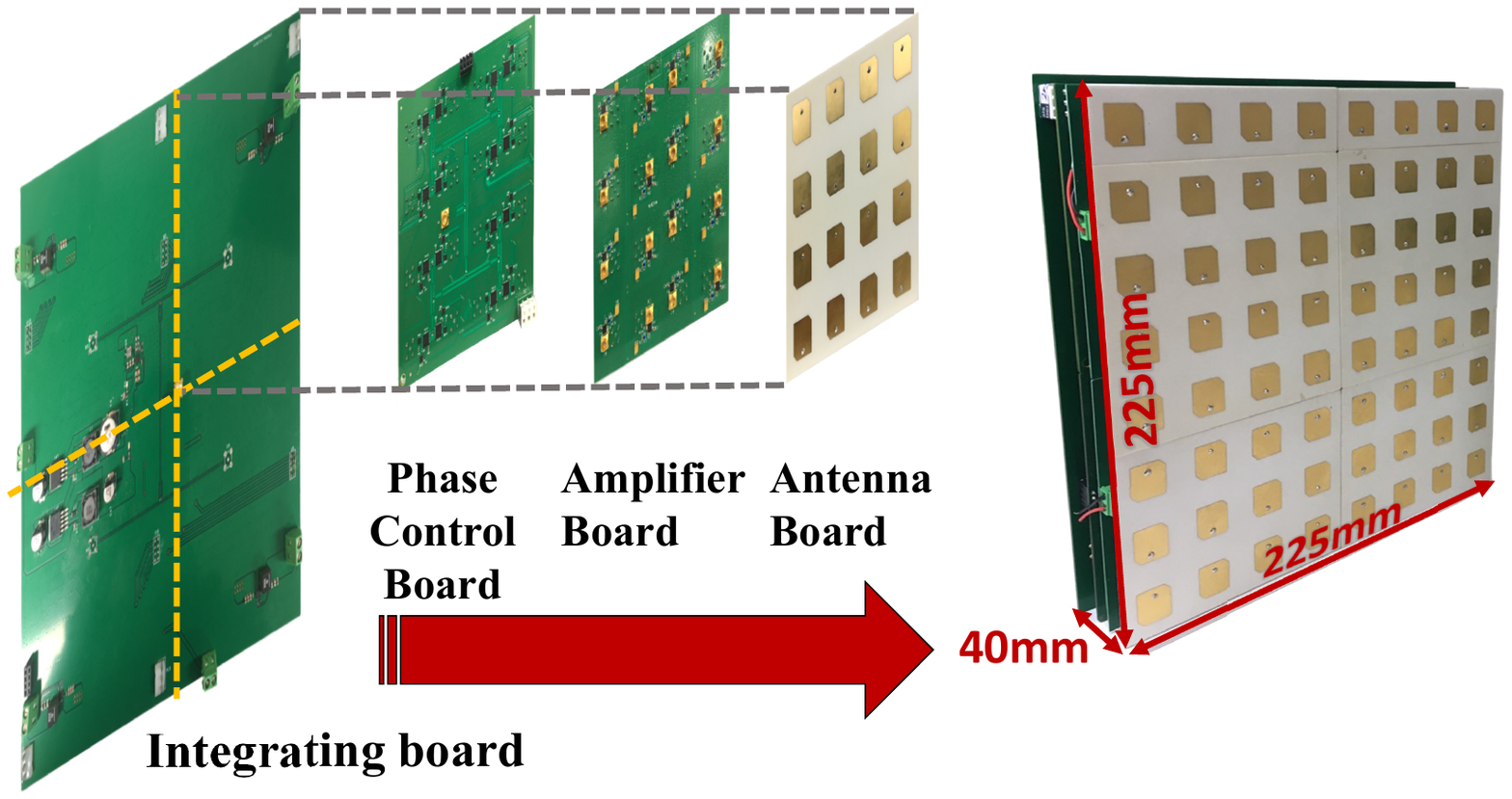}}
	\hspace{-0.5cm}
	\subfigure[]{\label{fig:rx_board}
		\includegraphics[trim={3in 1in 2in 1in},clip=true,width=0.16\textwidth]{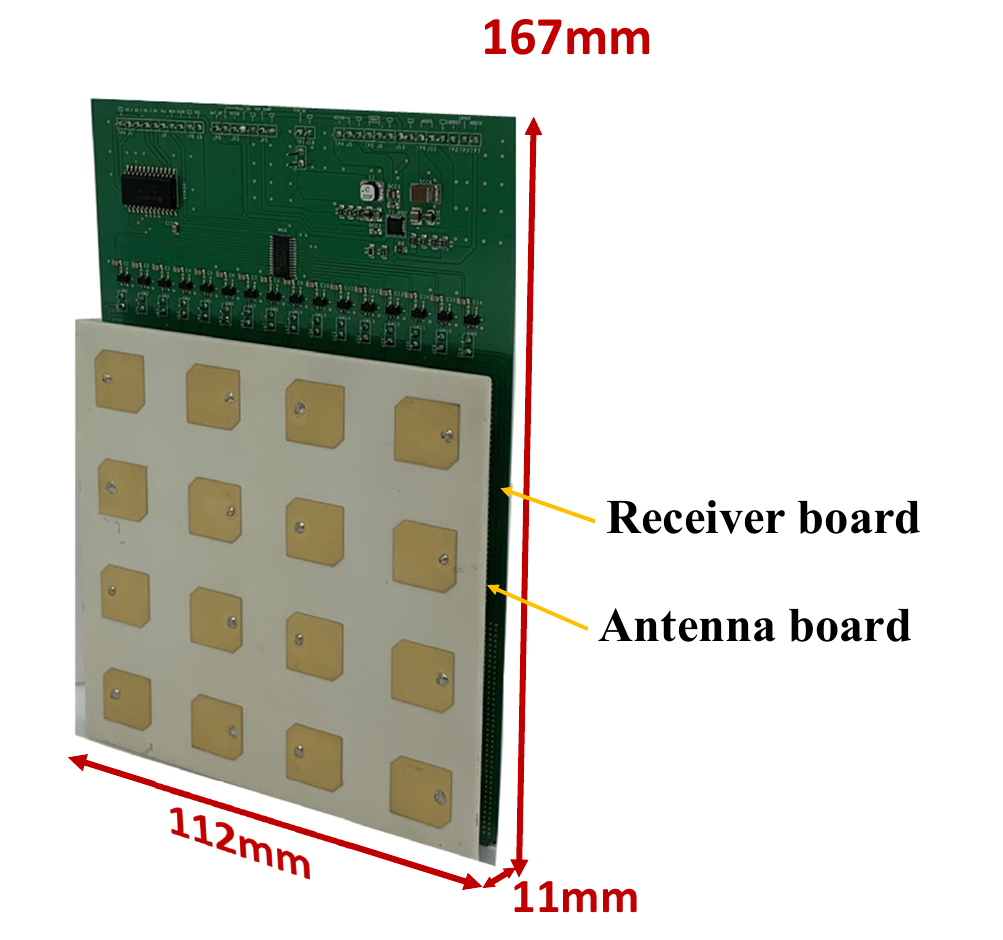}}
	\vfill
	\subfigure[]{\label{fig:experiment_setup_10}
		\includegraphics[trim={0.75in 0.75in 0.75in 0.75in},clip=true,width=0.23\textwidth]{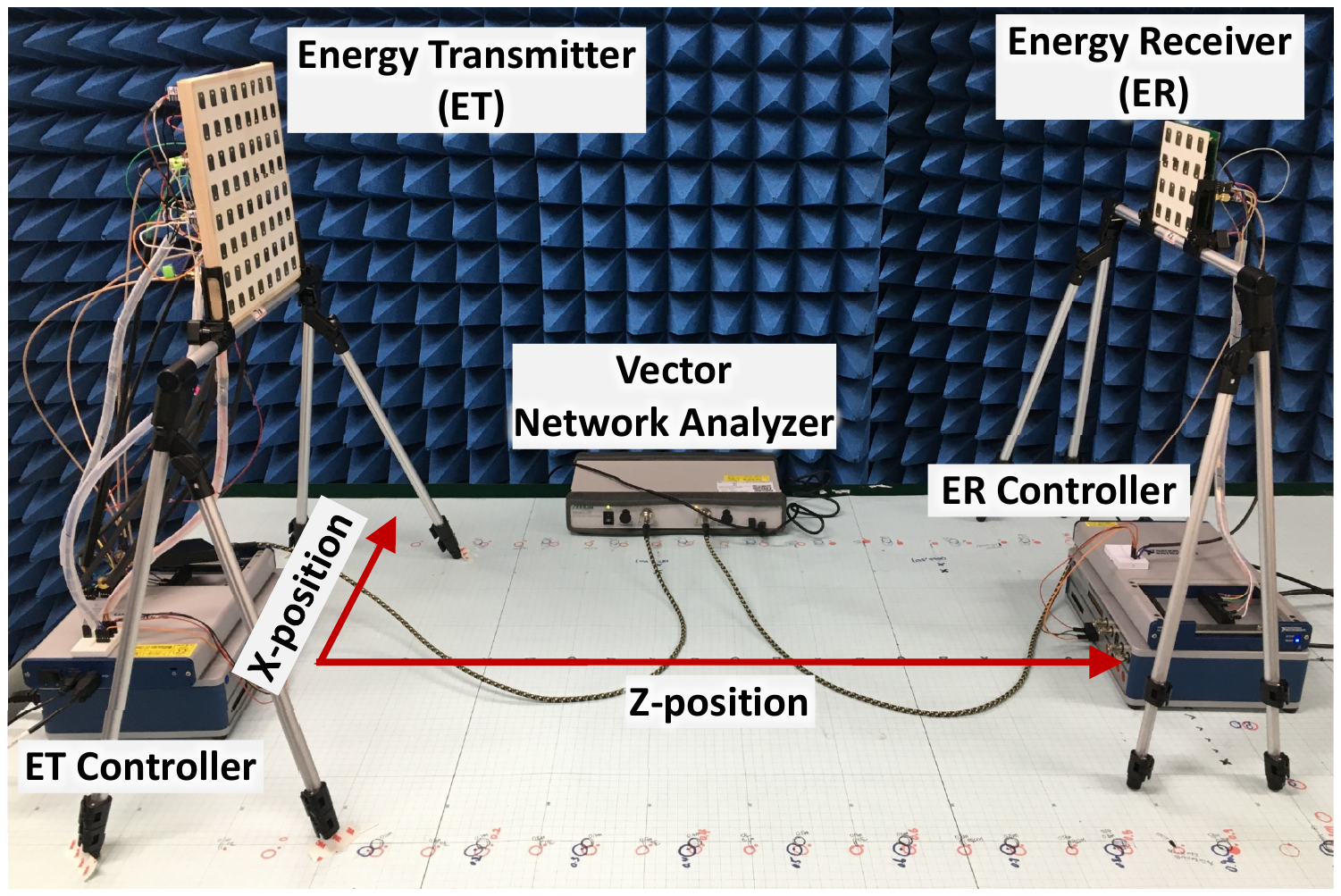}}
	\hspace{-0.05cm}
	\subfigure[]{\label{fig:eff_vs_dist_10}
		\includegraphics[trim={0in 0.25in 1in 0in},clip=true,width=0.22\textwidth]{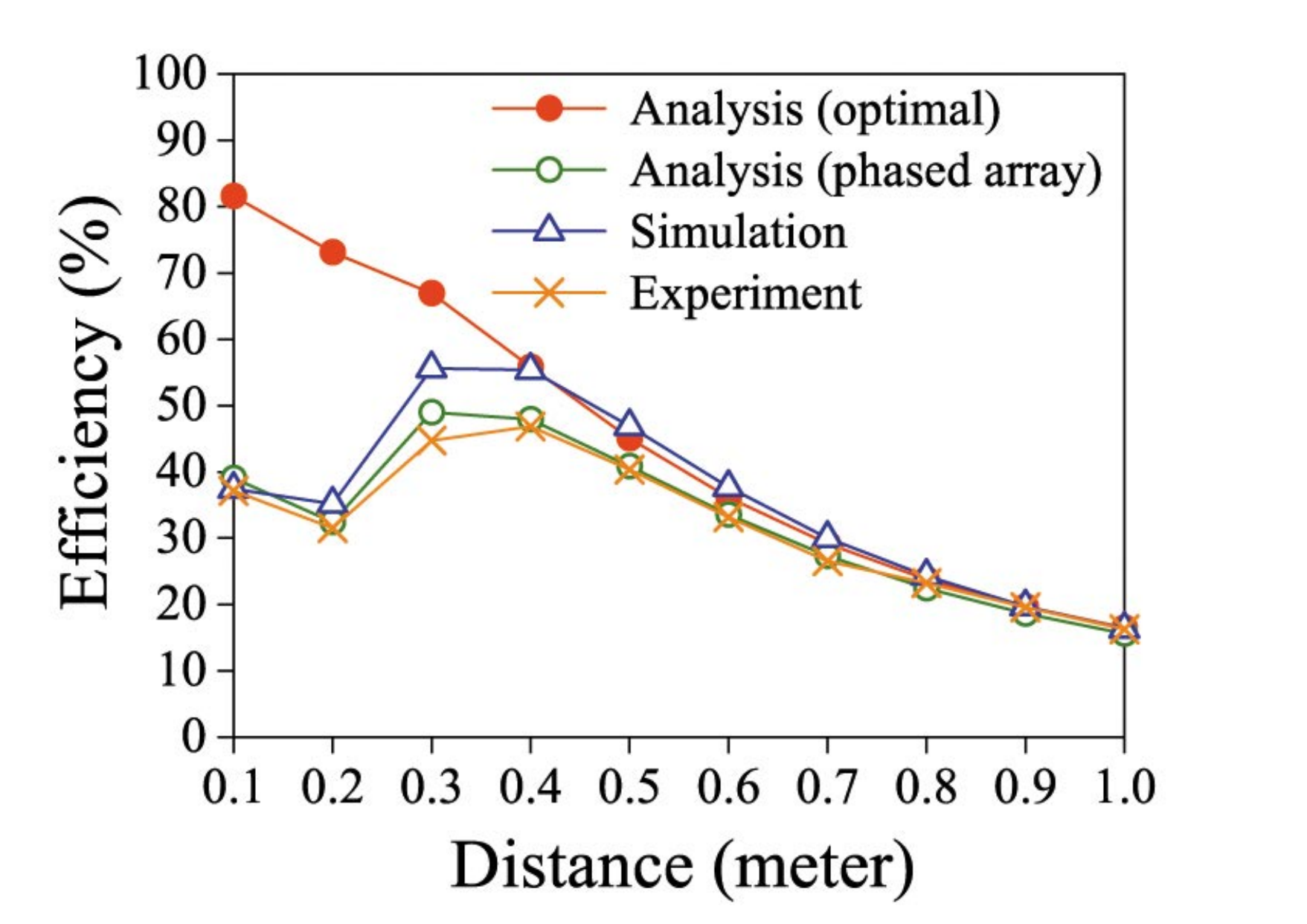}}
	\caption{Multi-antenna RF WET testbed and experimental result. (a) 5.8-GHz phased array ET. (b) The ER. (c) Experimental set-up. (d) PTE over distance.}
\end{figure}

\subsection{Energy Beamforming for Near-Field Phased Array-based WET}\label{subsec: Energy beamforming-near-field}
We first provide experimental results for energy beamforming solutions for near-field phased array WET. Specifically, we have implemented a real-life $5.8$-GHz phased array-based WET system and have verified the effectiveness of the proposed solutions \cite{Jehyeon:2020,Aziz:2019,Jehyeon:2021}. In \cite{Jehyeon:2020}, an efficient method is introduced to compute the PTE of RF WET with known S-parameter information between a multi-antenna TX and RX. We compared the obtained experimental results and full-wave EM simulation results to verify the accuracy of the proposed method. Figs. \ref{fig:Txphased_array} to \ref{fig:experiment_setup_10} display the prototype testbed, which comprises one energy TX (ET) and one EH RX (ER). Both the TX and the RX are equipped with circular polarization (CP) antenna arrays to maximize the PTE. We measured the PTE when moving the ER far away from the ET. From Fig. \ref{fig:eff_vs_dist_10}, we can notice a good agreement between the proposed ``analysis (phased array)'', ``simulation'', and ``experiment'' results. When the ER is close to the ET, the ``analysis (optimal)'' yields a higher PTE since the power losses at the edge of the phased array antenna are not taken into account. However, the proposed analysis can accurately predict the PTE, particularly when the separation distance between ET and ER increases beyond a certain value. Moreover, the proposed analysis can evaluate the PTE almost instantly, which usually requires several days with full-wave EM simulation tools. Despite our best efforts to maintain consistency in the model parameters and testbed configurations across analysis, simulation, and experiment, it is challenging to entirely eliminate all phenomena that may cause small discrepancies.

\begin{figure}[t]\centering
	\subfigure[]{\label{fig:2_testbed}
		\includegraphics[trim={0in 1in 0in 0.2in},clip=true,width=0.4\textwidth]{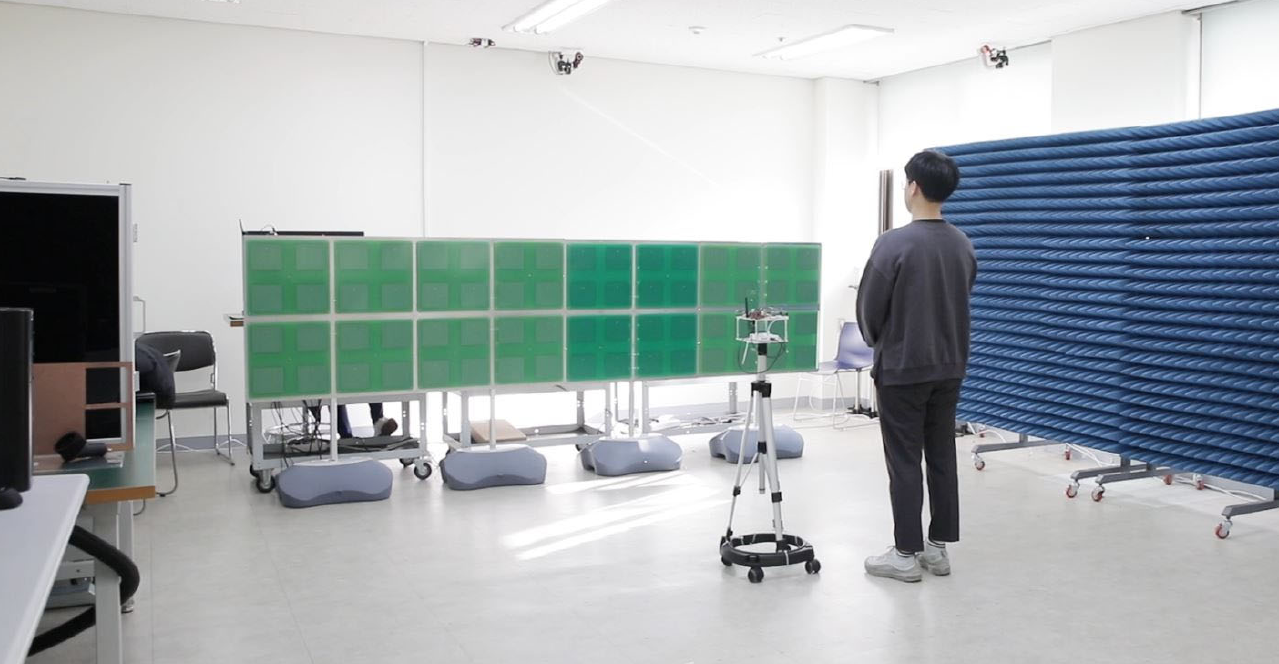}}
	\vfill
	\subfigure[]{\label{fig:2_eff_dist}
		\includegraphics[trim={1in 0.5in 1in 1in},clip=true,width=0.22\textwidth]{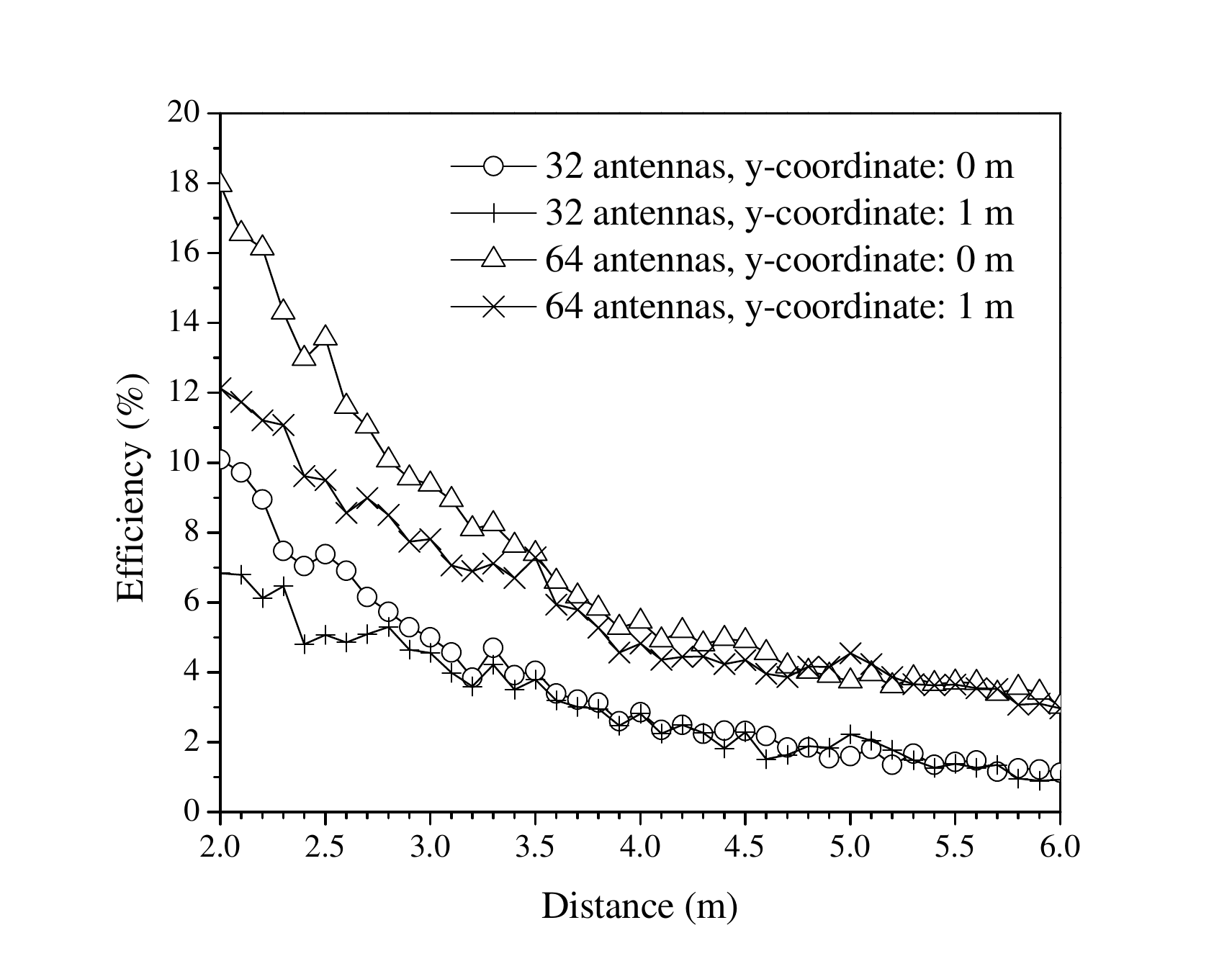}}
	\subfigure[]{\label{fig:2_powerbeam}
		\includegraphics[trim={0.25in 0.25in 0.25in 0in},clip=true,width=0.22\textwidth]{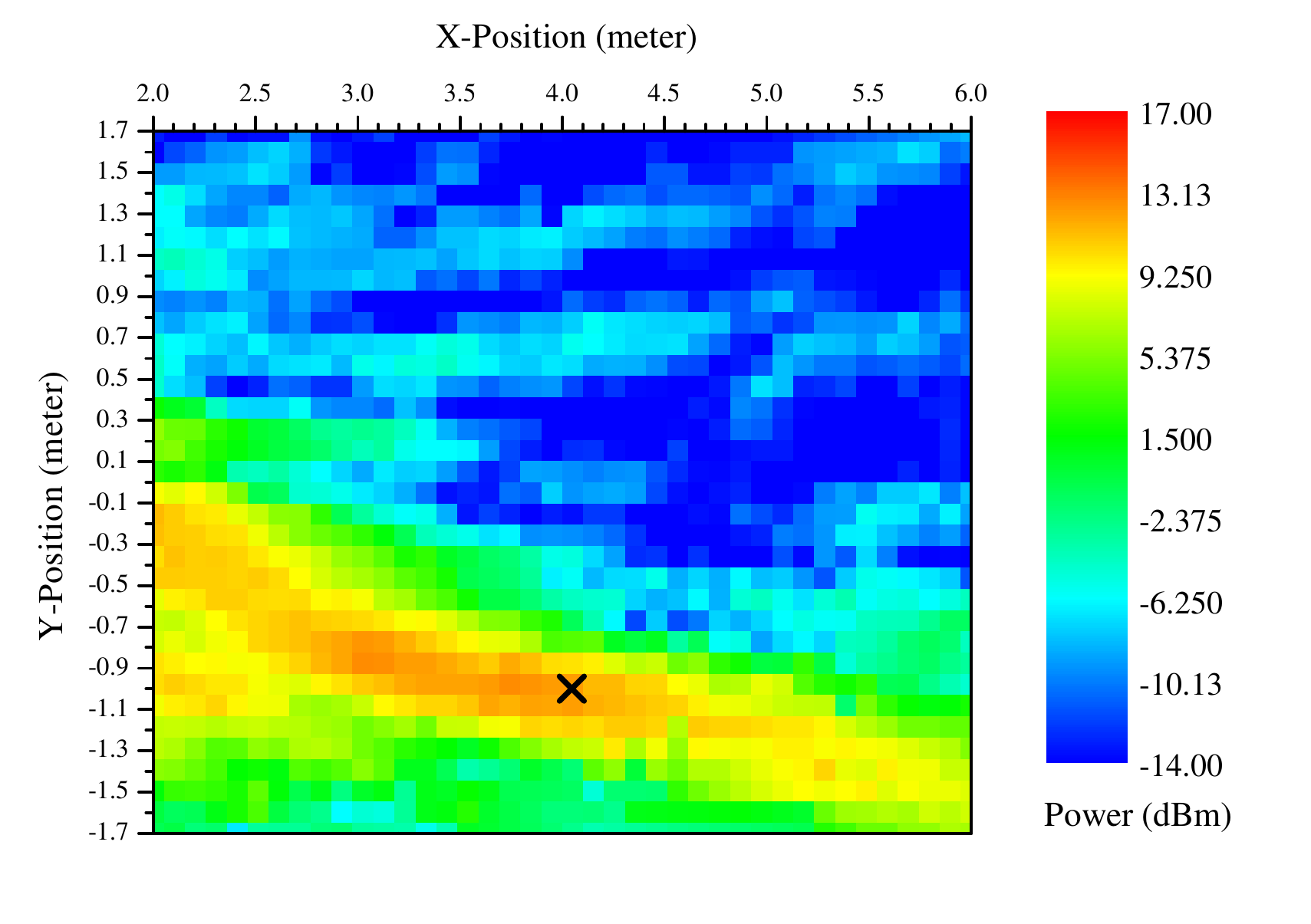}}
	\caption{Joint location tracking and WET experimental results. (a) System prototype. (b) PTE over distance. (c) Position tracking.}
\end{figure}

In \cite{Aziz:2019}, we proposed a phase estimation algorithm based solely on received power measurements to enable joint location tracking and WET. Specifically, we have designed a specific sequence of transmit phase patterns to estimate all the channel gains between the transmit antennas and receive antenna. Figs. \ref{fig:2_eff_dist} and \ref{fig:2_powerbeam} illustrate the experimental results obtained at $920$ MHz with the experiment testbed in Fig. \ref{fig:2_testbed}. After executing the algorithm, we measured the PTE as the ER was moved away from the ET, with either $32$ or $64$ antenna elements activated. We observe that the PTE peaks at $18$\% at $2$ meters distance when $64$ antennas are activated.

\begin{figure}[t]\centering
	\subfigure[]{\label{fig:rxpow_scanbeam}
		\includegraphics[trim={0in 1in 0in 0.5in},clip=true,width=0.4\textwidth]{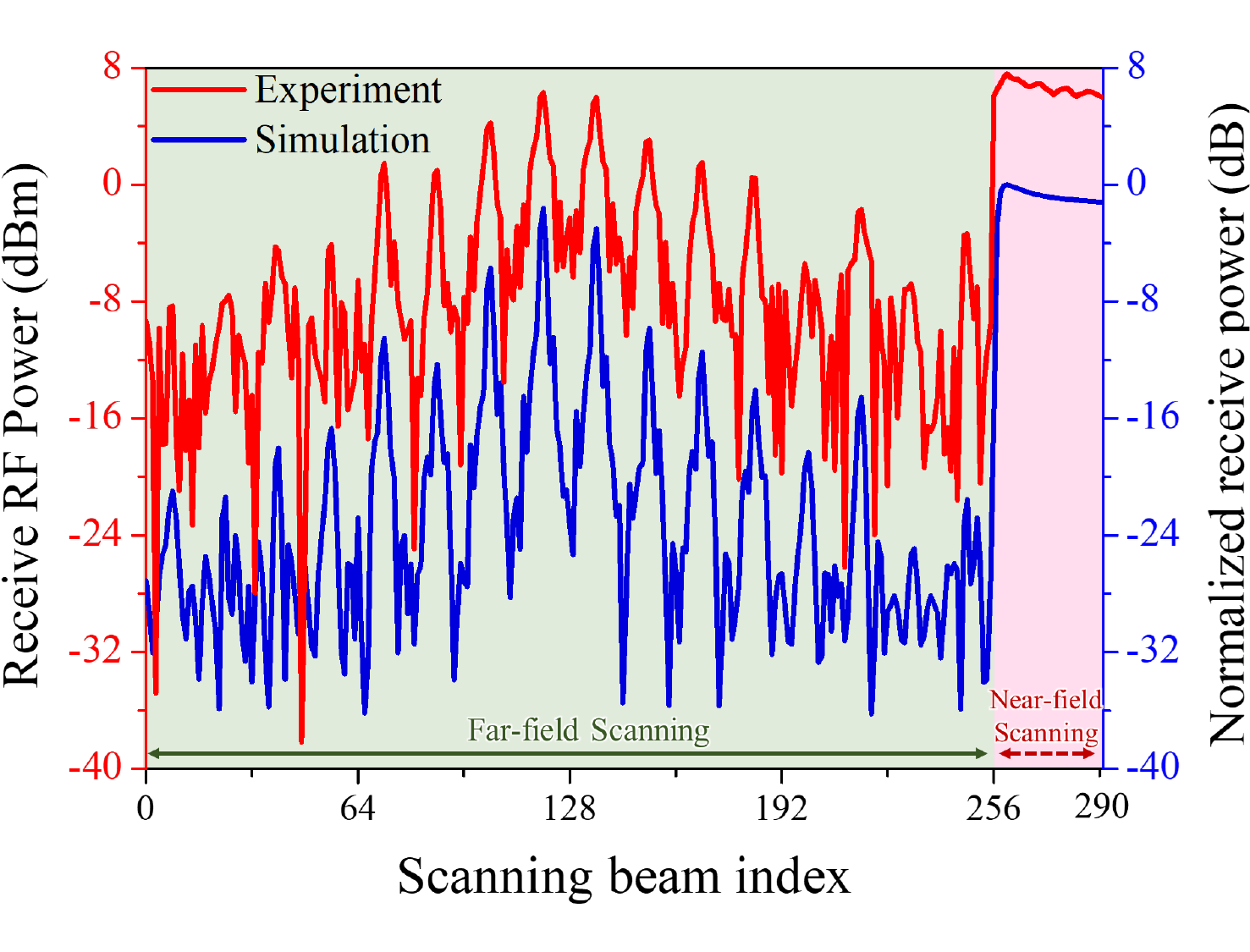}}
	\vfill
	\subfigure[]{\label{fig:exp_effvsdist}
		\includegraphics[width=0.225\textwidth]{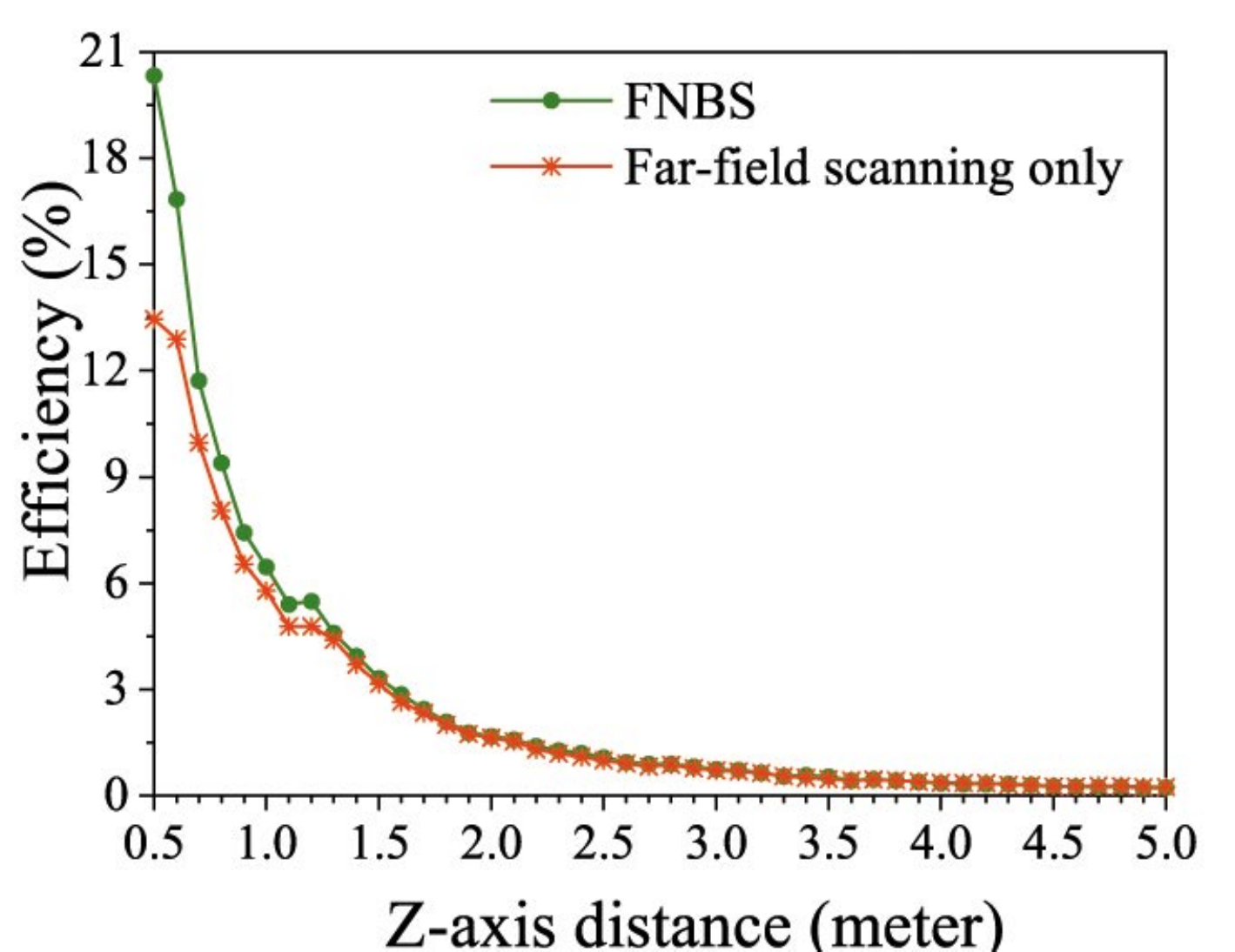}}
	\subfigure[]{\label{fig:exp_offset}
		\includegraphics[width=0.225\textwidth]{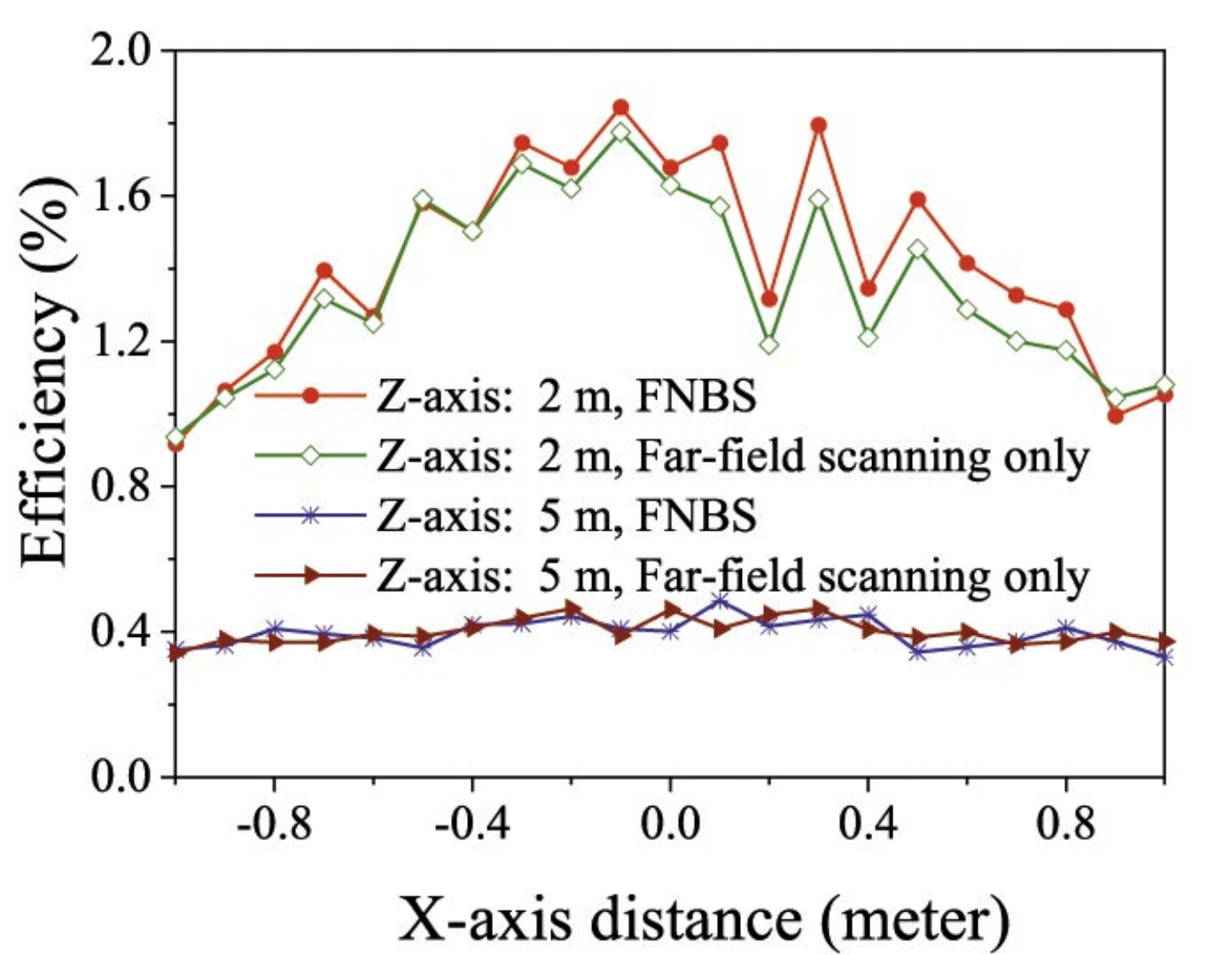}}
	\caption{Phased array-based WET experimental results. (a) Receive power according to scanning beams. PTE according to (b) z-axis distance. (c) x-axis distance.}
\end{figure}

Using the experimental set-up of Fig. \ref{fig:experiment_setup_10}, we verified a far-near beam scanning algorithm (FNBS), which can effectively focus the RF power on the ER located freely in both radiative near-field zone and far-field zone \cite{Jehyeon:2021}. Different from \cite{Aziz:2019}, we use a beam scanning method that uses near- and far-field pencil beams to realize the optimal transmit excitation phases at all the transmit antennas. Indeed, we first scan the ET with predefined far-field scanning beams to find the optimal far-field direction. After that, based on the far-field information, we adjust the ET to determine the optimal near-field beam. Fig. \ref{fig:rxpow_scanbeam} compares the received power according to each scanning beam between MATLAB simulation and experiment data when ER is placed at $0.5$ meters from ET. The experimental results correspond to the actual measured received RF powers (i.e., the left $y$-axis applies), whereas the simulation results are the normalized received powers (i.e., the right $y$-axis applies). We can clearly see a good agreement between both cases. In the far-field scanning phase, the $120$th scanning beam index is the optimal one which produces the highest received power for both cases. We can observe $1.61$ dB and $1.26$ dB improvements for simulation and experiment in the near-field scanning phase, corresponding to the $260$th near-field beam.

Furthermore, we compare the PTE between the proposed FNBS and conventional far-field scanning when moving the ER along the z-axis and x-axis, as shown in Figs.  \ref{fig:exp_effvsdist} and  \ref{fig:exp_offset}, respectively. The FNBS algorithm provides higher PTE than the counterpart when the ER is close to the ET. Almost the same performance is witnessed when the ER moves far away from the ET.

\subsection{Energy Beamforming for IRS-based WET}
\label{subsec: Energy beamfomring for IRS-based WET}
The prototype discussed in the previous subsection focuses on implementing active energy beamforming using a phased antenna array for near-field WET. As highlighted in Section \ref{sec:IRS}, the use of IRS promises to bring both active and passive beamforming gains to further boost WET and WIET performance.

\begin{figure}[t]\centering
	\subfigure[]{\label{fig:B1_IRSboard}
		\includegraphics[trim={2in 0.75in 2in 0.75in},clip=true,width=0.15\textwidth]{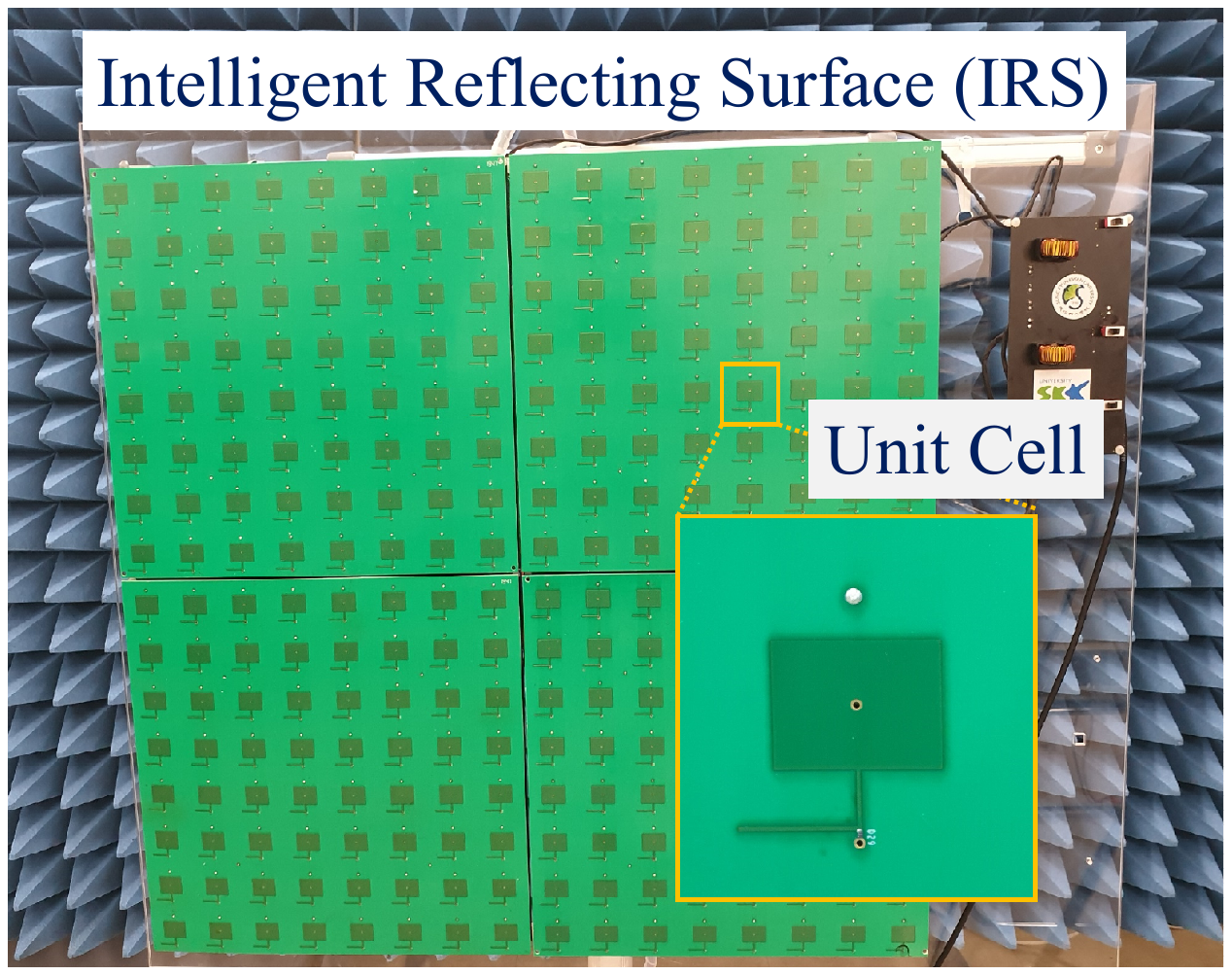}}
	\subfigure[]{\label{fig:B1_testbed}
		\includegraphics[trim={0.5in 0.75in 0in 0.5in},clip=true,width=0.3\textwidth]{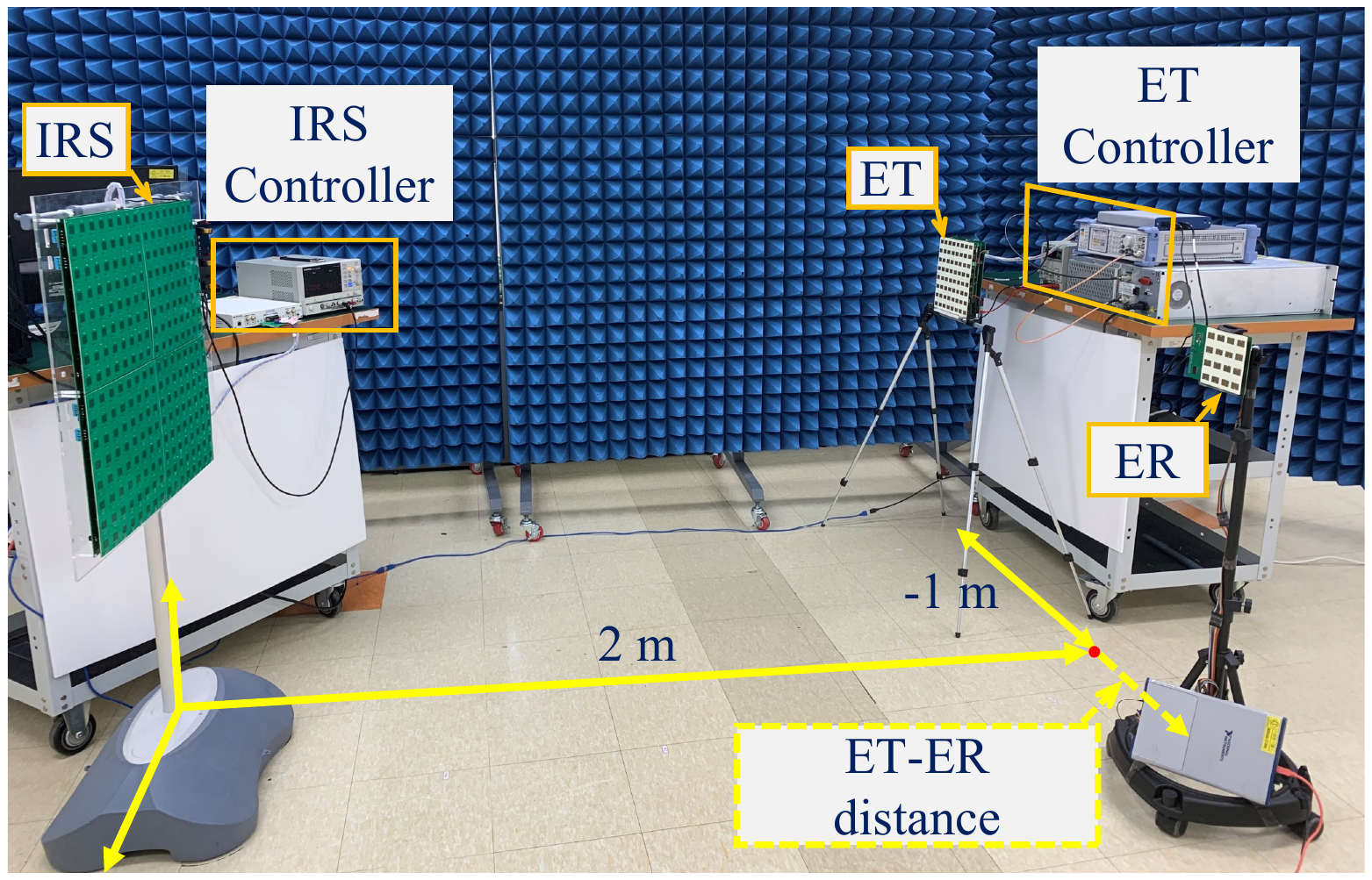}}     
	\vfill
	\subfigure[]{\label{fig:B1_exp_44}
		\includegraphics[trim={0in 0in 0in 0in},clip=true,width=0.225\textwidth]{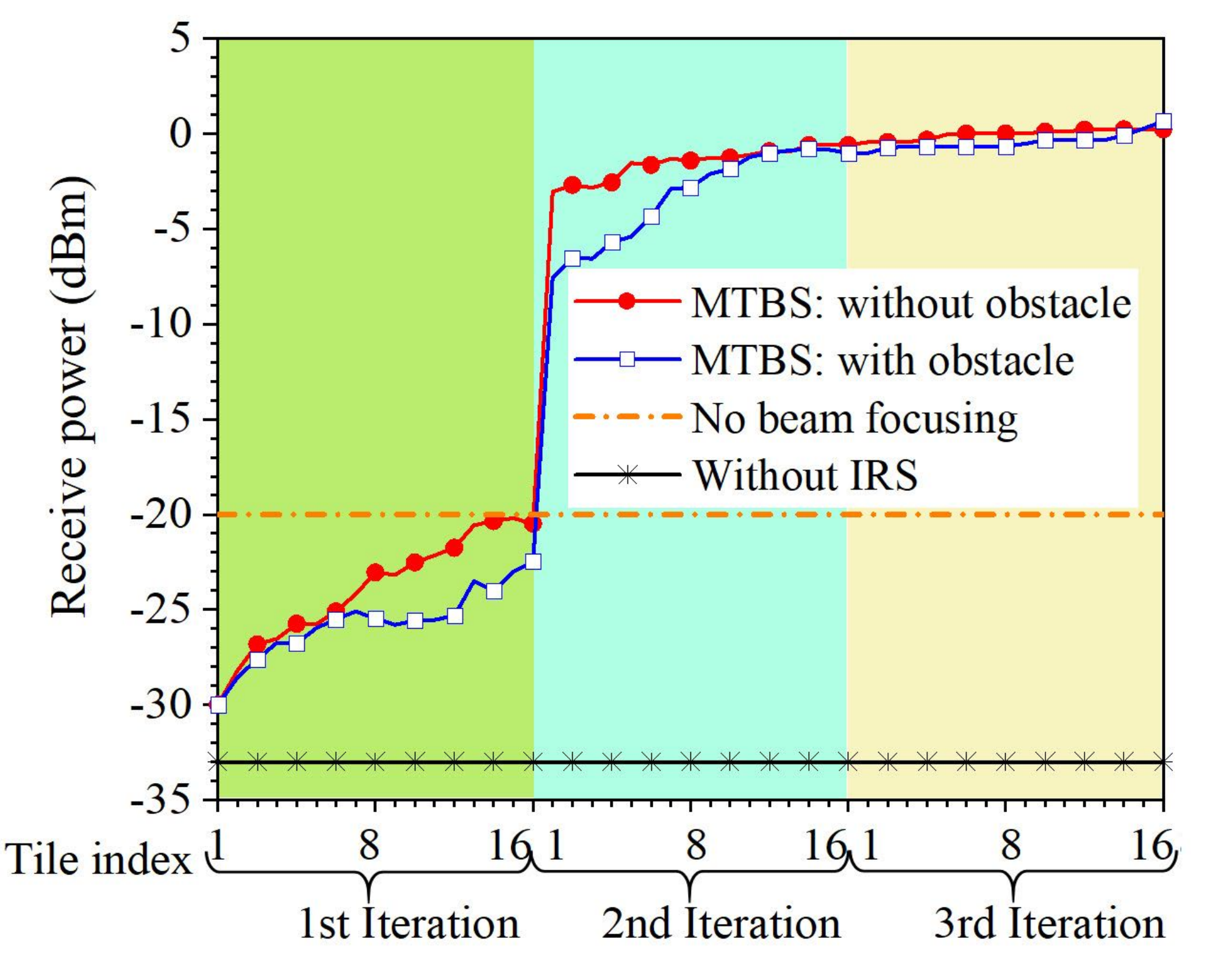}}
	\subfigure[]{\label{fig:B1_effvsdist}
		\includegraphics[trim={0.5in 0in 0.25in 0.75in},clip=true,width=0.225\textwidth]{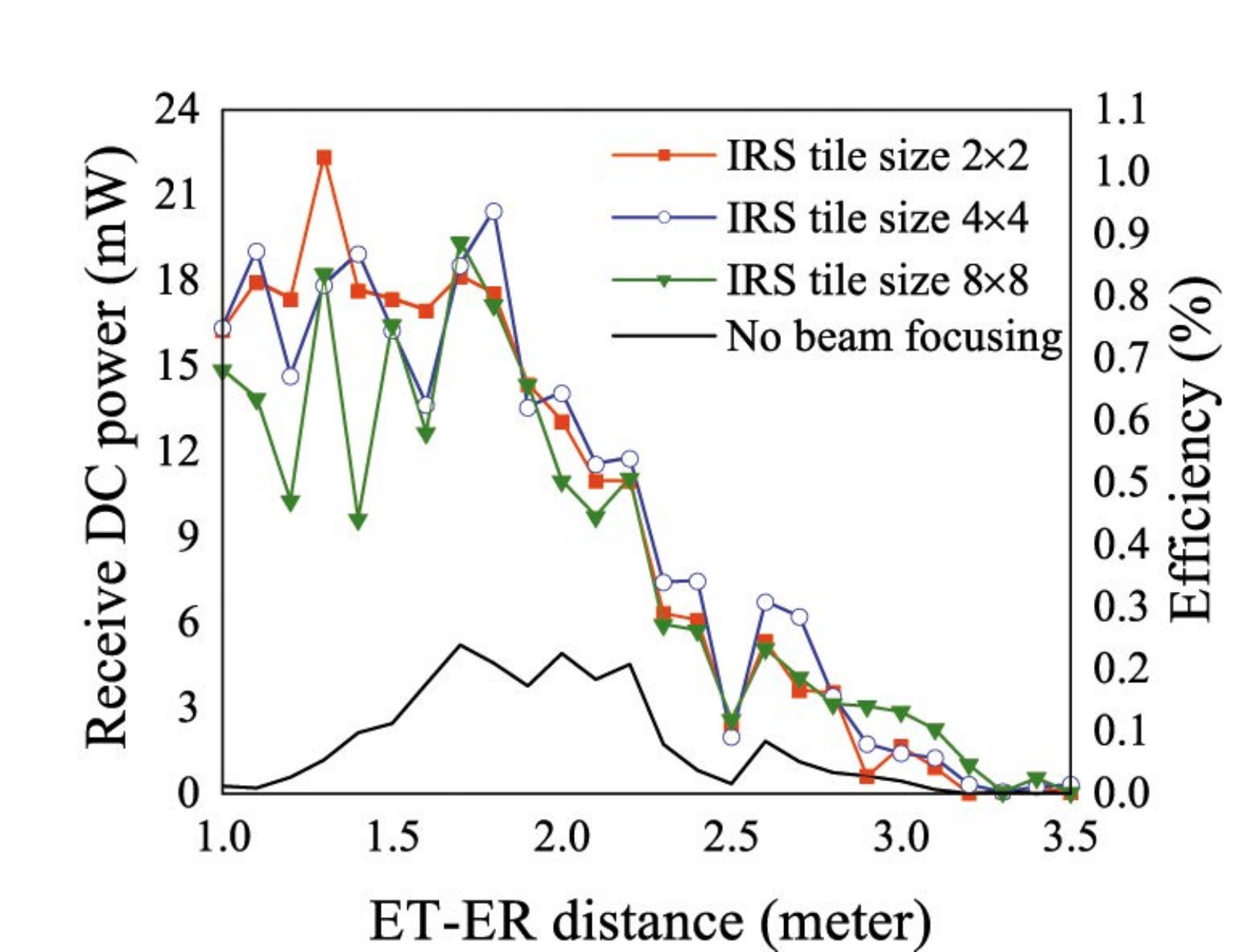}}
	\caption{Single focusing beam IRS-based WET. (a) Prototype testbed. (b) IRS board. (c) Receive power over the iteration (IRS tile size $4\times 4)$. (d) Receive power and PTE over distance.}
\end{figure}

We have extended the previous prototype testbed with an additional $16 \times 16$ 1-bit IRS, as shown in Fig. \ref{fig:B1_IRSboard}, and verified the benefits of using IRS with different near-field beam focusing algorithms based solely on the received power \cite{Tran:2021,Tran:2022}. In particular, we propose a multi-tile IRS beam scanning (MTBS) algorithm to enable a single-focusing beam to enhance the performance of WET in \cite{Tran:2021}. The algorithm estimates the channel based only on the receive power information at the ET. Indeed, to enable near-field beam focusing, we divide the entire IRS into small IRS tiles to ensure ET and ER lie within the far-field region of each tile. Then, while fixing the other IRS tiles, we iteratively scan a targeted IRS tile with pre-defined scanning beams to realize the optimal reflection coefficient vector of each IRS tile. We repeat the procedure for every IRS tile until the algorithm converges.

In our experiment, we repeatedly scanned the IRS in three iterations. In the first iteration, we activated one antenna element at the ET to reduce the complexity. After that, we activated all elements at the ET and started scanning to find the optimal beam at the ET. In the next iterations, we set the optimal beam at the ET. Fig. \ref{fig:B1_testbed} shows the experiment setup for verifying the proposed algorithm. We considered three different IRS tile sizes: $2\times 2$, $4\times 4$, and $8\times 8$. To avoid confusion, only the receive power over the scanning iteration of IRS tile size $4\times4$ is shown in Fig. \ref{fig:B1_exp_44}. Moreover, we added an obstacle between the ET and the ER to observe the effect of the IRS. As can be seen, after executing the algorithm, the receive power of the two cases ``MTBS: with obstacle'' and ``MTBS: without obstacle'' coincide. This result shows that the IRS can provide an alternative path to enhance the performance of the WET system when there is no LoS path between ET and ER. Compared to the case ``No beam focusing'' and ``Without IRS'', our proposed algorithm can achieve $20$ dB and $33$ dB improvement, respectively. Also, we moved the ER away from the ET and recorded the receive DC power and the corresponding PTE according to the ET-ER distance as presented in Fig. \ref{fig:B1_effvsdist}. The total transmit power is set to $2.1$ W. It is noticed that the maximum receive power is about $22.5$ mW, which corresponds to around 1.05\% PTE at $1.3$ m of ET-ER distance in case of a $2\times 2$ IRS tile size. The PTE is relatively low since the actual channel length (i.e., ET-IRS-ER) is almost $4.25$ m. However, the proposed algorithm consistently outperformed the ``No beam focusing'' scheme. The PTE gradually reduces to almost $0\%$ as the ET-ER distance approaches $3.5$ m.

\begin{figure}[t]\centering
	\subfigure[]{\label{fig:B2_testbed}
		\includegraphics[trim={1.5in 01in 1.5in 01in},clip=true,width=0.225\textwidth]{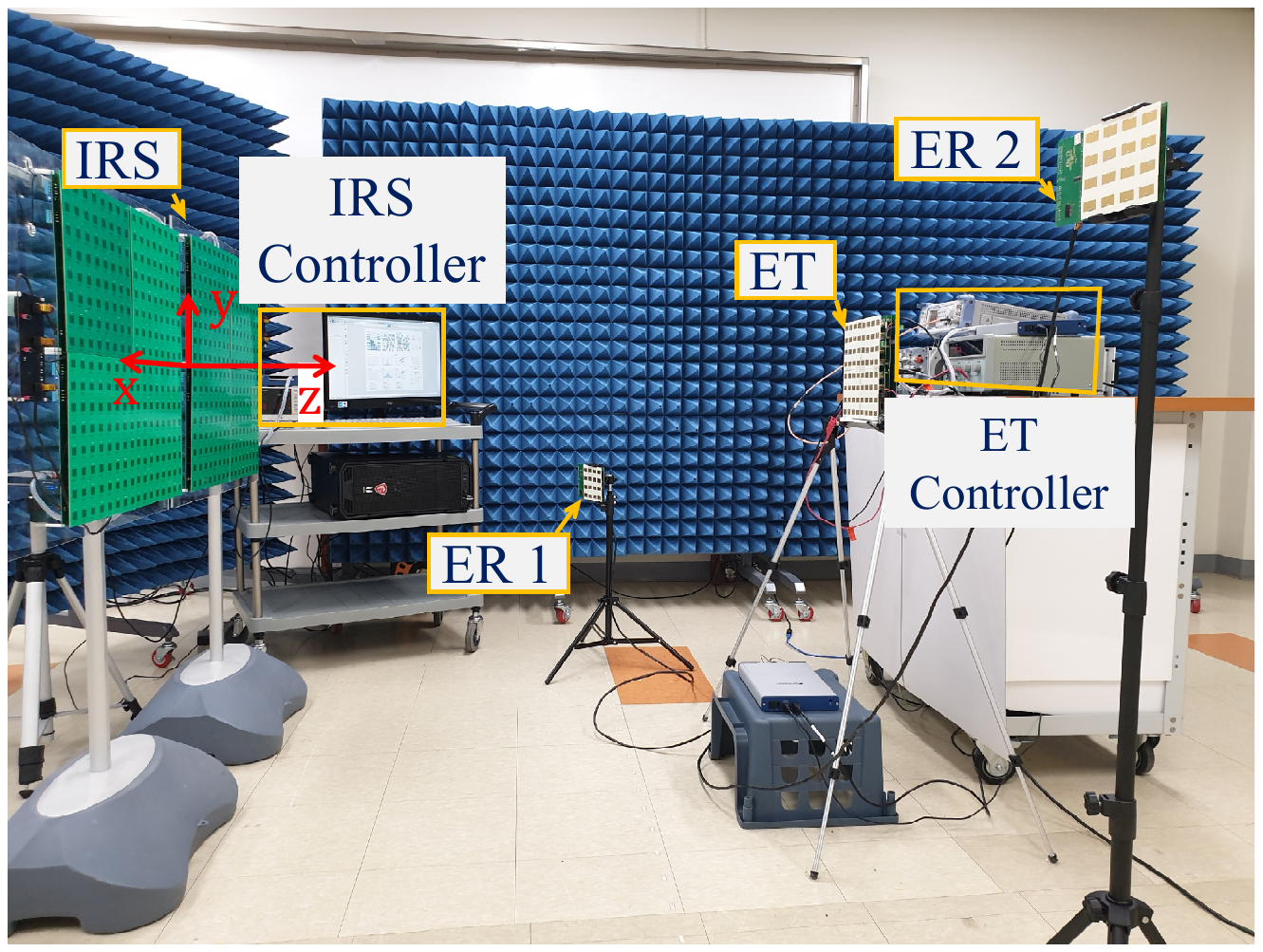}}
	\subfigure[]{\label{fig:B2_PA}
		\includegraphics[trim={0.2in 0in 0.25in 0.2in},clip=true,width=0.225\textwidth]{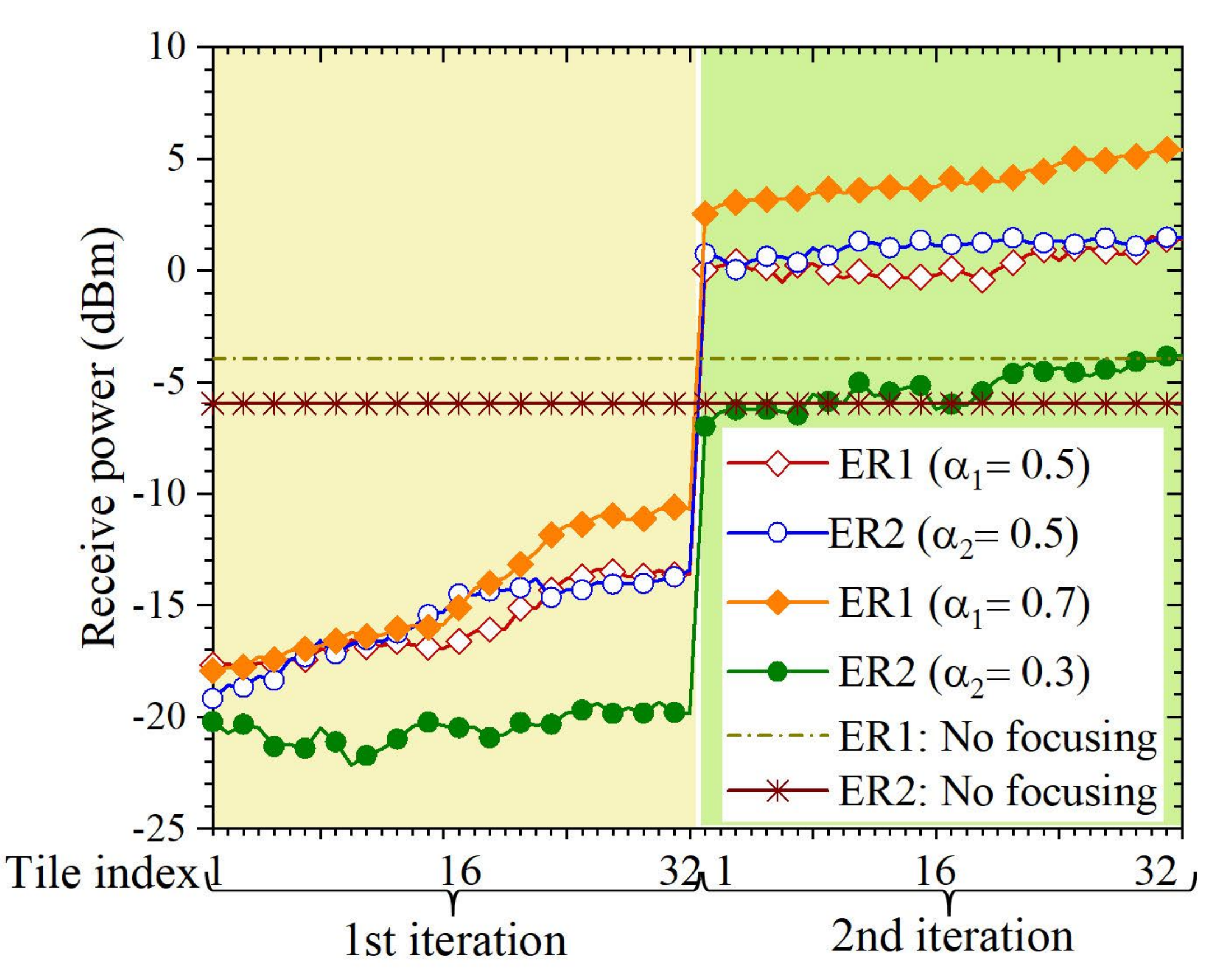}}
	\vfill
	\subfigure[]{\label{fig:B2_RUI}
		\includegraphics[trim={0.2in 0in 0.25in 0.2in},clip=true,width=0.225\textwidth]{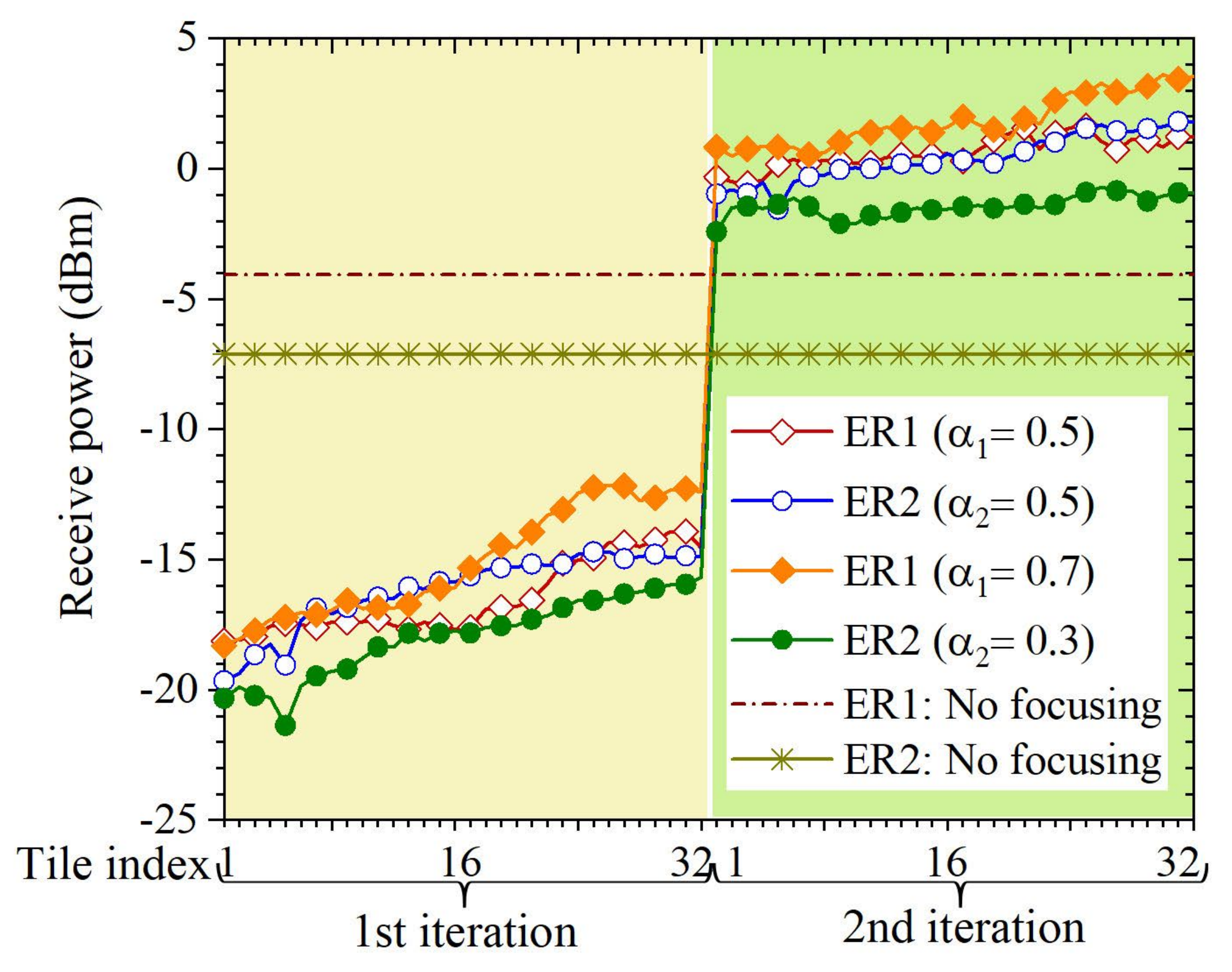}}
	\subfigure[]{\label{fig:B2_ITD}
		\includegraphics[trim={0.2in 0in 0.25in 0.2in},clip=true,width=0.225\textwidth]{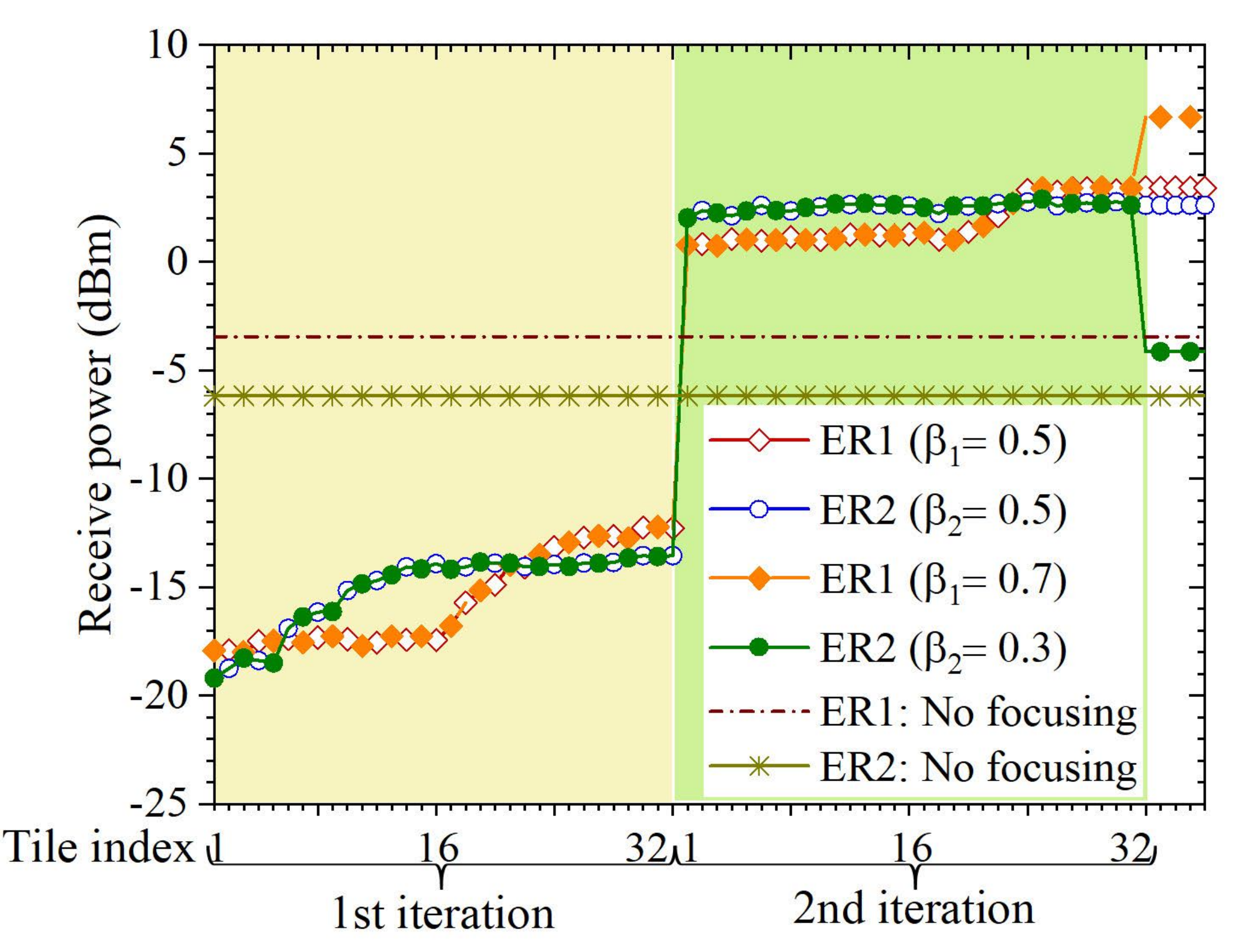}}
	\caption{Multi-focus IRS-aided WET. (a) Experimental setup. Receive power over each iteration with (b) PA. (c) RUI. (d) ITD.}
\end{figure}

Furthermore, the previous work was extended to three different multiple-focusing techniques: pattern addition (PA), random unit cell interleaving (RUI), and IRS tile division (ITD) \cite{Tran:2022}. Based only on the receive power information, these techniques manage to form multiple power beams to different focal points in the near-field region of the IRS. Moreover, we allocated a weight factor (i.e., $\alpha_l$ for $\text{ER}_l$) to control the power level distributed to each focal point. Then, we implemented a real-life multi-focus IRS-aided WET testbed (see Fig. \ref{fig:B2_testbed}) to verify these methods. Figs. \ref{fig:B2_PA}--\ref{fig:B2_ITD} illustrate the measured receive power over each iteration according to different techniques. All techniques effectively distribute a higher power level to the ER with a higher weight factor. With the proposed techniques, the two ERs receive higher power compared to the cases ``No focusing''. 

\begin{figure}[t]\centering
\subfigure[]{\label{fig:B2_testbed_c2}
	\includegraphics[trim={1.5in 0in 1.5in 0 in},clip=true,width=0.225\textwidth]{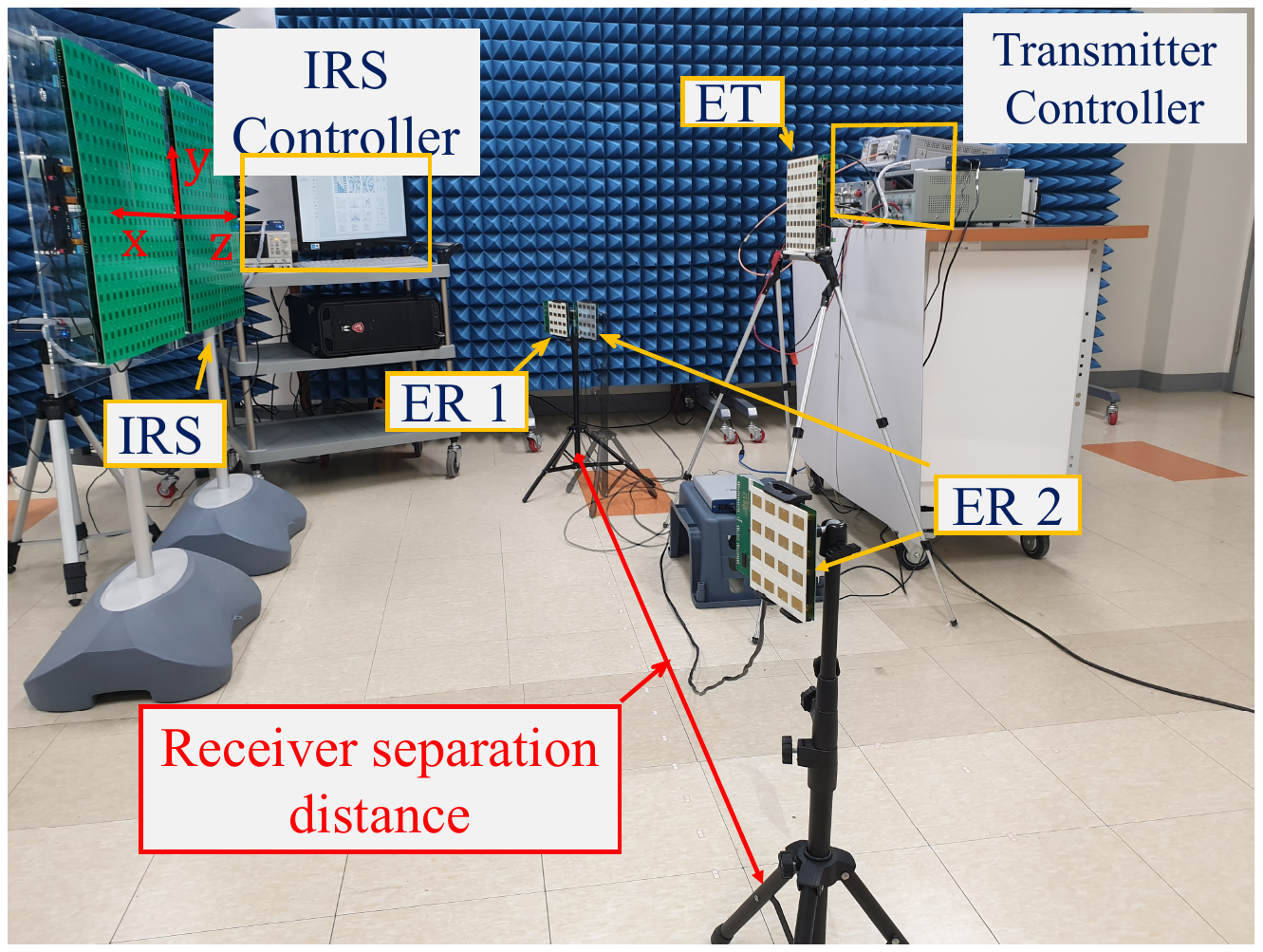}}
\subfigure[]{\label{fig:B2_rxpow_pte_2}
	\includegraphics[trim={0.25in 0.5in 0.25in 0.25in},clip=true,width=0.225\textwidth]{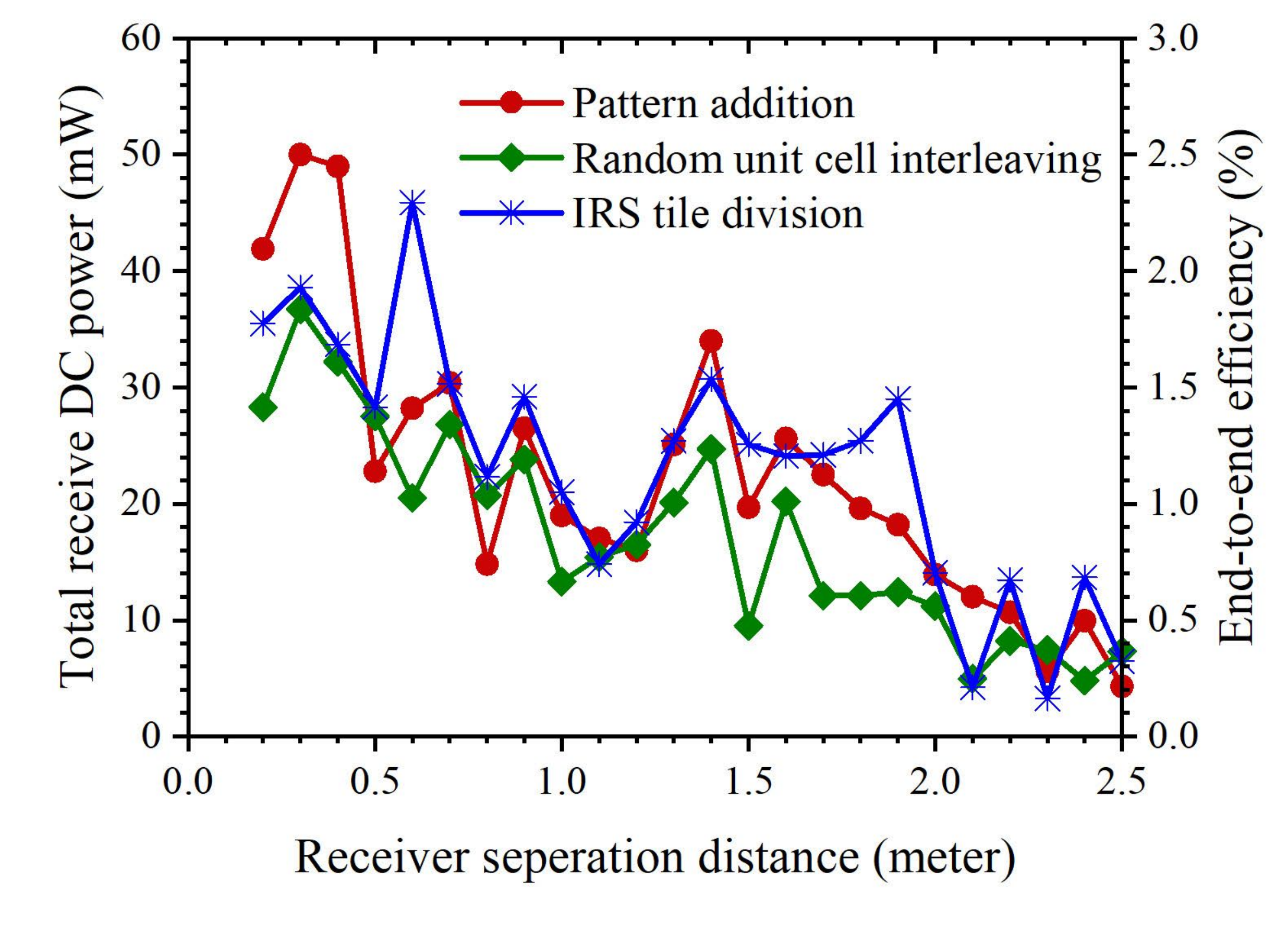}}
\caption{Performance according to distance. (a) Experimental scenario. (b) Total receive DC power and PTE over RX separation distance.}
\end{figure}

Additionally, we measured the total receive DC power and PTE of two ERs in which ER1 stays in a fixed location and ER2 moves away from ER1, as shown in Fig. \ref{fig:B2_testbed_c2}. We call the distance between two ERs ``Receiver separation distance'' and record the performance according to this distance. An equal weight factor set is given to the ERs, and the total transmit power is $1.84$ W. Fig. \ref{fig:B2_rxpow_pte_2} shows the measured data. The PA and RUI can provide the highest receive DC power of $51$ and $37$ mW, respectively, at $0.3$ m of receiver separation distance, and that of ITD is $46$ mW at $0.6$ m. As the ERs get closer, PA performs better than the other methods, while ITD outperforms its counterparts when the separation distance between ERs is longer. Note that the fluctuating results by the RUI method are due to the method's random approach.

\subsection{Beam Sharing for IRS-aided WIET}\label{subsec:Beam sharing IRS-aided WIET}
The prototypes so far were dedicated to WET. This section details the methodologies and experimental verification of an IRS-aided WIET application. Indeed, we experimentally verify the beam-sharing algorithm (BSA), which has been thoroughly presented in \cite{TranICASSP:2023}. 

\begin{figure}[t]\centering
\includegraphics[trim={0in 0.5in 0.5in 0.5in},clip=true,width=0.4\textwidth]{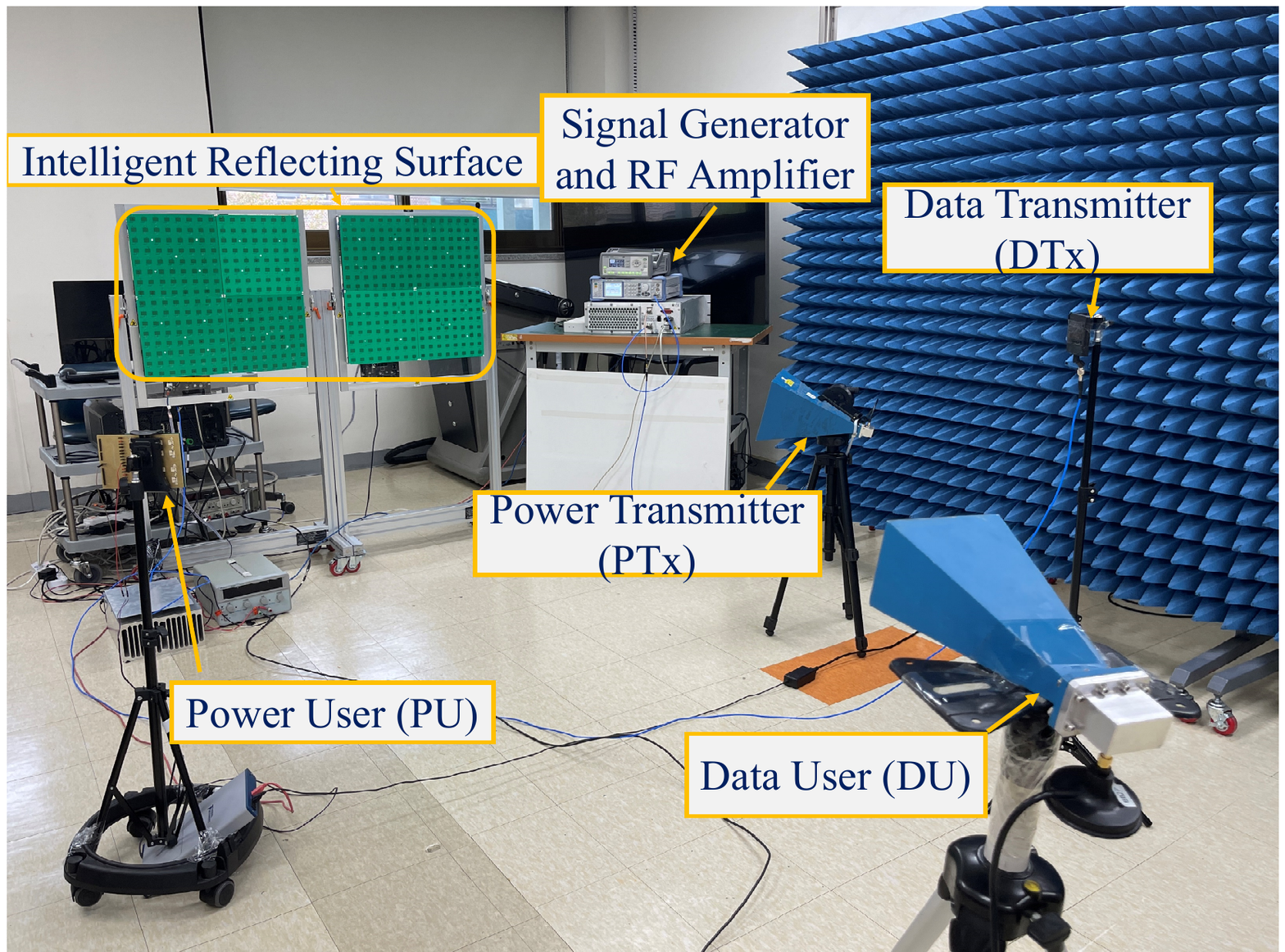}
\caption{IRS-aided WIET experimental setup.}\label{fig:C_testbed}
\end{figure}

The existing literature on WIET commonly considers two scenarios: a single transmitter serving two separate users - a data user (DU) and a power user (PU), and a single transmitter serving a single user. The former scenario may lead to high PAPRs, making practical implementations challenging. On the other hand, the latter scenario is more conducive for low-power devices such as sensors. The considered IRS-aided WIET system consists of one IRS, one data TX (Dtx), one power TX (Ptx), one DU, and one PU. Fig. \ref{fig:C_testbed} shows the experimental testbed for IRS-aided WIET. The Dtx sends an OFDM-modulated signal to the DU, while the Ptx transfers a high-power single-tone signal to the PU. The power required for the power transfer is significantly higher than that of the communication system. While data communication is guaranteed as long as the received power is higher than the noise floor (i.e., $-130$ dBm), powering an EH RX requires at least $0$ dBm \cite{KWChoi:2020}. Hence, the leakage of high-power signal from Ptx causes data distortion within the analog-to-digital converter of the DU which degrades the performance. Moreover, the low-noise amplifier in the DU might become saturated when it receives the data signal combined with the high-power signal leakage from PU. Given that the DU is equipped with delicate components (e.g., low-noise amplifier), high power leakage from the Ptx to the DU can cause fatal damage to the DU. Therefore,  the primary objective of this research is to enable the nulling capability of the IRS to minimize the leakage power from the Ptx to the DU. Meanwhile, the power delivered from the Ptx to the PU and the signal received at the DU from the Dtx should be maximized for optimal performance. To this end, we also divide the IRS into smaller IRS tiles to enable near-field beam focusing. After that, we consecutively execute the BSA algorithm to optimize each IRS tile.

\subsubsection{Algorithm Validation}\label{subsubsec:algoval}
In this experiment, we deployed a $16\times 32$ IRS divided into $8$ individual $8$-by-$8$ IRS tiles. The IRS was positioned at the origin of the global coordinate system. The Dtx and DU were located at coordinates ($-1$ m, $2.5$ m, $0$ m) and ($0.25$ m, $4$ m, $0$ m), respectively, while the Ptx and PU were positioned at ($-1.25$ m, $1.8$ m, $-0.3$ m) and ($1$ m, $1.8$ m, $-0.3$ m), respectively. We employed $16$-quadrature amplitude modulation (QAM) modulation at the Dtx. The transmitted power levels for Dtx and Ptx were set to $3$ mW and $62$ mW, respectively.

\begin{figure}[t]\centering
\subfigure[]{\label{fig:C1_rxpow}
\includegraphics[trim={0.5in 0in 0.75in 0.25in}, clip=true,width=0.35\textwidth]{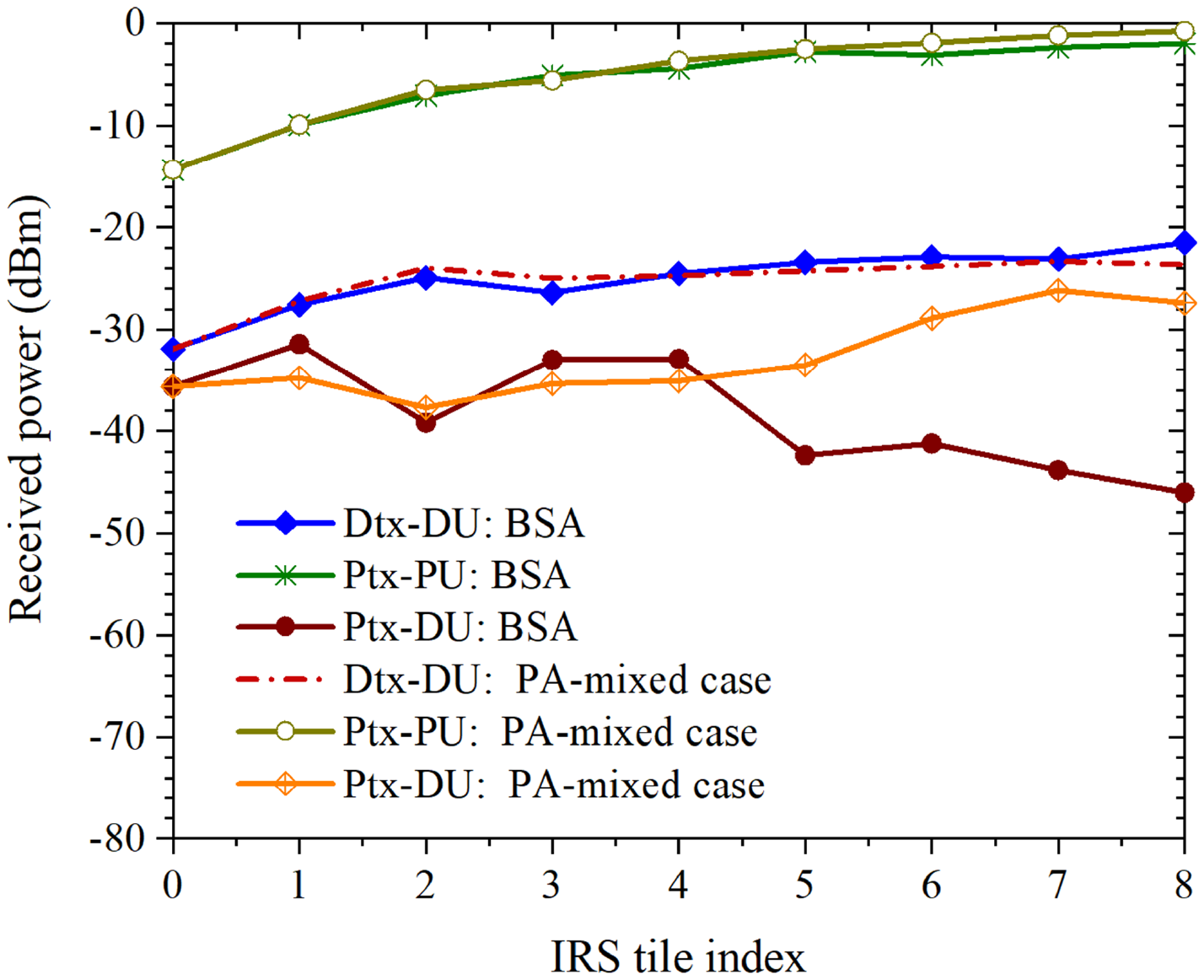}}
\vfill
\subfigure[]{\label{fig:C1_Const_OFF}
\includegraphics[trim={0.2in 0in 0.5in 0.25in},clip=true,width=0.225\textwidth]{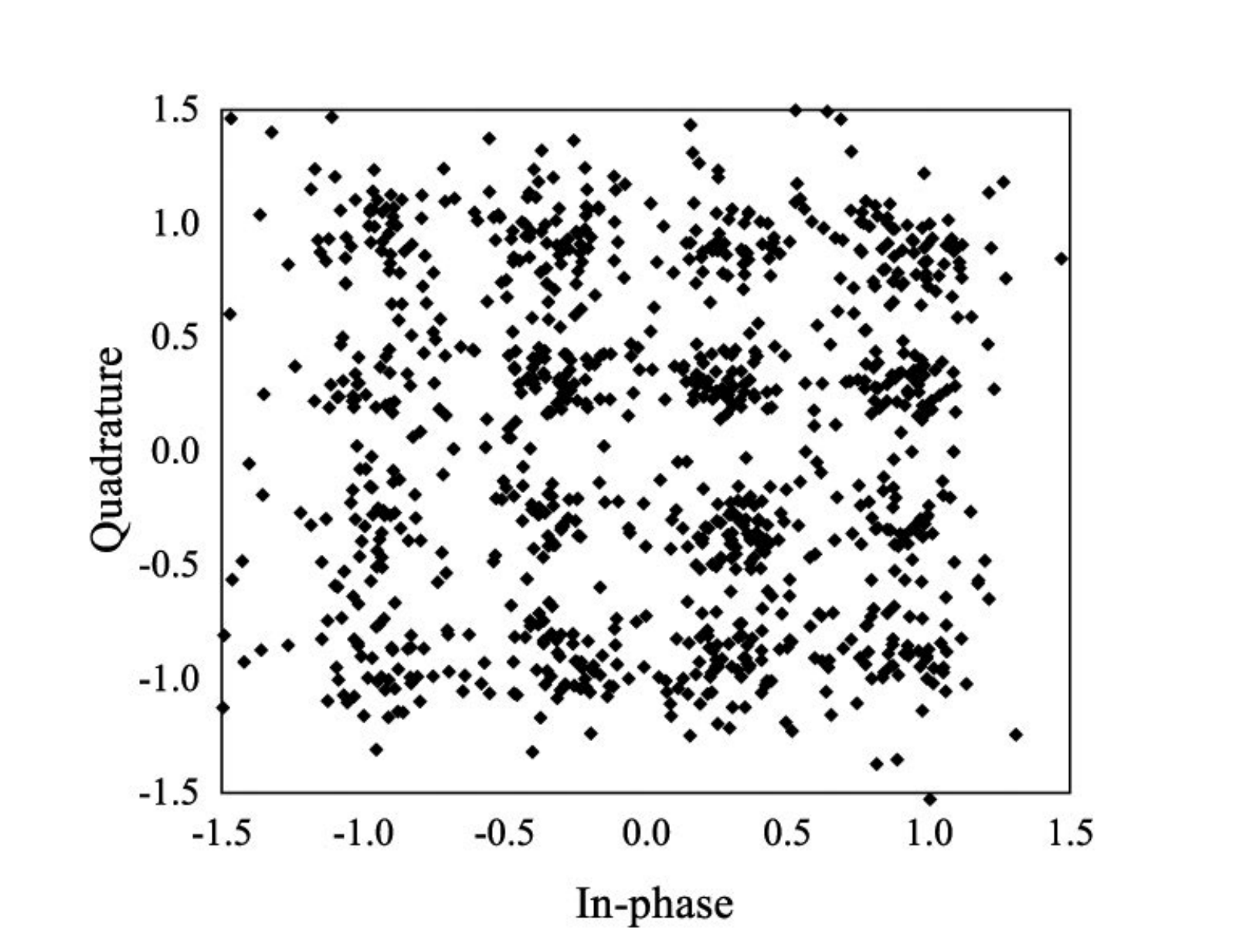}}
\subfigure[]{\label{fig:C1_Const_OPT}
\includegraphics[trim={0.2in 0in 0.5in 0.25in},clip=true,width=0.225\textwidth]{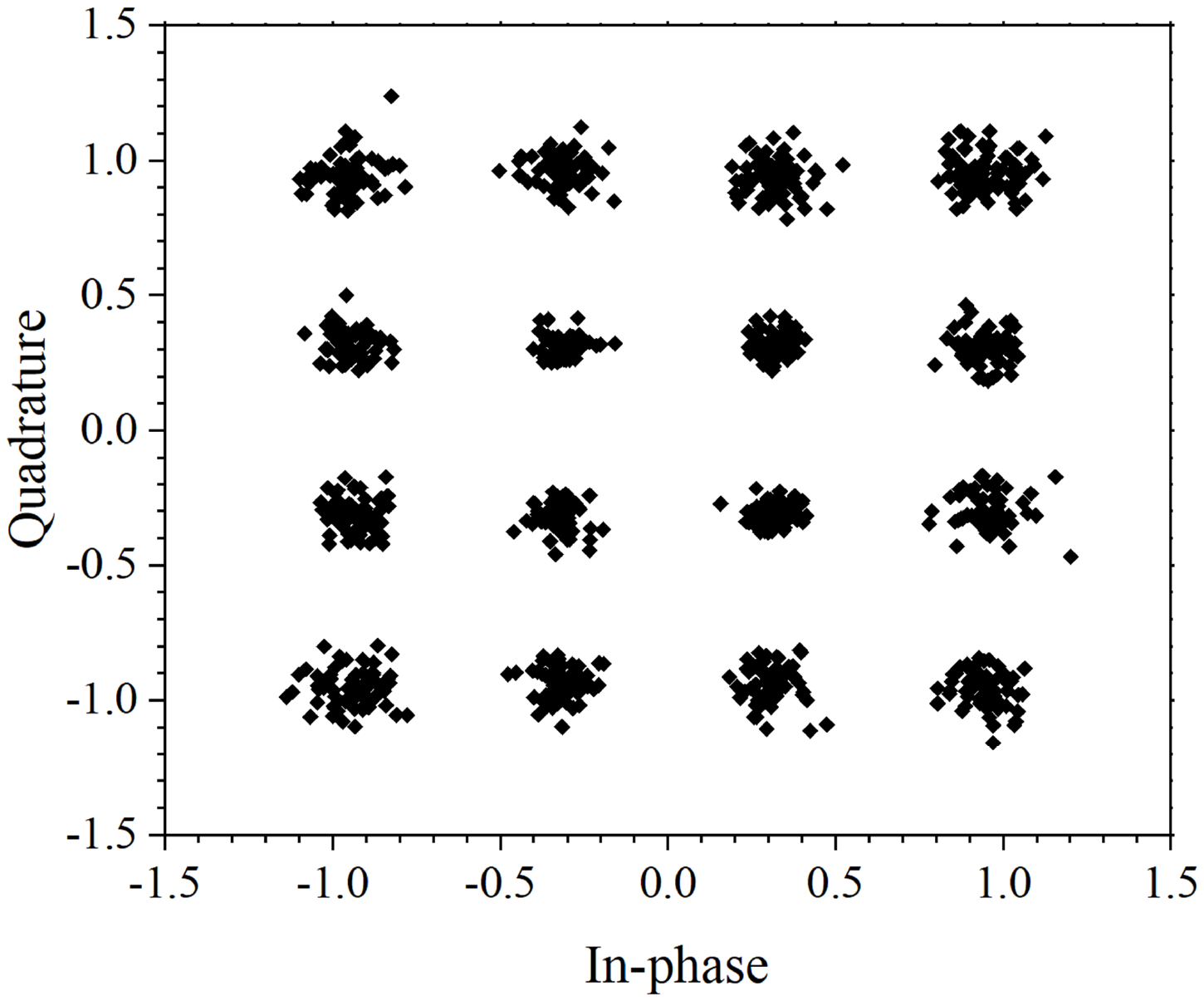}}
\vfill
\subfigure[]{\label{fig:C1_Const_M}
\includegraphics[trim={0.2in 0in 0.5in 0.25in},clip=true,width=0.225\textwidth]{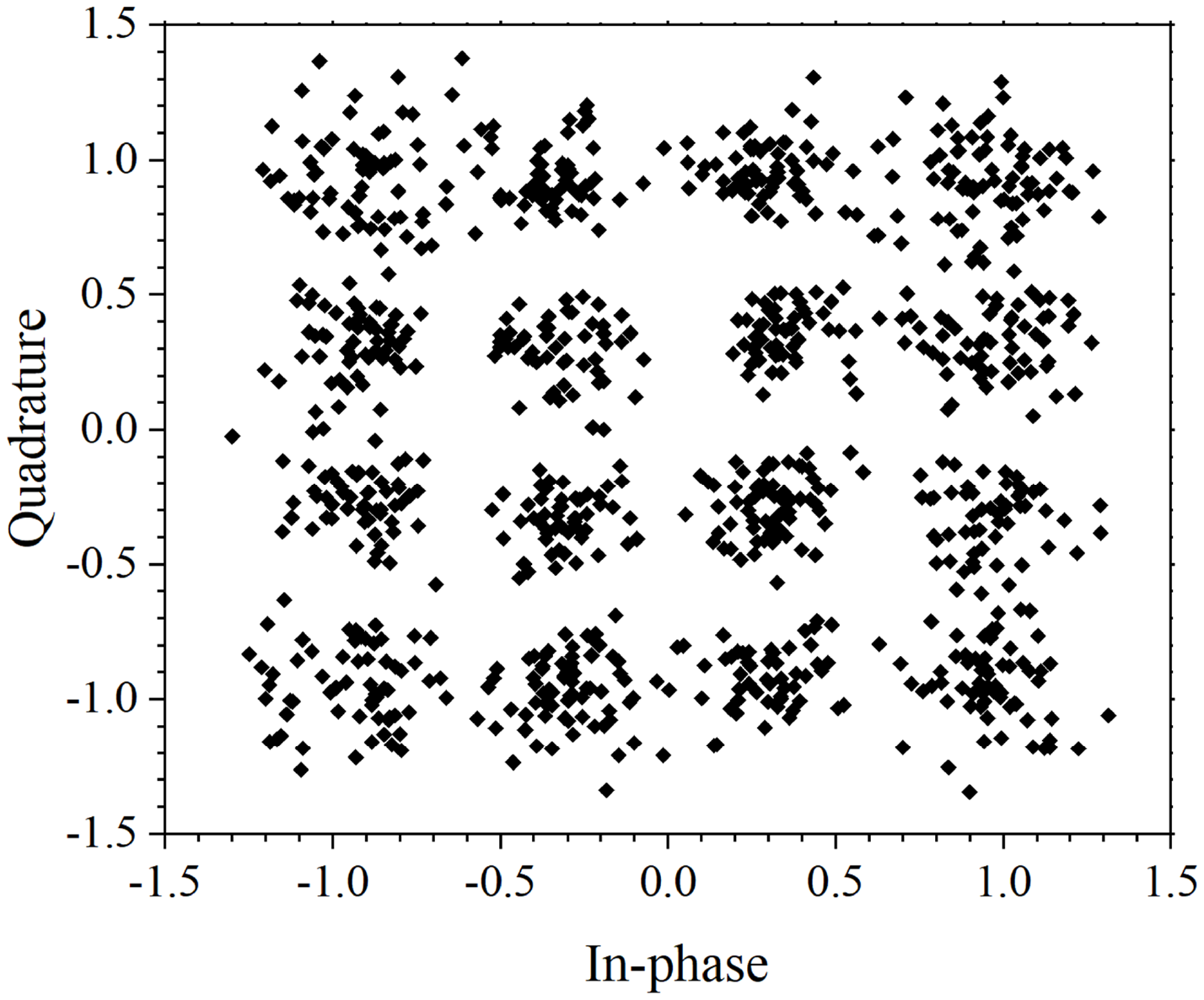}}
\subfigure[]{\label{fig:C1_Spec}
\includegraphics[trim={0.2in 0in 0.5in 0.25in},clip=true,width=0.225\textwidth]{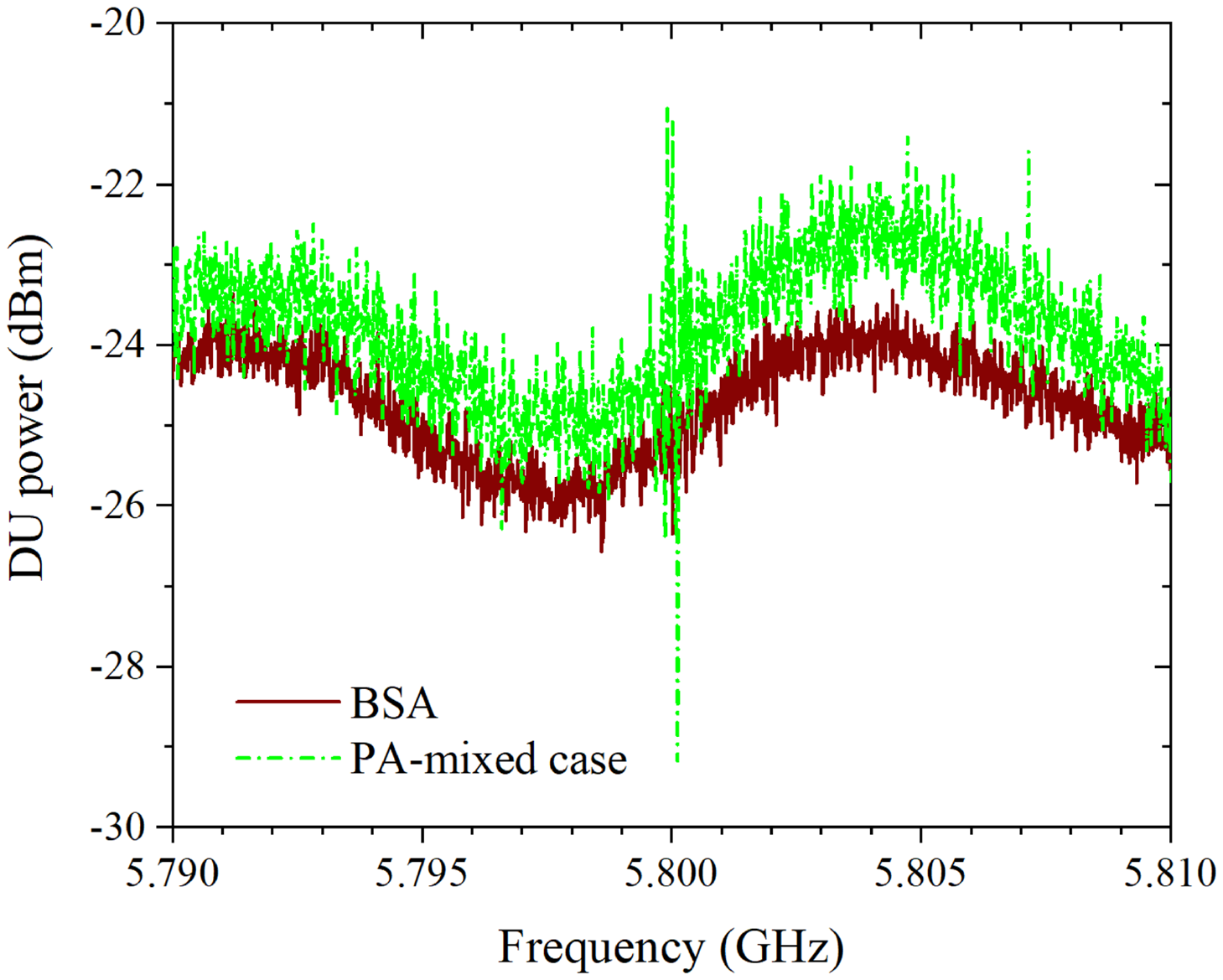}}
\caption{IRS-aided WIET experiment results. (a) Received power over the number of optimized IRS tiles. DU Constellation: (b) IRS inactive (c) with BSA (d) ``PA-mixed case'' (e) Receive power spectrum.}
\end{figure}

Fig. \ref{fig:C1_rxpow} illustrates the received power of DU and PU as a function of the number of optimized IRS tiles. Specifically, we monitored the power transmitted from Dtx to DU, denoted as ``Dtx-DU: with BSA,'' as well as the power from Ptx to PU/DU, denoted as ``Ptx-PU: with BSA'' and ``Ptx-DU: with BSA'', respectively. To highlight the effectiveness of the power suppression from Ptx to DU, we also recorded data where we used a multi-focus technique presented in Section \ref{subsec: Energy beamfomring for IRS-based WET} (i.e., the PA technique) to maximize only the Dtx-DU power and Ptx-PU power without canceling the Ptx-DU power. Unlike the previous multi-focus scenario, we apply the PA technique to the IRS to support two separate TXs (i.e., Dtx and Ptx) dedicated to two different RXs (i.e., DU and PU). Hence, we labeled these results as ``Dtx-DU: PA-mixed case'', ``Ptx-DU: PA-mixed case'', and ``Ptx-PU: PA-mixed case''. It is noteworthy that the proposed algorithm significantly suppresses the Ptx-DU power to $-48$ dBm, which is $20$ dB lower than that observed in the case ``PA-mixed case''. Additionally, the algorithm successfully maximizes both the Dtx-DU and Ptx-PU powers after its execution. The associated constellations for DU for ``IRS inactive'', ``with BSA'', and ``PA-mixed case'' schemes, can be observed in Figs. \ref{fig:C1_Const_OFF}, \ref{fig:C1_Const_OPT}, and \ref{fig:C1_Const_M} respectively. The strong interference from the Ptx in the ``PA-mixed case'' corrupted the data at the DU as shown in Fig. \ref{fig:C1_Const_M}. The clear constellation seen with BSA demonstrates the effectiveness of the proposed algorithm. The suppression effect of the proposed method can also be witnessed in the DU power spectrum as shown in Fig. \ref{fig:C1_Spec}.

\begin{figure}[t]\centering
	\subfigure[]{\label{fig:C2_PTE_scen}
		\includegraphics[trim={0.75in 1in 1in 0.5in},clip=true,width=0.235\textwidth]{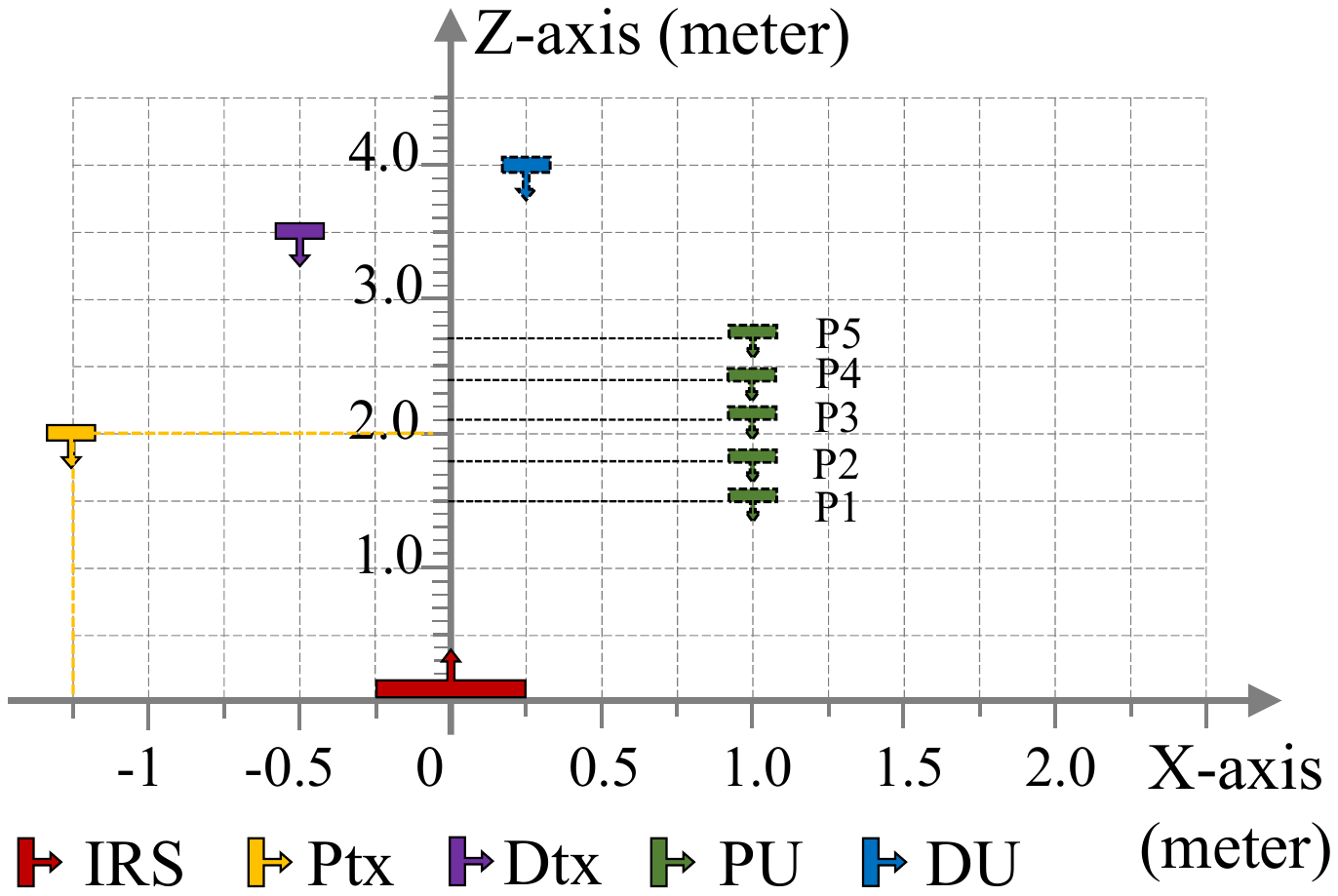}}
	\subfigure[]{\label{fig:C2_PTE}
		\includegraphics[trim={0.5in 0in 0.75in 0.25in},clip=true,width=0.22\textwidth]{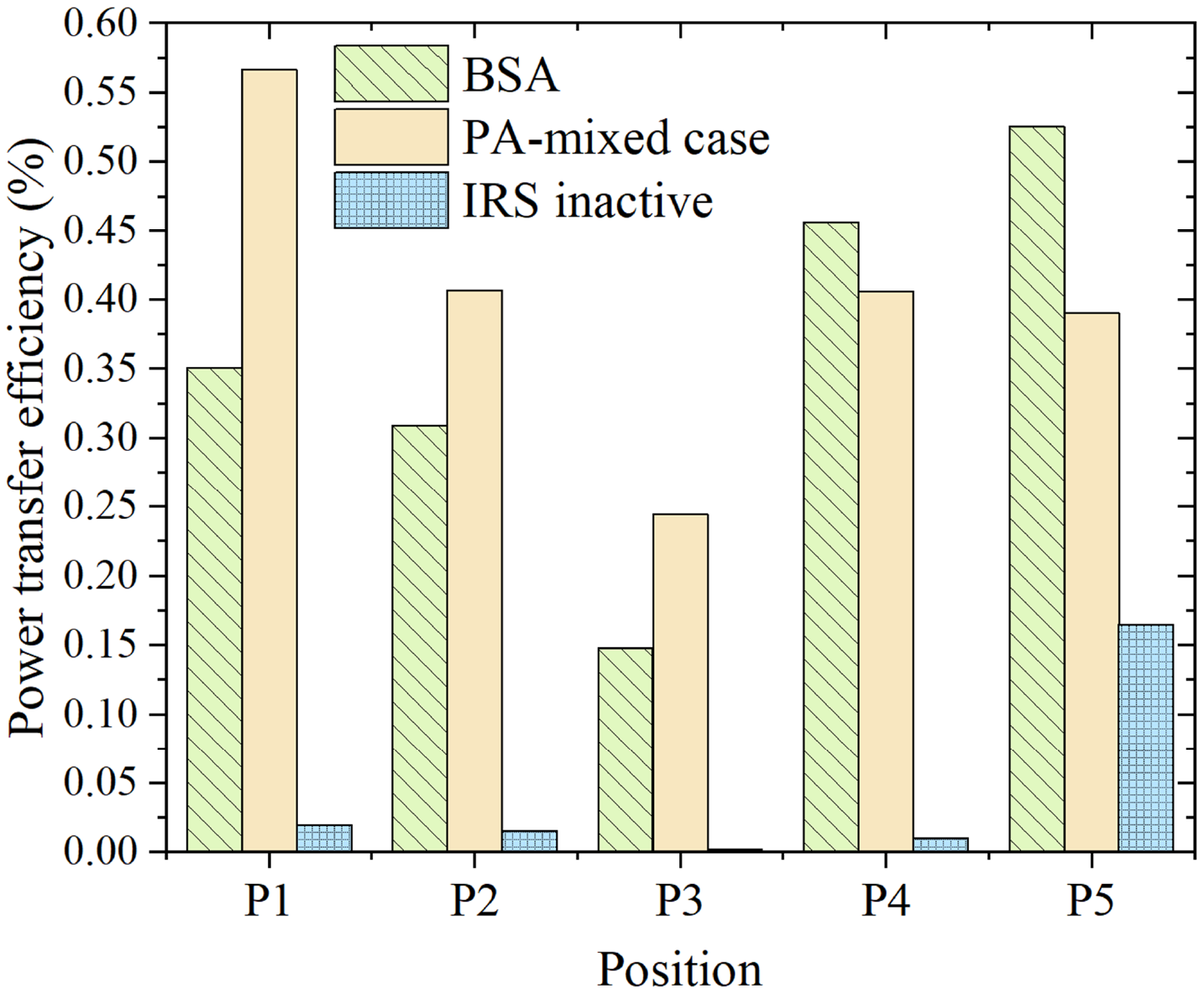}}
	\vfill
	\subfigure[]{\label{fig:C2_SNR}
		\includegraphics[trim={0.5in 0in 0.75in 0.35in},clip=true,width=0.225\textwidth]{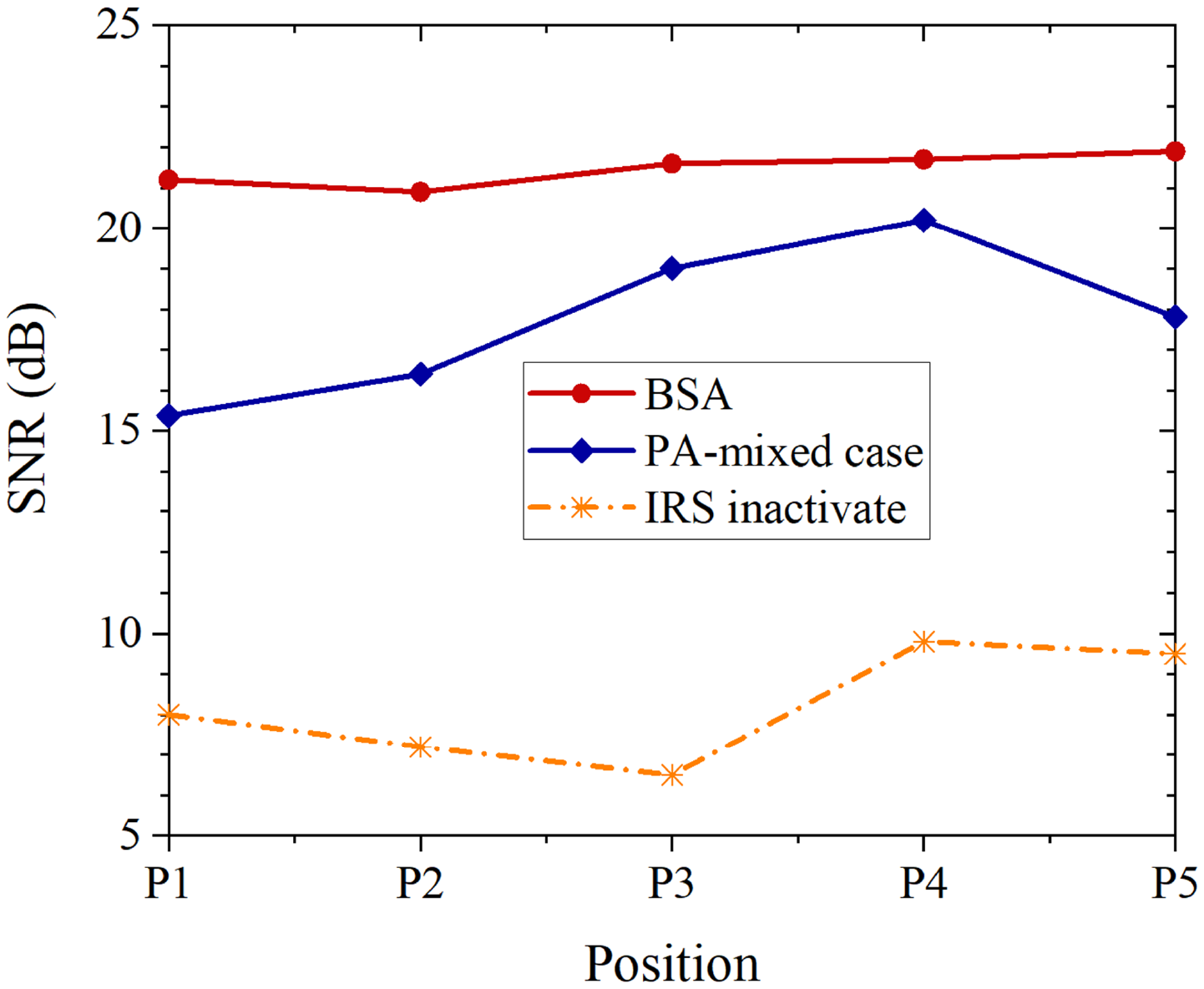}}
	\subfigure[]{\label{fig:C2_rxpow_pos}
		\includegraphics[trim={0.5in 0in 0.75in 0.35in},clip=true,width=0.225\textwidth]{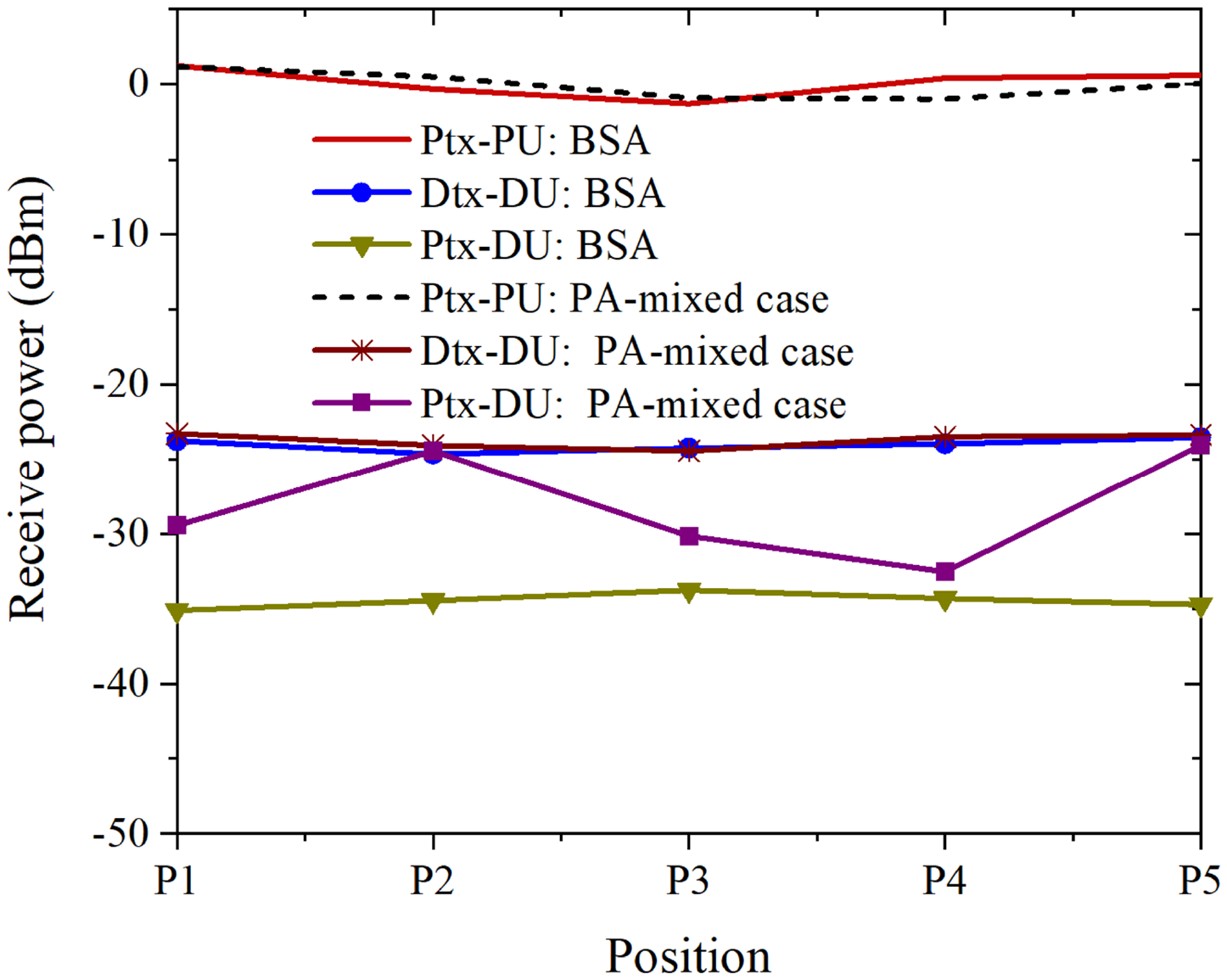}}
	\caption{IRS-aided WIET performance for various PU positions. (a) Experimental scenario. (b) PTE. (c) SNR of DU. (d) Comparison of PU's received power for the proposed and PA-mixed case.}
\end{figure}

\subsubsection{PTE Evaluation}\label{subsubsec:PTE}
We recorded the received power and calculated the PTE of the PU for different positions, from P1 to P5, in Fig. \ref{fig:C2_PTE_scen}. To guarantee a reasonable PTE, we increase the transmit power of the Ptx to $240$ mW, while the transmit power for the Dtx is again set to $3$ mW. As shown in Fig. \ref{fig:C2_PTE}, the ``PA-mixed case'' yields higher PTE at the PU than the BSA. This is because we should meet the coexistence condition of Dtx/Ptx with the BSA so the information transmission is reliably performed by suppressing the strong interference from the Ptx to the DU below a certain threshold. With this condition being met, we attempt to maximize the power delivered to the PU. In this sense, the BSA is not always optimal, in that the power delivered to the PU is not always maximum as displayed in Fig. \ref{fig:C2_PTE}, where the ``PA-mixed case'' (used in the multi-focus technique of Section \ref{subsec: Energy beamfomring for IRS-based WET} and designed for WET only) outperforms the BSA at PU positions P1, P2, and P3. However, the former may violate the coexistence condition as the interference from the Ptx is not suppressed enough to guarantee reliable information transmission, which is validated by Figs. \ref{fig:C2_SNR} and \ref{fig:C2_rxpow_pos}. The SNR acquired with the proposed algorithm consistently outperforms the ``PA-mixed case'' in all positions. Fig. \ref{fig:C2_rxpow_pos} confirms that the BSA can effectively suppress the interference from the Ptx to the DU up to $12$ dB to support the coexistence of information and power transmission, while maximizing the power delivered to the PU.

\section{Conclusion \& Future Directions}\label{sec:conc}
The integration of WIET in 6G networks is essential towards the realization of an advanced communication infrastructure, which will facilitate the co-existence of information and energy transfer to low-power devices. WIET technology will bring about new services, products, and applications, e.g., for smart cities, the health care sector (implantable biomedical devices), and consumer electronics, but will also have a positive environmental impact as it will reduce the use and replacement of batteries. Therefore, it is not surprising, that there is increasing interest from industry in WIET-related technologies \cite{powercast, ossia, energous}. There are also activities on standardization, with AirFuel Alliance leading the way with the AirFuel RF wireless power specification \cite{airfuel}, and 3GPP's working group ``Service and System Aspects'' has initiated discussions on ambient EH devices \cite{3gpp}.

In this paper, we have provided a comprehensive overview of WIET systems in the 6G era, from both the theoretical and experimental standpoints. It has been clearly demonstrated that the effective co-design of information and power signals is imperative when it comes to resource allocation optimization, waveform design, the impact of the propagation characteristics of the EM wavefronts, the employed antenna technology, and the considered frequency band. Such a dedicated design framework is needed for achieving the QoS requirements of low-power sensor and IoT devices with respect to data rates and EH. Despite the recent research advances on all of the topics discussed in this paper, many challenges and open questions still exist, which are elaborated in detail below.

\subsection{WIET Modeling \& Communication Aspects}
Firstly, more accurate mathematical models for rectenna circuits are needed, aiming towards a better balance between the ``simplistic'' communication theory models and the ``complicated'' microwave theory models. This, in turn, will facilitate obtaining insights for the design of WIET systems. Indeed, as illustrated in Subsection \ref{sec:waveform}, a potential mismatch in the model can lead to a resource allocation optimization with significant performance losses. Moreover, WIET designs should consider specific aspects of the rectenna circuit that have an impact on EH. For instance, a design should take into account the effect of various rectenna components and non-idealities such as the matching network \cite{Ayir21} and the cut-off frequency of the RC filter \cite{psomas2022} but also the effect of different technologies such as complementary metal-oxide semiconductor (CMOS) rectifiers \cite{Elazar2023} and RTD diodes \cite{Shanin2023d}. As shown in Section \ref{sec:thz}, in contrast to conventional Schottky diodes, the non-monotonic behavior of RTDs implies that maximizing the average input power at the RX does not maximize the EH, which certainly influences the fundamental information-theoretic limits but also the resource allocation and waveform design.

All current waveforms for WIET mainly refer to specific signal shapes, e.g., multisine, chaotic signals, etc., in order to simplify their design and optimization (see Subsection \ref{sec:waveform}). Hence, a novel mathematical framework/formulation is needed for the design of WIET waveforms without these limitations. In addition, we need to design suitable waveforms that take into account frequency and bandwidth aspects, spectral leakage due to non-linearities, etc., \cite{Ayir21}, which will facilitate more efficient WIET. Safety constraints with respect to human exposure to EM fields are also an important direction for future investigation. Although some initial works on WIET exploiting 6G technologies showed that operating in the near-field \cite{lopez2023} and over higher frequency bands \cite{psomas2022pieee} can be favorable in terms of EMF exposure, further research is required for developing proper WIET designs under safety constraints. For IRS-assisted wireless systems, the positive effect of proper phase shift design on EMF exposure has been demonstrated \cite{ibraiwish2022} but the resulting impact on the performance of IRS-aided WIET is still unclear. Furthermore, the potential benefits and/or limitations of H-MIMO and RAs under EMF exposure constraints are open questions.

\subsection{Antenna Technologies for WIET}
The design of H-MIMO WIET systems requires novel mathematical optimization tools compared to those utilized in MIMO and M-MIMO networks. The channel modeling for H-MIMO should take into account the electrical coupling between the TX antenna elements due to the negligibly small antenna spacing \cite{Yuan2023}. Furthermore, the hardware architecture of H-MIMO transceivers differs substantially from that of traditional MIMO and may impose new physical constraints, which are not present in MIMO/M-MIMO \cite{Huang2020}. To fully exploit the benefits of H-MIMO, the transmit and receive signals are processed in the analog EM domain, which requires a thorough analysis and optimization of EM waves as well as the synergy of classical communication theory with EM wave theory to develop advanced H-MIMO signal processing approaches \cite{Gong2023a}. Since continuous holographic surfaces have large apertures, the RXs may be located in the radiative near-field, where the far-field planar wave assumption does not hold and the spherical-wave propagation of realistic EM waves should be taken into account \cite{An2023P1, Demir2022} (as discussed in detail in Subsection \ref{sec:nearfield}). The impact of the spherical nature of the EM wavefronts has far-reaching consequences on the design of WIET systems operating in the Fresnel region. While large performance gains are possible through beam focusing, several challenges compared to conventional far-field WIET systems arise. An important challenge that remains an open problem is a statistical channel model to design WIET systems that captures the multipath fading effects in the near-field. Moreover, while the additional distance information contained in the amplitude variations of the wireless channel facilitates a performance gain via beam focusing in the Fresnel region, which is exploited for efficient WIET, it increases the difficulty of estimating the channel. Therefore, conventional far-field channel estimation approaches may fail or become infeasible since, in contrast to the far-field channel, the near-field channel is not sparse in the angular domain. Finally, as the wireless propagation environment is fundamentally different in the near-field compared to the far-field, the design of the waveform is an important issue to be addressed.

The performance of passive IRS-based systems relies on perfect CSI knowledge in order to achieve beamforming towards the RX(s). However, channel estimation becomes challenging when the number of elements and/or the number of RXs increases \cite{10159024}. What is more, in WIET systems consisting of low-power devices with low-computational capabilities, the transmission of pilots can be impractical. As such, novel estimation techniques, based on machine learning tools or signal processing techniques (e.g., compressive sensing), are critical for the efficient operation of IRS-assisted WIET systems. Scenarios with multiple IRSs are also important as they can provide massive gains in both information and energy transfer. The appropriate design for the optimal deployment of the multiple IRSs, the optimization of their phase shifts as well as the consideration of near-field propagation effects can increase the performance of WIET RXs but also increase the overall energy efficiency of WIET systems. Moreover, the fully-connected active IRS architecture, which integrates reflect-type power amplifiers with the antenna elements in order to compensate for the product path loss of the cascaded IRS-aided channels, has been studied to a limited extend in the context of WIET \cite{IRSAct5,IRSAct6,IRSAct7}. Likewise, the utilization of other novel IRS structures, such as the sub-connected active \cite{Subconnected} and hybrid \cite{HybridActPasRIS,NtougiasHyb} IRS designs, which make use of a limited number of reflect-type power amplifiers to strike a more desirable balance between performance and power consumption, or the STAR-IRS architecture \cite{STAR-IRS}, where the RF signal incident on the IRS panel is subject to both reflection and transmission, require further exploration as they could help unlock additional gains for WIET.

Regarding RAs, despite some initial works on their performance, the potential gains with respect to WIET have not been fully investigated, yet. In order to fully exploit the extra degrees of freedom provided by RAs, more sophisticated signal processing and communication techniques that enhance the data rate and EH are required. As illustrated in Subsection \ref{sec:ras}, no configuration (i.e. position of the liquid) maximizes both information and energy transfer \cite{kit2023b}. Therefore, it is important to develop schemes that satisfy the WIET constraints, either in the time domain (through TS) or in the power domain (through PS). Moreover, multi-RA architectures at the RX can further improve WIET performance with the employment of AS. Since an RA can have a large number of configurations, a significant challenge in RA systems is channel estimation. As such, low-complexity channel estimation algorithms will play a critical role in attaining the expected performance of RAs in WIET systems. Initial works on movable RAs have focused on a finite set of configurations, which may not be true for liquid-based RAs. The theoretical investigation of a continuous set of configurations is challenging but will provide important insights and opportunities in WIET scenarios \cite{Psomas2023b}. The design and analysis of WIET-based RA MIMO is also a significant issue and remains an open problem in the literature. Due to the complexity that arises from RA MIMO setups, usually leading to NP-hard optimization problems (e.g., select the configurations that maximize the capacity and/or EH), low-complexity and near-optimal signal processing and communication techniques are essential.

\subsection{THz WIET Systems}
Even though THz antennas can be very small, the development of micro-scale THz transceivers and signal processing units and their on-chip integration with microscopic antennas is challenging \cite{Villani2021}. A few prototypes based on graphene, SiGe, and GaAs have recently been reported in the literature \cite{Rana2008, Momeni2011, Ma2019_1, Asada2008} but still suffer from significantly lower stability and higher phase noise compared to modern devices operating at lower frequencies \cite{Yi2021}. Furthermore, the design of microscopic energy storage units for buffering harvested energy, e.g., rechargeable batteries or super-capacitors, is also necessary for the development of micro-scale IoT WIET systems. Due to the microscopic size of IoT devices, the computational capability of their signal processing units is rather low compared to their needs \cite{Tataria2021, Sarieddeen2021}. This limitation may not allow the practical implementation of the complicated source and channel coding methods and beamforming algorithms that are utilized for WIET at lower frequencies (typically, below $\SI{6}{\giga\hertz}$) \cite{Vaezi2022, Shanin2021a, Shanin2023}. Additionally, the phase noise and instability of practical THz local oscillators, signal mixers, and modulators further limit the applicability of complex signal modulation and multiplexing schemes, e.g., QAM and OFDM \cite{Yi2021}. Therefore, low-complexity schemes, such as unipolar ASK modulation, have been proposed for THz band communications \cite{Sarieddeen2021}. In principle, the huge path loss of THz channels can be compensated by focusing the transmit signal on the RX. However, the use of large aperture antennas at both the TX and RX may not be desired in future micro-scale 6G IoT networks \cite{Yi2021}. Furthermore, due to the distance-dependent path loss, efficient WIET is possible only if the RX is located in the range of $\SI{1}{\centi\meter} - \SI{100}{\centi\meter}$ from the TX \cite{Villani2021, Rong2017}. Since the Fraunhofer distance scales with frequency, the spherical wavefront of EM waves has to be considered, when the RX is in close proximity of the TX \cite{Zhang22}. Finally, as the number of objects located near the TX is typically small and longer alternative paths to the LoS are strongly attenuated, scattering in wireless THz channels is low and can often be neglected \cite{Zhang22}. Thus, the electrical characteristics of THz band communication channels are different from those at lower frequencies and must be taken into account when designing THz WIET systems.

\bibliographystyle{IEEEtran}
\bibliography{references,references2}
\end{document}